\pdfoutput=1

\documentclass[twocolumn,twoside,nofootinbib,showpacs,prd,aps,tightenlines,10pt]{revtex4-1}

\usepackage{amsmath}
\usepackage{amssymb}
\usepackage{slashed}
\usepackage{graphicx}
\usepackage{xspace}
\usepackage{braket}
\usepackage{multirow}
\usepackage[hyperfootnotes=false,bookmarks=false]{hyperref}

\newcommand{\eq}[1]{Eq.~\eqref{eq:#1}}
\newcommand{\eqs}[2]{Eqs.~\eqref{eq:#1} and \eqref{eq:#2}}
\renewcommand{\sec}[1]{Sec.~\ref{sec:#1}}

\newcommand{\subsec}[1]{Sec.~\ref{subsec:#1}}
\newcommand{\subsecs}[2]{Secs.~\ref{subsec:#1} and \ref{subsec:#2}}
\newcommand{\fig}[1]{Fig.~\ref{fig:#1}}

\def\nn{\nonumber \\ }

\def\abs#1{\left| #1 \right| }

\newcommand{\ord}[1]{{\mathcal O}(#1)}

\newcommand{\tr}{\mathrm{tr}}

\def\df{{\rm d}}
\def\rd{{\rm d}}
\newcommand{\img}{\mathrm{i}}
\def\bn{\overline n}
\def\nslash{\slashed{n}}
\def\bnslash{\slashed{\bn}}
\def\lra{\leftrightarrow}
\newcommand{\sdt}{\!\cdot\!}
\def\bq{\overline q}

\def\al{\alpha}
\def\bt{\beta}
\def\ga{\gamma}
\def\Ga{\Gamma}
\def\de{\delta}
\def\De{\Delta}
\def\eps{\epsilon}
\def\La{\Lambda}
\def\si{\sigma}
\def\om{\omega}

\newcommand{\sieff}{\si_\mathrm{eff}}
\newcommand{\GeV}{\:\mathrm{GeV}}
\newcommand{\lqcd}{\Lambda_\mathrm{QCD}}

\newcommand{\HOPPET}{\textsc{HOPPET}\xspace}

\allowdisplaybreaks[1]

\begin{document}

\title{A QCD Analysis of Double Parton Scattering:\\  Spin and Color Correlations, Interference Effects and Evolution}

\author{Aneesh V.~Manohar}

\author{Wouter J.~Waalewijn}

\affiliation{Department of Physics, University of California at San Diego,
  La Jolla, CA 92093\vspace{4pt} }

\begin{abstract}
 We derive a factorization formula for the double Drell-Yan  cross section in terms of double parton distribution functions (dPDFs). Diparton flavor, spin and color correlations and parton-exchange interference terms contribute, even for unpolarized beams. Soft radiation effects are nontrivial for the color correlation and interference contributions, and are described by non-perturbative soft functions. We provide a field-theoretic definition of the quark dPDFs and study some of their basic properties, including discrete symmetries and their interpretation in a non-relativistic quark model. We calculate the renormalization group evolution of the quark dPDFs and of the soft functions. The evolution  receives contributions from both ultraviolet  and rapidity divergences.  We find that color correlation and interference effects are Sudakov suppressed, greatly reducing the number of dPDFs needed to describe double parton scattering at high energy experiments.
\end{abstract}

\maketitle
\tableofcontents

\section{Introduction}

\begin{figure}
  \centering
  \includegraphics[width=0.35\textwidth]{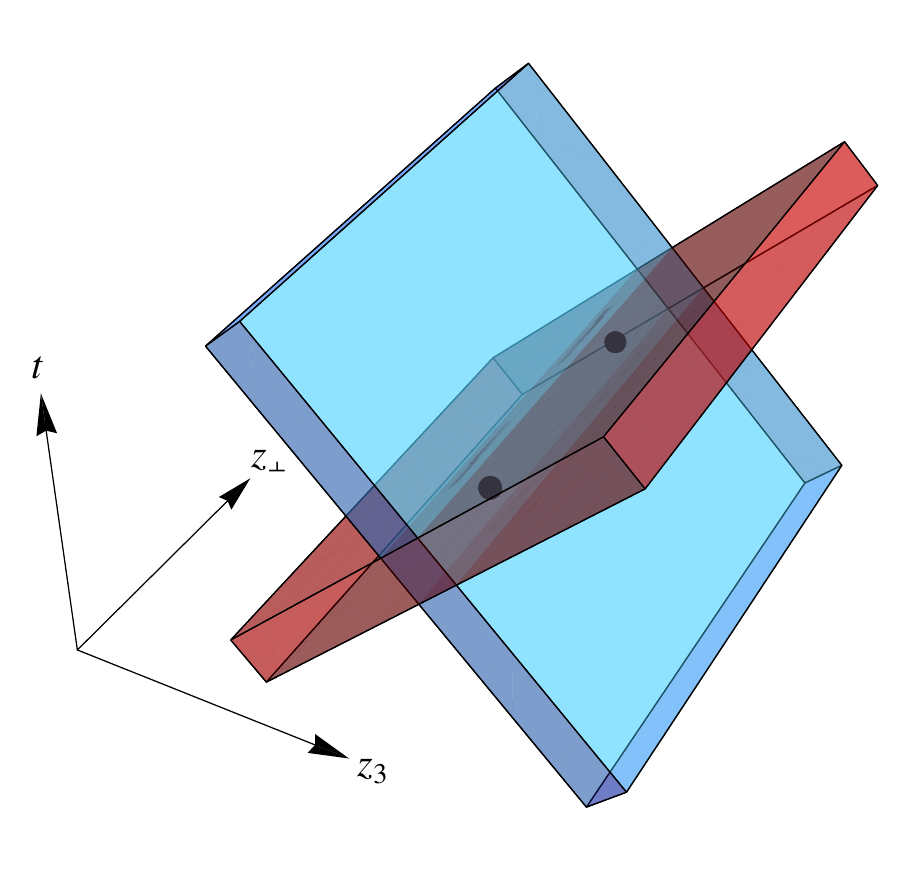}
\caption{Double parton scattering in space-time. The two hadrons have transverse size of order $1/\lqcd$ and longitudinal size of order $1/(\gamma \lqcd)$, where $\gamma \gg 1$ is the boost. The two hard interactions are shown by black dots. They have longitudinal and time separation $1/(\gamma \lqcd)$, and transverse separation $1/ \lqcd$. }
\label{fig:dps_3d}
\end{figure}

In high-energy hadronic collisions, one parton from each hadron can collide via a hard interaction to produce a final state with a large invariant mass. The classic example of such single parton scattering (SPS) is Drell-Yan production, $p_1 + p_2 \to  \ell^+ \ell^-$.\footnote{We will denote the beams by their momenta $p_{1,2}$, without specifying the hadron. The most common cases are $p \overline p$ collisions (Tevatron) or $pp$ collisions (LHC).}
 In some hadronic collisions, two partons in one hadron can have simultaneous hard interactions with two partons from the other hadron. This process is called double parton scattering (DPS).\footnote{
At the Tevatron and LHC one needs to be careful to distinguish double parton scattering from pile-up, i.e.~two single parton scatterings involving different pairs of hadrons   during the same bunch crossing, since this would produce a similar signal. Separating the two relies on identifying the vertex of the hard collision.} 
A representation of DPS in space-time is shown in \fig{dps_3d}. 
DPS was first considered in Ref.~\cite{Landshoff:1978fq,Goebel:1979mi,Takagi:1979wn,Politzer:1980me} and was subsequently studied in the context of jet production~\cite{Paver:1982yp,Humpert:1983pw,Ametller:1985tp}, double Drell-Yan~\cite{Mekhfi:1983az,Halzen:1986ue} and $W$+jets~\cite{Godbole:1989ti}.
Two examples we will study are double Drell-Yan, $p_1+ p_2 \to \ell_1^+ \ell_1^- \ell_2^+ \ell_2^-$ (where $\ell_1$ and $\ell_2$ could be the same flavor), and same sign $W$ pair production, $p_1+ p_2 \to W^+ W^+$.

DPS is higher twist, i.e.~it is suppressed by order $\lqcd^2/Q^2$ compared to single parton scattering (SPS), where $Q$ is the scale of the short-distance interactions. Heuristically, this arises because the two partons which collide in the second hard interaction have to be within a transverse area of order $1/Q^2$ of each other, whereas they could each be anywhere within the incoming hadrons (of transverse area $\sim1/\lqcd^2$).

Experimentally, DPS has been studied in four-jet events by the AFS collaboration~\cite{Akesson:1986iv} at $\sqrt{s} = 63\, \GeV$, the UA2 collaboration~\cite{Alitti:1991rd} at $\sqrt{s}=630\, \GeV$ and the Tevatron~\cite{Abe:1993rv}. At the Tevatron, DPS has also been studied in $\ga+3$ jet events \cite{Abe:1997xk,Abazov:2009gc} and there is an analysis using early LHC data for DPS in $W+2$ jets~\cite{ATLAS-CONF-2011-160}.  
In these experiments DPS is quantified using an effective cross section $\sieff$, defined in Eq.~(\ref{sigmaeff}).  The measured values  of $\sieff$ range from 5 to  15$\:\mathrm{mb}$. 

DPS is an important background for light Higgs searches in the channel $pp \to WH \to \ell \nu b\bar b$ \cite{DelFabbro:1999tf,Hussein:2006xr,Bandurin:2010gn,Berger:2011ep}. A clean channel for studying DPS at the LHC is provided by same-sign lepton searches, i.e.\ via $p_1+ p_2 \to W^+ W^+$ with $W^+ \to \ell^+  \nu$, since SPS is suppressed~\cite{Kulesza:1999zh,Cattaruzza:2005nu,Maina:2009sj,Gaunt:2010pi}. The maximum incoming partonic charge in SPS is $+1$ from $u \overline d$, so the conservation of electric charge requires the presence of at least two additional jets in the final state via $u \overline d \to W^+ W^+ \overline u d$, leading to a large suppression of SPS by $[\alpha/(4\pi)]^2[\alpha_s/(4\pi)]^2$ relative to single Drell-Yan. There is no corresponding suppression of DPS, which can proceed via $u u \overline d \overline d \to W^+ W^+$. After typical cuts the DPS cross section is of fb order~\cite{Maina:2009sj,Gaunt:2010pi}, making it more of a long-term goal at the LHC.

In the original DPS formalism, the cross section is described as~\cite{Paver:1982yp}
\begin{eqnarray} \label{eq:si_old}
  \df \si &=& \frac{1}{S} \sum_{i,j,k,l} \int\! \df^2 \mathbf{z_\perp}\, F_{ij}(x_1,x_2,\mathbf{z_\perp},\mu) F_{kl}(x_3,x_4,\mathbf{z_\perp},\mu)  \nn
  && \times \hat \si_{ik}(x_1 x_3 \sqrt{s},\mu) \hat \si_{jl}(x_2 x_4 \sqrt{s},\mu)
\,.\end{eqnarray}
Each incoming hadron is described by a double parton distribution function (dPDF) $F$ and the short-distance processes are described by partonic cross sections $\hat \si$, in analogy with SPS. $F_{ij}(x_1,x_2,\mathbf{z_\perp})$  is the number density for simultaneously finding two partons with flavors $i,j=g,u, \bar u, d, \dots$, longitudinal momentum fractions $x_1,x_2$ and transverse separation $\mathbf{z_\perp}$ inside the hadron. Our convention is that in formul\ae\ such as Eq.~(\ref{eq:si_old}), the first (second) dPDF is for the beam with momentum $p_1$ ($p_2$). The $\hat \si_{ik}(x_1 x_3 \sqrt{s})$ is the partonic cross section for partons $i, k$ going to the desired final state. $S$ is a symmetry factor that can arise if there are identical particles in the final state.

\begin{figure}
  \centering
  \includegraphics[width=0.47\textwidth]{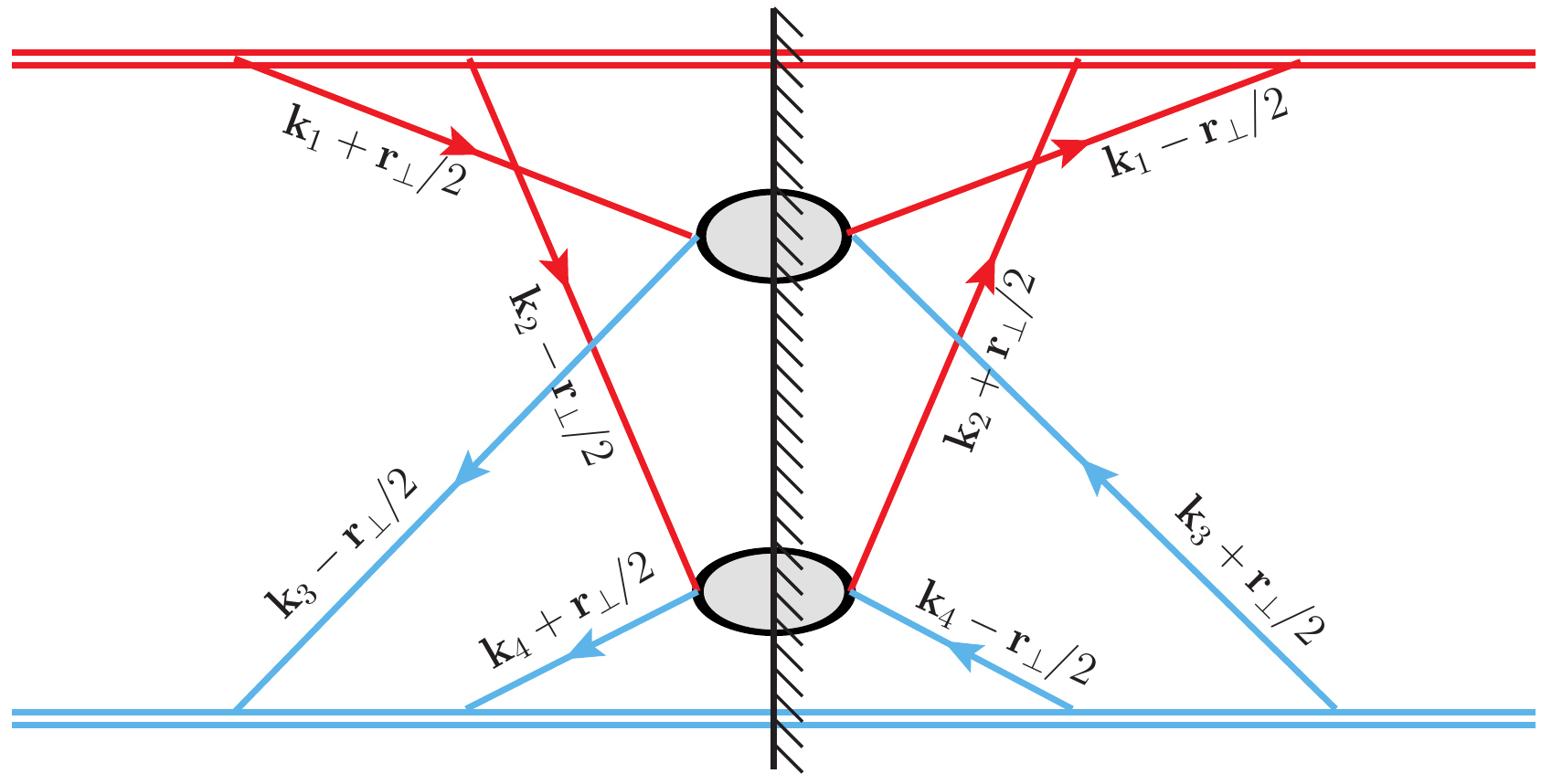}
\caption{Double PDFs in momentum space (forward diagram).}
\label{fig:dps_rperp}
\end{figure}

The dPDF in momentum space is defined by\footnote{We follow the convention of Refs.~\cite{Diehl:2011tt,Diehl:2011yj}, where $\mathbf{r}_\perp$ is a transverse momentum, not a coordinate.}
\begin{eqnarray} 
\!\!\!\!\! F_{ij}(x_1,x_2,\mathbf{r_\perp},\mu)  &=& \!\int\! \rd^2 \mathbf{z_\perp} e^{-i \mathbf{r_\perp \cdot z_\perp} } F_{ij}(x_1,x_2,\mathbf{z_\perp},\mu)
\,.\end{eqnarray}
It should be clear from the context whether the third argument refers to position or momentum. The  dPDF in momentum space is shown in \fig{dps_rperp}. The figure also shows that the dPDF is not the squared absolute value of an amplitude, since the partons on the two sides of the cut have different momenta.

It is commonly assumed that the dependence on the transverse separation is uncorrelated with the momentum fractions or parton flavors, 
\begin{equation}
  F_{ij}(x_1,x_2,\mathbf{z_\perp},\mu) = F_{ij}(x_1,x_2,\mu) G(\mathbf{z_\perp},\mu)
  \label{eq:2}
\,.\end{equation}
In addition, a factorized ansatz is made
\begin{equation} \label{eq:dPDFfactor}
  F_{ij}(x_1,x_2,\mu) = f_i(x_1,\mu) f_j(x_2,\mu)\, \theta(1-x_1-x_2) (1-x_1-x_2)
\,,\end{equation}
where $f$ denotes a single PDF and the last factors smoothly impose the kinematic constraint $x_1+x_2 \leq 1$. For small momentum fractions $1-x_1-x_2 \approx 1$ these factors can be neglected, and the cross section in \eq{si_old} becomes
\begin{equation}\label{sigmaeff}
  \si = \sum_{i,j,k,l} \frac{\si_{ik}\, \si_{jl}}{S\, \sieff}
  \,, \quad
  \sieff = \bigg[\int\! \df^2 \mathbf{z_\perp}\, G(\mathbf{z_\perp},\mu)^2\bigg]^{-1}
\,.\end{equation}
The effective cross-section is a measure of the area of the proton $\sim1/\lqcd^2$, consistent with the fact that DPS is higher twist.

The expression in Eq.~(\ref{eq:si_old}) is an example of a factorization formula, where the hadronic cross-section is written as the convolution of hard-scattering partonic cross-sections which are target-independent, and non-perturbative distribution functions which depend on the hadronic target. We will see that there are several important modifications to the naive expressions above. In particular, the product forms of \eqs{2}{dPDFfactor} are spoiled by QCD radiative corrections. There are also several different spin and color structures that enter the factorization formula, even for unpolarized beams~\cite{Mekhfi:1985dv,Diehl:2011tt,Diehl:2011yj}. The product form in \eq{dPDFfactor} also ignores interesting diparton flavor correlations. For example, in a naive quark model, the proton has quark constituents $uud$. One would therefore expect that the $dd$ dPDF, which measures the probability to simultaneously find two $d$ quarks in the proton, is suppressed relative to the product $d(x_1)d(x_2)$ of the $d$-quark PDFs, each of which measures the probability to find a single $d$ quark in the proton.

In processes such as double Drell-Yan, one can measure the invariant mass and transverse momentum of each lepton pair. Double Drell-Yan can arise from a double parton process, $(q \overline q \to \gamma^*)+(q \overline q \to \gamma^*)$. The transverse momentum of each incoming parton is the intrinsic transverse momentum of a quark in a hadron, and is typically of order $\lqcd$. Double Drell-Yan can also arise from a single parton process such as $(g \to q \overline q) + (g \to q \overline q)$ where a quark from one gluon and an antiquark from the other gluon annihilate into a virtual photon. In this case, the transverse momentum of the quarks (and thus the lepton pair) can be of order the hard scale $Q$. There is an inevitable overlap between the SPS and DPS contributions to a physical process such as double Drell-Yan in the region where the transverse momentum of the lepton pair (not the individual leptons) is small. The total SPS contribution to double Drell-Yan is leading twist, but the contribution to the small transverse momentum region is $\lqcd^2/Q^2$ suppressed (because it is a fraction $\lqcd^2/Q^2$ of the total phase space) and is the same order as DPS in this region (see e.g.~Ref.~\cite{Blok:2011bu} for a more detailed discussion).

\begin{figure}
  \centering
  \includegraphics[width=8cm]{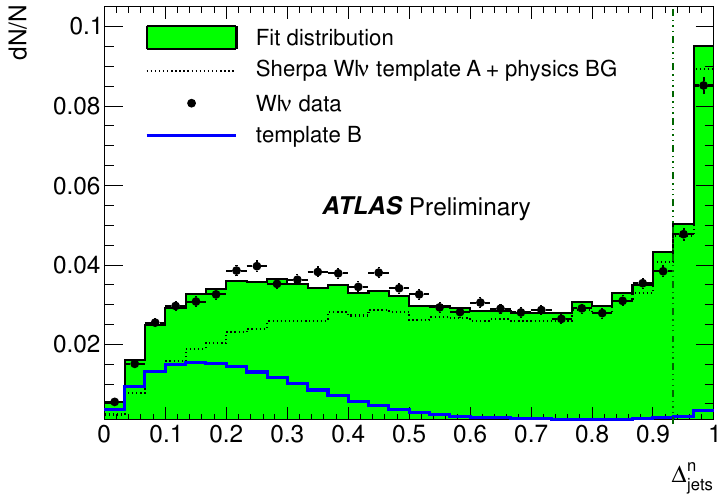}
\caption{Extraction of the double parton scattering contribution to $pp \to W+2$ jets by the Atlas collaboration \cite{ATLAS-CONF-2011-160}. The extraction is performed by comparing the observed spectrum of the normalized total transverse momentum of the two jets $\De_\text{jets}^n = |\mathbf{p}_{1\perp}+\mathbf{p}_{2\perp}|/(|\mathbf{p}_{1\perp}|+|\mathbf{p}_{2\perp}|)$ to templates for SPS and DPS obtained from Monte Carlo programs. (ATLAS Experiment \textcopyright \ 2012 CERN.)}
\label{fig:atlas}
\end{figure} 
\begin{figure}[t]
  \centering
  \includegraphics[width=0.3\textwidth]{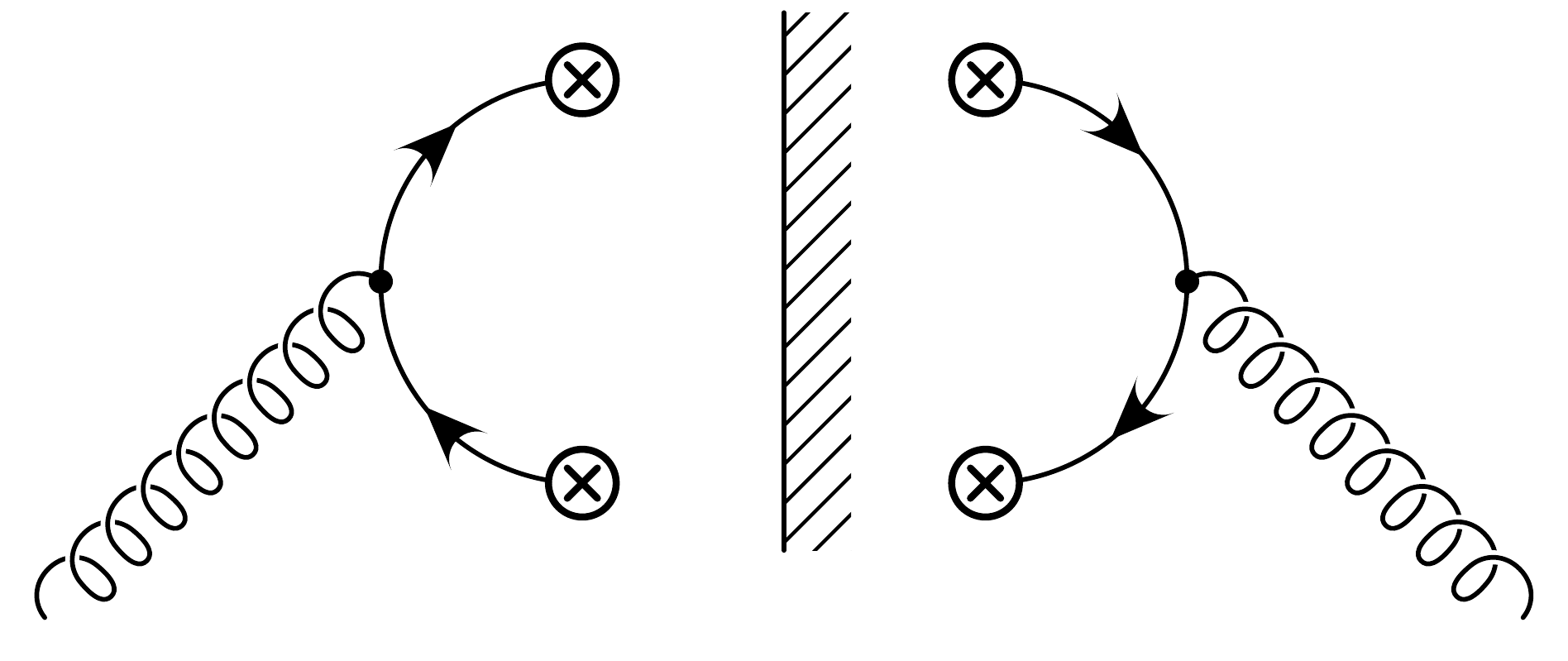}
\caption{Mixing between the gluon single PDF and the $q \overline q$ double PDF.}
\label{fig:mixing}
\end{figure} 

In light of the above discussion, observables that are sensitive to the regions of phase space where SPS and DPS are of the same order, are used to separate SPS and DPS contributions~\cite{Humpert:1984ay}.  In a recent analysis by the Atlas collaboration~\cite{ATLAS-CONF-2011-160}, DPS was studied in $pp \to W + 2$ jets, for a leptonically decaying $W$. The normalized total transverse momentum of the two jets, $\De_\text{jets}^n = |\mathbf{p}_{1\perp}+\mathbf{p}_{2\perp}|/(|\mathbf{p}_{1\perp}|+|\mathbf{p}_{2\perp}|)$, was used to separate DPS from SPS, and is shown in Fig.~\ref{fig:atlas}.

The dPDF is a new non-perturbative object. It has a renormalization group (RG) evolution similar to that for the conventional (single) PDF. The evolution of $F_{ij}(x_1,x_2,\mu)$ was determined a  long time ago in Refs.~\cite{Kirschner:1979im,Shelest:1982dg}. It has recently been extended to include the $\mathbf{z_\perp}$ dependence~\cite{Diehl:2011tt,Diehl:2011yj}. The RG evolution for the $q \overline q$ dPDF is
\begin{widetext}
\begin{eqnarray}
  \mu \frac{\df}{\df \mu} F_{q\bar q}(x_1,x_2,\mathbf{r_\perp},\mu) &=& \frac{\al_s C_F}{\pi} \bigg\{\sum_{i=q,\bar q,g} \int\! \frac{\df z}{z}\, \bigg[P_{qi}\Big(\frac{x_1}{z}\Big) F_{i \bq}(z,x_2,\mathbf{r_\perp},\mu) 
 + P_{qi}\Big(\frac{x_2}{z}\Big) F_{qi}(x_1,z,\mathbf{r_\perp},\mu)\Big] \nn
  && + P_{qg}\Big(\frac{x_1}{x_1+x_2}\Big) \frac{f_g(x_1+x_2,\mu)}{x_1+x_2} \bigg\}
\,,\end{eqnarray}
\end{widetext}
where $P$ denotes the usual  PDF splitting functions.  The first two terms describe the independent evolution of each of the partons with the standard (single) PDF kernel. The second line is the contribution of the single gluon PDF feeding into the dPDF via $ g \to q \overline q$, shown in \fig{mixing}. 
This term has no $\mathbf{r_\perp}$ dependence, and leads to a $\mathbf{r}_\perp$ independent contribution to $F(x_1,x_2,\mathbf{r_\perp},\mu)$, or equivalently to a $\delta^{(2)}(\mathbf{z_\perp})$ contribution to $F(x_1,x_2,\mathbf{z_\perp},\mu)$. This $\delta$-function in transverse position space leads to a divergence in the cross section Eq.~(\ref{eq:si_old}),
as discussed in Refs.~\cite{Diehl:2011tt,Diehl:2011yj}. The resolution of this singularity is related to the issue of double counting between SPS and DPS mentioned earlier, and will be discussed in a subsequent paper~\cite{Manohar:DPS2}. The double-counting problem only enters through the mixing between single and double PDFs, which we therefore also postpone to Ref.~\cite{Manohar:DPS2}.  The factorized form in \eq{dPDFfactor} for the dPDF is not preserved by the evolution~\cite{Snigirev:2003cq,Korotkikh:2004bz}. Sum rules for dPDFs were derived and used to find a new ansatz that satisfies the evolution equation reasonably well~\cite{Gaunt:2009re}. 

In DPS, the two partons extracted out of the proton can be correlated in spin and color, as was first discussed in Ref.~\cite{Mekhfi:1985dv}. These correlations are not present in phenomenological models and were recently revisited in Refs.~\cite{Diehl:2011tt,Diehl:2011yj} for transverse momentum dependent dPDFs.

\begin{figure}
  \centering
  \includegraphics[width=0.4\textwidth]{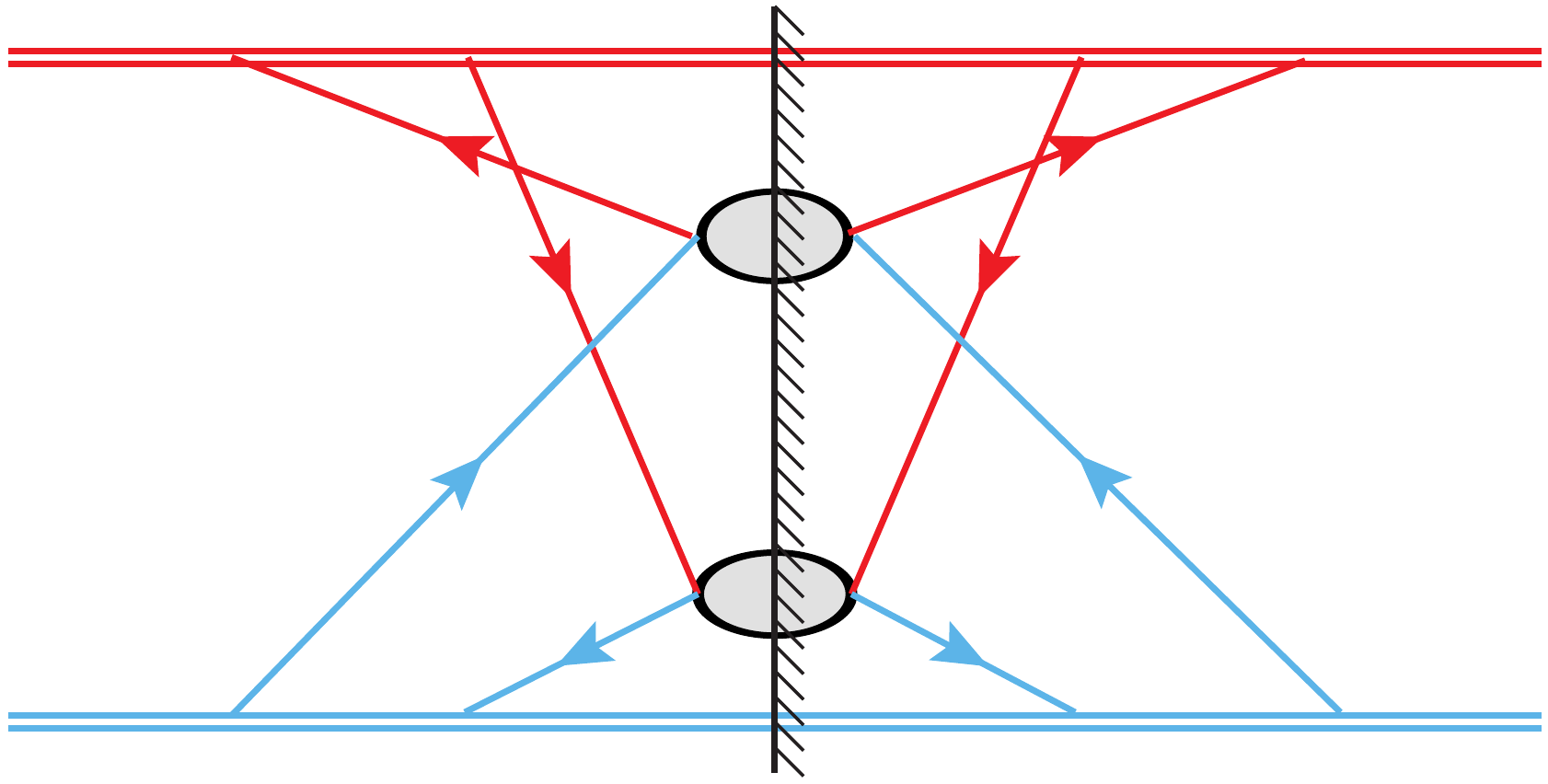}
\caption{Interference contribution to DPS. Note the orientation of the fermion lines.}
\label{fig:int}
\end{figure}

In this paper our main focus will be on formulating the QCD factorization theorem for double Drell-Yan and related processes in terms of dPDFs, and on studying the RG evolution of the quark dPDFs. We find that the color-correlated dPDFs contain rapidity divergences, which are tied to the presence of large rapidity logarithms of $\mathbf{r_\perp^\mathrm{2}}/Q^2$.  We will treat the rapidity divergences and resum the corresponding series of logarithms using the recently introduced rapidity renormalization group~\cite{Chiu:2011qc,Chiu:2012ir}. We  find that the effects of soft gluon exchange do not cancel for the color-correlated dPDF: the color-correlated dPDFs contribute to the cross-section in combination with a non-perturbative soft function. The RG evolution shows that the color-correlated dPDF is Sudakov suppressed, in agreement with Ref.~\cite{Mekhfi:1988kj}. In addition, we will also study the interference contributions such as shown in \fig{int}, which were first considered in Refs.~\cite{Diehl:2011tt,Diehl:2011yj}. We find that all interference dPDFs are Sudakov suppressed. These conclusions greatly reduce the number of possible parton distributions that contribute to DPS at high energies. There is  significant overlap of Secs.~\ref{sec:fact} and \ref{sec:dPDF} with the topics covered in a recent paper~\cite{Diehl:2011yj} which appeared while this work was in progress. Ref.~\cite{Diehl:2011yj} focuses on double parton scattering where the transverse momenta of the final state particles are measured. The evolution of the standard dPDFs, which we calculate in \sec{RGE}, differs from the evolution of transverse-momentum-dependent dPDFs obtained in Ref.~\cite{Diehl:2011yj}. This is similar to the situation in regular Drell-Yan, which is described by PDFs if the lepton transverse momentum is not measured (or integrated over), and by transverse-momentum-dependent PDFs if it is. The two are not simply related to each other, because they are renormalized differently.

In \sec{pheno}, we discuss the phenomenological aspects and implications of our results. The remainder of the paper contains the more technical aspects of our work. We present a systematic derivation of the factorization formula for the DPS contribution to double Drell-Yan production in \sec{fact}, after reviewing the steps that lead to the well-known factorization theorem for single Drell-Yan. In contrast to \eq{si_old}, we will include spin and color correlations and interference contributions, which naturally arise. In \sec{dPDF} we define the quark dPDF, classify its spin and color structures, study its properties under discrete symmetries and give  an interpretation in the context of a quark model. The calculation of the RGE for the quark dPDFs and the corresponding soft functions is given in \sec{RGE}. It includes a brief introduction to the topic of rapidity divergences and the technology of the rapidity renormalization group with explicit examples. All spin and color correlations as well as interference effects are considered.  We conclude in \sec{conc}. 

\section{Phenomenology}
\label{sec:pheno}

In this section we give an overview of our work and its phenomenological implications. The technical details can be found in the remainder of the paper. As mentioned in the introduction, flavor, spin and color correlations can appear for DPS. Already in SPS there are different spin dependent quark PDFs. For example, the polarized distribution function $\Delta q(x)$ measures the number of right minus the left-handed quarks with momentum fraction $x$, i.e.~the longitudinal polarization. It contributes to the polarized structure function $g_1(x)$ that is measured in polarized deep-inelastic scattering. The transversity distribution $h_1(x)$ measures the transverse quark polarization $\delta q(x)$. It is a chiral-odd distribution, and does not contribute to polarized deep-inelastic scattering, but does contribute to polarized Drell-Yan with transversely polarized beams~\cite{Ralston:1979ys,Jaffe:1991ra}.

The quark spin in a proton is correlated with the hadron spin, so that both $\Delta q(x)$ and $\delta q(x)$ vanish for unpolarized proton targets. By contrast, in DPS the spins of the two partons can be correlated with each other, and so nontrivial spin structures exist even if the proton is unpolarized. In addition to the usual unpolarized distribution $F_{qq}$, there are $F_{\Delta q \Delta q}$, and $F_{\delta q \delta q}$, which measure longitudinal and transverse spin correlations between two partons. $F_{\delta q \delta q}$ is chiral odd in each $\delta q$, but has overall chirality zero. The dependence of the dPDF on $\mathbf{z}_\perp$ allows for several more spin structures, but these do not contribute to the cross sections we consider.

In addition, dPDFs can also have color correlations which have no analog in SPS. The regular quark PDF is (schematically) the hadron matrix of the quark bilinear $\overline q q$. For dPDFs, there are two possible color structures $\overline q q\, \overline q q$ and $\overline q T^A q\, \overline q T^A q$. Both objects are overall color singlets, but they give information on parton color correlations in the hadron target. We will refer to $1\otimes1$ and $T^A \otimes T^A$ as the color-summed and color-correlated dPDFs. The two operators can also be written as linear combinations of $\overline q q\, \overline q q$ and $\overline q^\alpha q_\beta \, \overline q^\beta q_\alpha$, which can be thought of as  color-direct and color-exchange contributions.

Two quarks can be in a $\mathbf{6}$ or $\mathbf{\overline{3}}$ color representation, which contribute to two different dPDFs $F_{qq}^{(6)}$ and $F_{qq}^{(\overline{3})}$, measuring the color $\mathbf{6}$ and $\mathbf{\overline{3}}$ diquark distributions. In terms of the $\overline q q\, \overline q q$ and $\overline q T^A q\, \overline q T^A q$ basis,
\begin{align} \label{eq:F63}
 F_{qq}^1 &= 6 F_{qq}^{(6)} + 3 F_{qq}^{(\overline{3})} \,, \nn
 F_{qq}^T &= 4 (F_{qq}^{(6)} - F_{qq}^{(\overline{3})}) \,, 
\end{align}
where $F^1_{qq}$ is the color-summed dPDF ($1\otimes 1$)  and $F^T_{qq}$ is the color-correlated dPDF ($T^A\otimes T^A$). (The factor of 4 on the second line is due to an arbitrary normalization.) 
Since $F_{qq}^{6}$ and $F_{qq}^{\overline{3}}$ are proportional to the probabilities to find diquarks in a color $\mathbf{6}$ or $\mathbf{\overline{3}}$ and are positive, \eq{F63} implies $- 4F_{qq}^1 /3 \le  F_{qq}^T \le 2  F_{qq}^1/3$.
Similarly the $q \overline q$ dPDFs $ F_{q \overline q}^1$, $ F_{q \overline q}^8$ measure the singlet and octet  $q \overline q$ distributions. 
The complete classification of  dPDFs in terms of spin and color structures is discussed in \subsec{class}. $F^1_{qq}$ and $F^T_{qq}$ evolve differently with energy, so the color correlations are energy dependent. 

In addition to spin and color correlations, there are also interference contributions to the cross section, an example of which is shown in \fig{int}. The interpretation of this as an interference contribution becomes clear in the context of a non-relativistic quark model, as discussed in \subsec{int}. There we find that the regular dPDF $F_{q\bq}$ and the interference dPDF $I_{q\bq}$ are (roughly) given by
\begin{equation}
 F_{q\bq}^1 \sim |\phi_{q\bq}(k_1,k_2)|^2
 \,, \quad
 I_{q\bq}^1 \sim \phi_{\bq q}(k_1,k_2)^* \phi_{q\bq}(k_1,k_2)
\,,\end{equation}
where $\phi$ is the $q \overline q$ wave function and $k_1$ and $k_2$ are the momenta of the quark and antiquark. The interference dPDFs do not have a nice probabilistic interpretation. Whereas regular dPDFs $F(x_1,x_2,\mathbf{z}_\perp)$ are positive (or real, in the case of spin and color correlations), the interference dPDF does not even have to be real, as we argue in \subsec{sym}. Only the contribution of the interference dPDFs to the cross section is real. In \subsec{sym} we also discuss the properties of dPDFs under discrete symmetries. For example, charge conjugation invariance leads to $F_{qq/P} = F_{\bq\bq/\overline{P}}$, where the spin and color structures match up on both sides of this equation.

To see how spin and color correlations and interference effects contribute to the cross section, requires a refinement of \eq{si_old}. In \sec{fact} we systematically derive the formula for the DPS cross section for double Drell-Yan production, which results in the following leading-order factorization theorem
\begin{widetext}
\begin{eqnarray} \label{eq:si_new}
\frac{\df \si^\text{DPS}}{\df q_1^2\, \df Y_1\, \df q_2^2\, \df Y_2}
&=& \Big(\frac{4\pi \alpha^2 Q_q^2}{3N_c\, s}\Big)^2 \frac{1}{q_1^2 q_2^2} \int\! \rd^2 \mathbf{z}_\perp \bigg\{\big[(F^{1}_{qq}F^{1}_{\bq\bq} + F^{1}_{\Delta q \Delta q} F^{1}_{\Delta \bq \Delta \bq}) + (F^{1}_{q\bq}F^{1}_{\bq q} + F^{1}_{\Delta q \Delta \bq} F^{1}_{\Delta \bq \Delta q})\big] \nn
&& + \frac{2N_c}{C_F} \big[(F^{T}_{qq}F^{T}_{\bq\bq} + F^{T}_{\Delta q \Delta q} F^{T}_{\Delta \bq \Delta \bq}) + (F^{T}_{q\bq}F^{T}_{\bq q} + F^{T}_{\Delta q \Delta \bq} F^{T}_{\Delta \bq \Delta q})\big] S^{TT} \nn
&& + \frac{1}{2} \Big[(I^{1}_{\bq q} + I^{1}_{\Delta \bq \Delta q})(I^{1}_{q \bq} + I^{1}_{\Delta q \Delta \bq}) + I^{1}_{\delta \bq \delta q} I^{1}_{\delta q \delta \bq}\Big] S_I^{11}
\nn
&& + \frac{N_c}{2} \Big[(I^{T}_{\bq q} + I^{T}_{\Delta \bq \Delta q})(I^{1}_{q \bq} + I^{1}_{\Delta q \Delta \bq}) + I^{T}_{\delta \bq \delta q} I^{1}_{\delta q \delta \bq} + (1 \lra T) \Big] S_I^{T1}
\nn
&& + \frac{N_c}{C_F} \Big[(I^{T}_{\bq q} + I^{T}_{\Delta \bq \Delta q})(I^{T}_{q \bq} + I^{T}_{\Delta q \Delta \bq}) + I^{T}_{\delta \bq \delta q} I^{T}_{\delta q \delta \bq}\Big] S_I^{TT} + (q \lra \bq) \bigg\}
\,, \end{eqnarray}
\end{widetext}
Here $q^2_i$ and $Y_i$ are the total invariant mass and rapidity of each lepton pair, and $Q_q$ is the quark charge. We suppressed the arguments of all functions in \eq{si_new} for brevity. The arguments of the dPDFs are the momentum fractions and transverse separation $\mathbf{z}_\perp$, just as in \eq{si_old}. In addition to the dPDFs, the last four lines involve soft functions $S$, describing the effects of soft radiation. The soft functions only depend on the large $\mathbf{z}_\perp$ separation, since soft radiation does not resolve the short distances associated with the momentum fractions. The second line of \eq{si_new} contains the contribution from color correlations and the third through fifth line contains the interference contributions. \eq{si_new} agrees with the terms in the factorization theorem presented in Ref.~\cite{Diehl:2011yj} that were explicitly written out. The analogous expression for the cross section for $WW$ production in terms of double PDFs is given in \subsec{WW}.

The position space dPDF $F(\mathbf{z_\perp})$ is dimension two, and of order $\lqcd^2$. The momentum space dPDF $F(\mathbf{r_\perp})$ is dimensionless, and of order unity. The dPDF terms in \eq{si_new}
\begin{eqnarray}
\sim \int \rd^2 \mathbf{z}_\perp F(\mathbf{z}_\perp)F(\mathbf{z}_\perp)
\end{eqnarray}
are of dimension two, and order $\lqcd^2$, since the $\mathbf{z}_\perp$ integral produces a $1/\lqcd^2$. This shows that the dPDF cross-section is $\lqcd^2/Q^2$ suppressed.

Clearly the large number of functions that appear in \eq{si_new} is worrisome for the prospect of measuring them in experiments. The good news is that color-correlations and interference contributions are Sudakov suppressed at high energies, leaving only the first line of this equation, which is one of the main conclusions of our paper. The $F_{\delta q \delta q}$ dPDF does not enter in the leading order expression in \eq{si_new}.

\begin{figure}[t]
  \centering
 \includegraphics[width=0.35\textwidth]{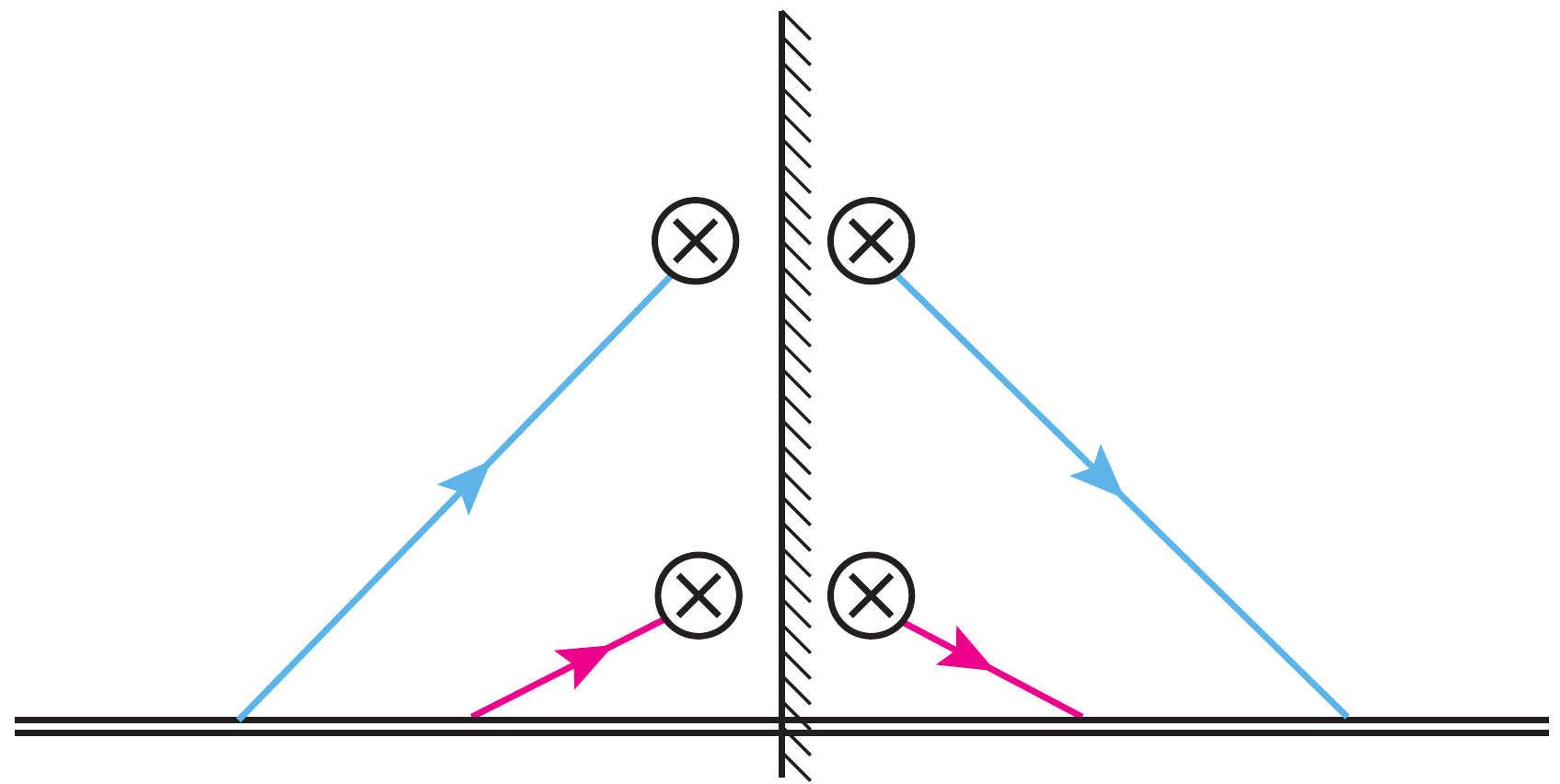}
   \includegraphics[width=0.35\textwidth]{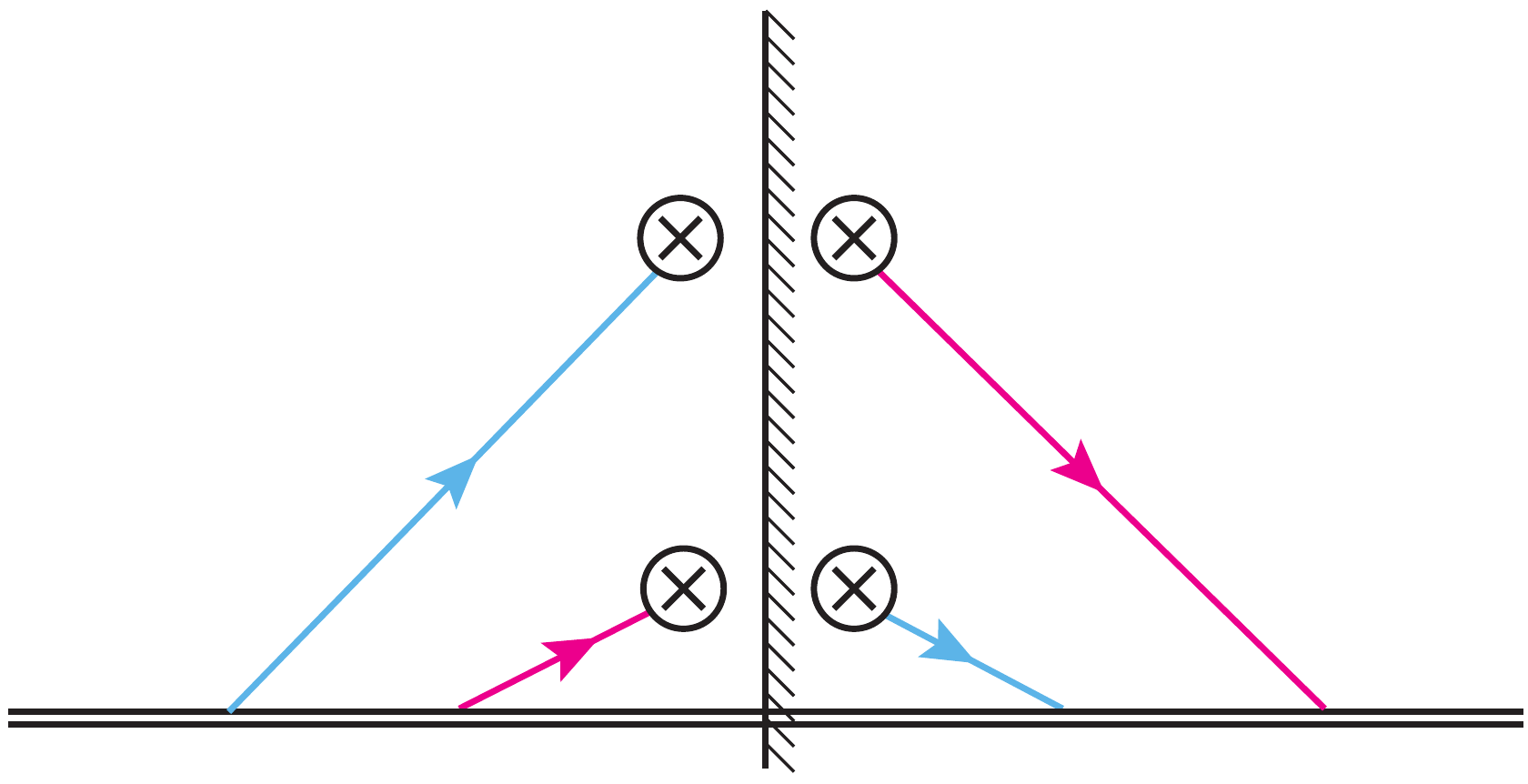}  
\caption{Color flow in the color-summed dPDF $F^1_{qq}$ and color-correlated dPDF $F^T_{qq}$. The vertical separation of the vertices $\otimes$ is $\mathbf{z_\perp}$.}
\label{fig:color}
\end{figure}

\begin{figure}
  \centering
  \includegraphics[width=0.35\textwidth]{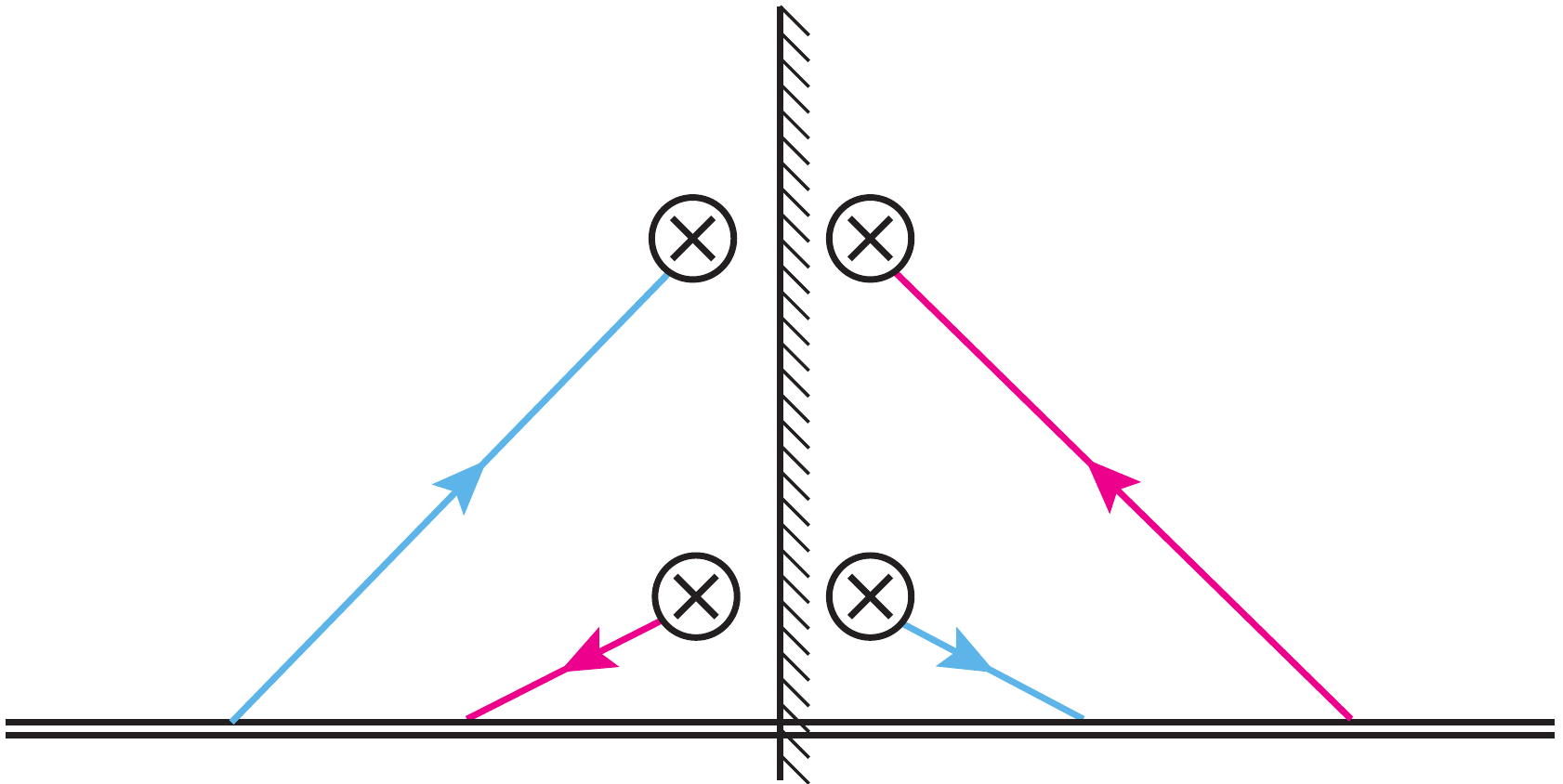} \\[0.2ex]
  \includegraphics[width=0.35\textwidth]{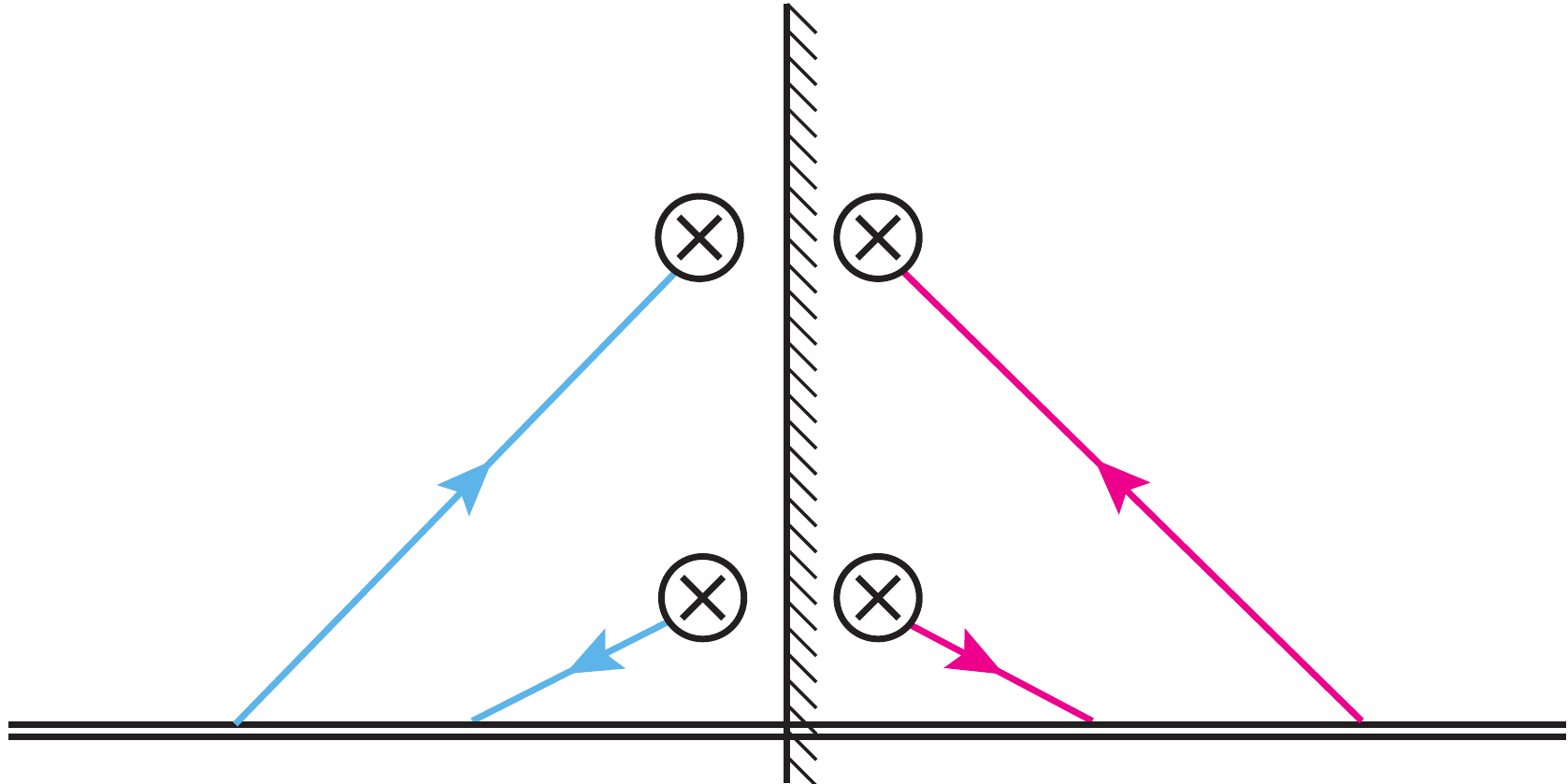}  
\caption{Color flow in the interference dPDFs $I^1_{q\bq}$ and $I^T_{q\bq}$. Both have long-range color correlations. The vertical separation of the vertices $\otimes$ is $\mathbf{z_\perp}$.}
\label{fig:int_color}
\end{figure}

Intuitively, the Sudakov suppression of color-correlated and interference dPDFs can be understood as a consequence of long range ($\mathbf{z}_\perp \sim 1/\lqcd$) color correlations. This is illustrated in \fig{color}, where the color flow is shown for both the color-summed and color-correlated dPDF. In \fig{int_color} we show the color flow for the interference dPDFs $I^1$ and $I^T$, which both involve long range color correlations. While color is conserved in the hadron matrix element, the color-correlated and  interference dPDFs move color a distance of order $1/\lqcd$ within the hadron.

More formally, the Sudakov suppression follows from our study of the anomalous dimensions of the dPDFs and soft functions in \sec{RGE}. In the remainder of this section, we summarize our results for the RGE evolution and Sudakov suppression of $F^T$. Similar results hold for the interference dPDFs, and are presented in Sec.~\ref{sec:RGE}.

At this point we need to briefly discuss rapidity divergences, which we encounter in our calculations. These divergences cancel between the soft function and dPDFs and are thus absent in the cross-section. However, they need to be regulated and there is a corresponding series of large (single) logarithms that needs to be summed for reliable predictions. We achieve this using the recently introduced rapidity renormalization group~\cite{Chiu:2011qc,Chiu:2012ir}, whose workings are similar to that of dimensional regularization for UV divergences. Just as $1/\eps$ UV divergences lead to a $\mu$ anomalous dimension, $1/\eta$ rapidity divergences lead to a $\nu$ anomalous dimension. The new renormalization scale $\nu$ is then used to sum the rapidity logarithms. (An introduction to the rapidity renormalization group and sample calculations involving the rapidity regulator are provided in \subsecs{RRGE}{soft_RGE}.)

\begin{figure}
  \centering
  \includegraphics[width=0.4\textwidth]{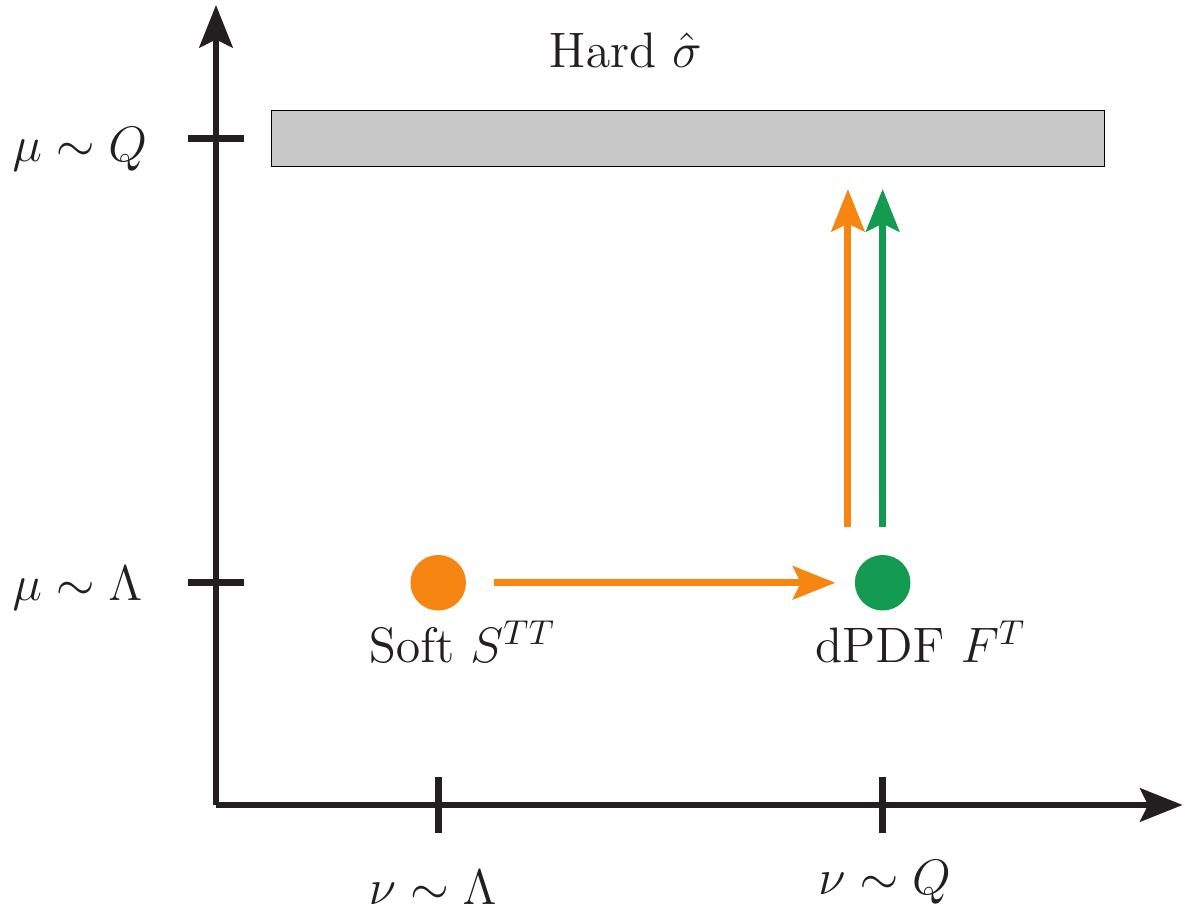}
\caption{Natural scales of the dPDF $F$ and soft function $S$.}
\label{fig:munu}
\end{figure} 

The anomalous dimensions for $F^T$ and $S^{TT}$ are calculated in \subsec{dPDF_RGE} and \subsec{soft_RGE} and are given by
\begin{eqnarray}  \label{eq:gaT}
\gamma_\mu^{F^{T}} 
&=&  \frac{\alpha_s(\mu)}{\pi}\bigg[\left(C_F-\frac12C_A\right) P_{qq}(x_1) +  C_A \Big( \ln \frac{\nu}{p_1^-} + \frac34\Big)\nn
&& \times \delta(1-x_1)\bigg] 
\delta(1-x_2)\de^{(2)}(\mathbf{r}_\perp) + (1 \leftrightarrow 2)
\,,\nn
\gamma_\nu^{F^{T}} &=& -\frac{\alpha_s(\mu) C_A}{\pi^2}  \frac{1}{\mu^2} \frac{1}{(\mathbf{r}_\perp^2/\mu^2)}_+
\de(1-x_1) \de(1-x_2) \,,\nn
  \ga_\mu^{S^{TT}} &=& \frac{2\al_s(\mu)C_A}{\pi} \ln \frac{\mu^2}{\nu^2}\, \de^{(2)}(\mathbf{r}_\perp)\,,\nn
  \ga_\nu^{S^{TT}} &=& \frac{2\al_s(\mu) C_A}{\pi^2}\, \frac{1}{\mu^2} \frac{1}{(\mathbf{r}_\perp^2/\mu^2)}_+ 
\,,\end{eqnarray}
written in terms of the usual (one-dimensional) plus distributions
\begin{equation} \label{eq:plusdef}
\frac{1}{u}_+  = \lim_{\xi \to 0} \biggl[ \frac{\theta(u- \xi)}{u} +  \delta(u - \xi) \, \ln\xi \biggr]
\,.\end{equation}
In \eq{gaT} we explicit showed the dependence on all variables, where a $\de(1-x_1)$ or $\de^{(2)}(\mathbf{r}_\perp)$ means that the evolution does not affect the $x_1$ or $\mathbf{r}_\perp$ dependence, respectively.
From \eq{gaT} we read off that the natural scales for $F^T$ are $(\mu_F,\nu_F) \sim (|\mathbf{r}_\perp|,p^-) \sim (\Lambda,Q)$ and for $S^{TT}$ are  $(\mu_S,\nu_S) \sim (\Lambda,\Lambda)$. $\Lambda$ is a  scale of order $\lqcd$ which we take to be 1.4~GeV in our numerical analysis. By evaluating $F^T$ and $S^{TT}$ at these scales, and running them to a common scale using the $\mu$ and $\nu$ RGE, the large logarithms in the cross-section are summed. The natural scales and our running strategy are summarized in \fig{munu}.
As we mentioned, the rapidity divergences, and thus the corresponding $\nu$ evolution, must cancel between the dPDFs and the soft function. Indeed,
\begin{equation}
  \ga_\nu^{F^T} + \frac{1}{2} \ga_\nu^{S^{TT}} \de(1-x_1) \de(1-x_2) = 0
\,,\end{equation}
which provides a consistency check on our calculations.

\begin{figure}[t]
  \centering
  \includegraphics[width=0.45\textwidth]{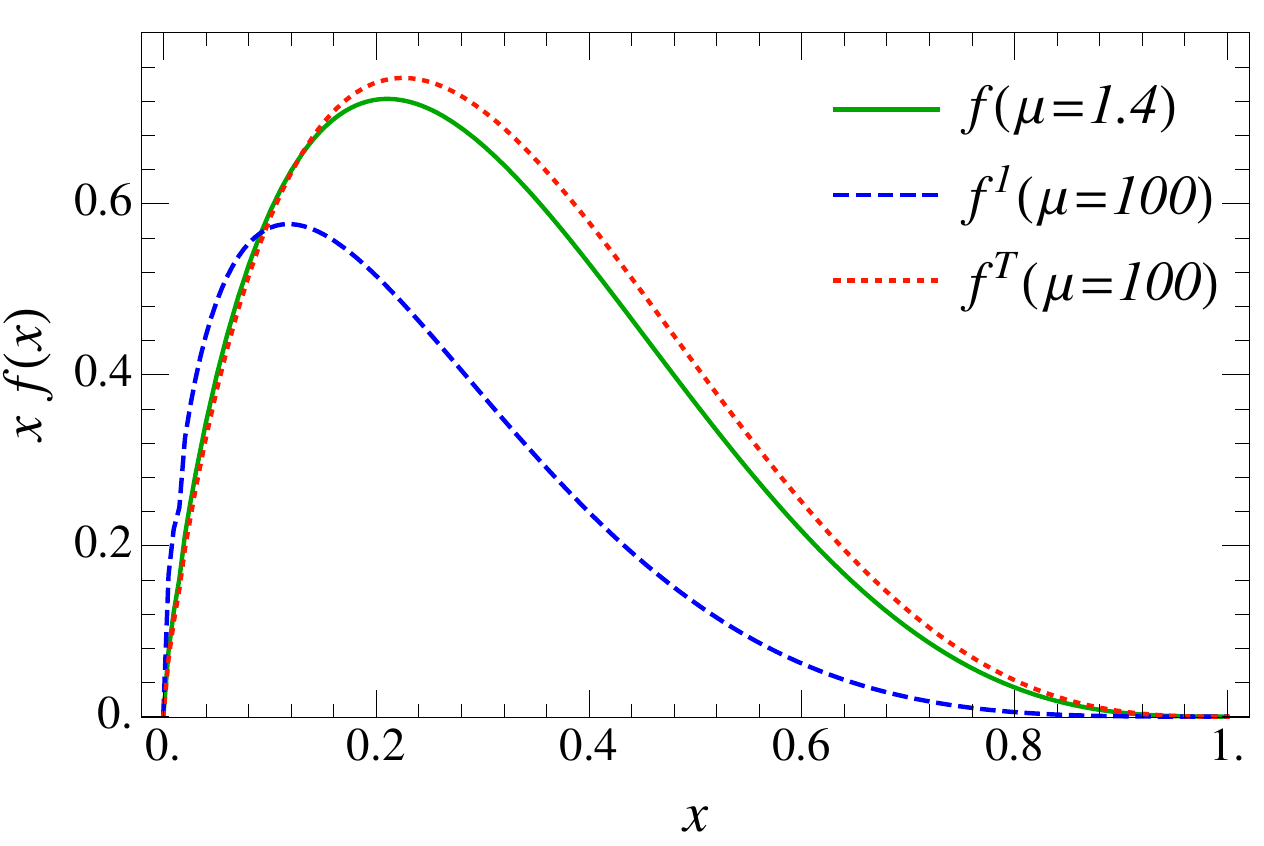}
\caption{Comparing the $x$-dependent part of the evolution of $F^1$ and $F^T$ assuming a common initial PDF. The three curves correspond to the initial PDF at $\mu=1.4$~GeV, and the evolution of $F^1$ and $F^T$ to $\mu=100$~GeV. }
\label{fig:fx}
\end{figure}

We will now show that the contribution of the color-correlated dPDF to the cross section is Sudakov suppressed. For both $F^T$ and $S^{TT}$, $\mu \sim \La$ is their natural scale, from we simultaneously evolve them to the hard scale $\mu=Q$. Combining the $\mu$-evolution of the dPDFs and the soft function, we find
\begin{eqnarray} \label{eq:gamu}
&&\gamma_\mu^{F^{T}} + \frac{1}{2} \gamma_\mu^{S^{TT}} \de(1-x_1) \de(1-x_2) \nn
&&=  \frac{\alpha_s}{\pi}\bigg[\left(C_F\!-\!\frac12C_A\right) P_{qq}(x_1) +  C_A \Big( \ln \frac{\mu}{p_1^-} \!+\! \frac34\Big)\delta(1\!-\!x_1)\bigg]\nn
&& \quad \times 
\delta(1-x_2)\de^{(2)}(\mathbf{r}_\perp) + (1 \leftrightarrow 2) 
\end{eqnarray}
for each dPDF. The $\mu$ anomalous dimension has an interesting structure. The $x$-dependent piece is the usual splitting function, but with a modified color factor of  $-1/6$ rather than $4/3$, so the $x$-dependent part of the anomalous dimension can be interpreted as a slow ``reverse" evolution due to the change in magnitude and sign of the color factor. The effect of this reverse evolution is evaluated using \HOPPET~\cite{Salam:2008qg} and shown in \fig{fx} for a sample initial PDF. It would be interesting to see this effect experimentally. 

\begin{figure}
  \centering
  \includegraphics[width=0.44\textwidth]{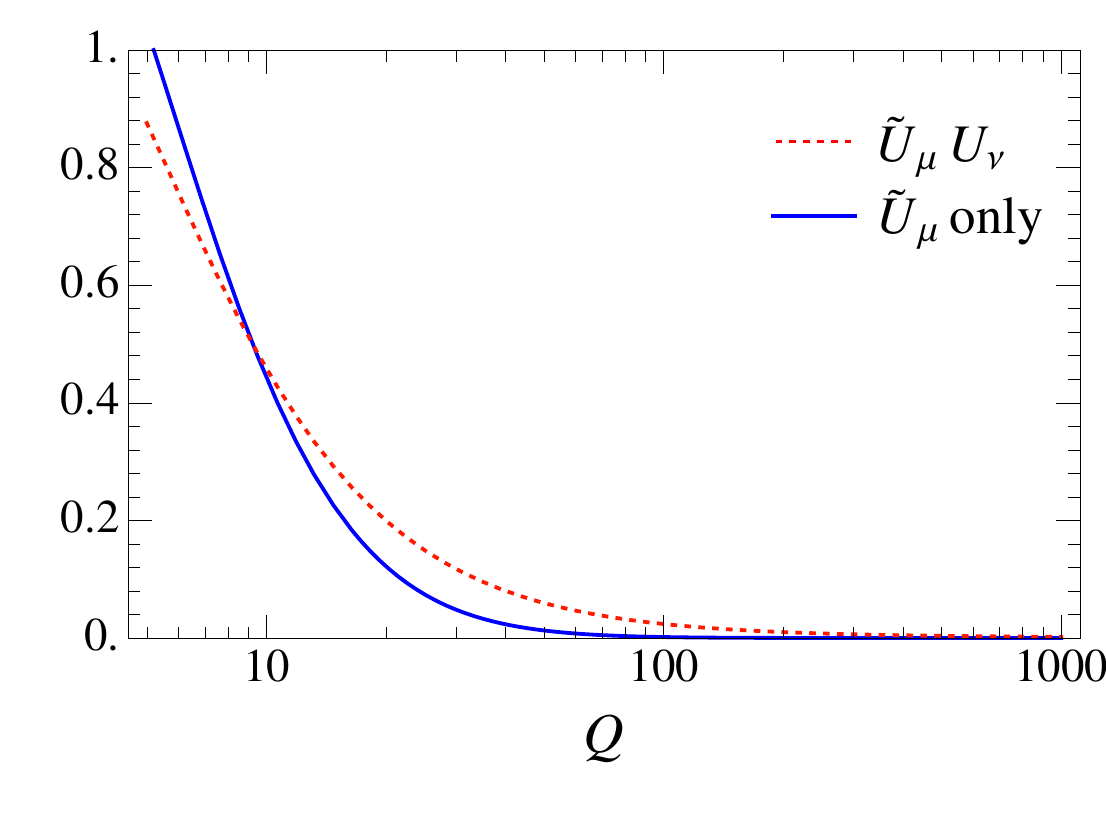}
\caption{The Sudakov suppression factor with and without rapidity resummation, running from $\La = 1.4$ GeV to $Q$. The evolution kernels $\widetilde U_\mu$ and $U_\nu$ are discussed in the text.}
\label{fig:sud}
\end{figure}

There is also the second term in \eq{gamu}, which is $x$-independent and thus does not change the shape of the dPDF. Its evolution (combining both dPDFs) from a low scale $\La$ to the hard scale $Q \sim Q_1 \sim Q_2$ is given by an overall multiplicative factor
\begin{equation} \label{eq:ga_sud}
  \widetilde U_\mu(\La,Q) = \exp\Big( -\frac{\al_s C_A}{2\pi} \ln^2 \frac{Q^2}{\La^2} \Big)
\,,\end{equation}
at leading-logarithmic accuracy. Color correlations are thus Sudakov suppressed and can be neglected for $Q \gg \La$. \eq{ga_sud} agrees with Ref.~\cite{Mekhfi:1988kj}, which arrived at this conclusion by studying the color factors in the real and virtual Sudakov form factor. Ref.~\cite{Mekhfi:1988kj} did not attempt to factorize the cross section and does not discuss the interference case, which we find is also Sudakov suppressed.  The effect of the Sudakov suppression is shown in \fig{sud} at next-to-leading-logarithmic accuracy and taking the running of $\al_s$ into account [which was neglected in \eq{ga_sud}]. 

We also need to perform the $\nu$-evolution to sum the rapidity logarithms. In \subsec{nu_evo} we calculate the evolution kernel $U_\nu$. We estimate the effect of the $\nu$-evolution on the cross section by (schematically)
\begin{eqnarray}
  \si &\sim& \int \df^2 \mathbf{p}_\perp \df^2 \mathbf{q}_\perp \df^2 \mathbf{s}_\perp F^T(\mathbf{p}_\perp,\nu_F) F^T(\mathbf{q}_\perp,\nu_F) \nn
  && \times U_\nu(\mathbf{s}_\perp,\nu_F,\nu_S) S^{TT}(-\mathbf{p}_\perp-\mathbf{q}_\perp-\mathbf{s}_\perp,\nu_S) \nn
&\sim& \int^{|\mathbf{p}_\perp|\leq \La} \df^2 \mathbf{p}_\perp U_\nu(\mathbf{p}_\perp,Q,\La) 
\,,\end{eqnarray}
Here we assumed that $F^T$ and $S^{TT}$ have a width of order $\La$ at their natural scales $\nu_F \sim Q$, $\nu_S \sim \La$. In \fig{sud} we take $\La=1.4 \GeV$, and normalize our above estimate for the effect of the $\nu$ evolution such that it is 1 when $Q = \La$. The resummation of the rapidity logarithms increases the cross section, as shown in \fig{sud}. However, it is only a single-logarithmic series and thus has a smaller effect than the Sudakov suppression. At high energies the first line in \eq{si_new} thus dominates.

\section{Factorization Theorem at Leading Order}
\label{sec:fact}

This section discusses the factorization theorem for DPS. We start by reviewing the usual derivation of the factorization theorem for single Drell-Yan, and then repeat the analysis for double Drell-Yan. Other derivations of factorization for double parton scattering have been presented in Ref.~\cite{Diehl:2011yj} and for scalar partons in Ref.~\cite{Paver:1982yp}.

\subsection{Single Drell-Yan}

\begin{figure}
\includegraphics[width=2.15cm]{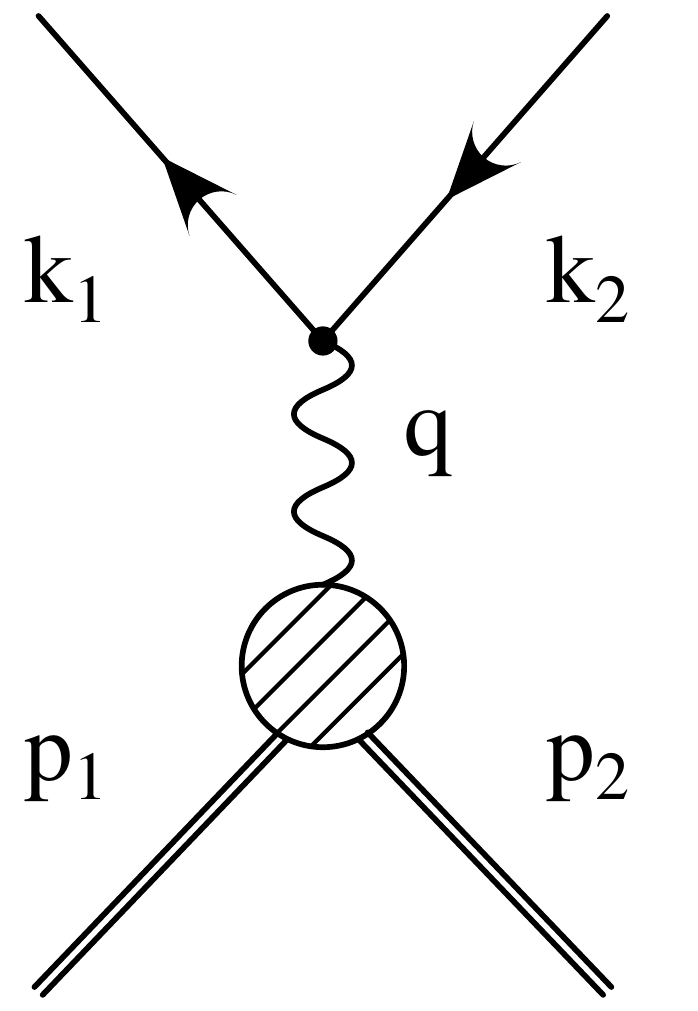}
\caption{Drell-Yan production, $p_1+p_2 \to \ell^+ \ell^-$.}
\label{fig:DY}
\end{figure}
The  single Drell-Yan process is shown in \fig{DY}.  The Drell-Yan cross-section is
\begin{eqnarray}
\rd \sigma &=& \frac{1}{2E_{p_1}}\frac{1}{2E_{p_2}} \frac{1}{v_\mathrm{rel}} \int \rd^4 q\, \frac{\rd^3 k_1}{(2\pi)^3 2E_{k_1}}\frac{\rd^3 k_2}{(2\pi)^3 2E_{k_2}} \\
&& (2\pi)^4 \delta^{(4)}(q-k_1-k_2) \delta^{(4)}(q+p_X-p_1-p_2) \sum_X \abs{A}^2 
\nonumber \end{eqnarray}
where the sum is over hadronic final states $X$. The momenta are shown in \fig{DY}.
The amplitude for leptons produced through a virtual photon is
\begin{eqnarray}
A &=& -\img e\, \overline u(k_1) \gamma^\mu v(k_2)\, \frac{-\img g_{\mu \nu}}{k^2}\, \img Q_q
\braket{X | J^\nu | p p}
\,,\end{eqnarray}
where $Q_q$ is the electric charge of the quark.
Performing the sum over the lepton spins
\begin{eqnarray}
&&\sum_\mathrm{spins} \bar u(k_1) \gamma^\mu v(k_2)\, \bar v(k_2) \gamma^\nu u(k_1) \nonumber \\[-1ex]
&&\qquad = 4 \left(k_1^\mu k_2^\nu + k_2^\mu k_1^\nu - g^{\mu\nu}k_1 \sdt k_2\right)
\,,\end{eqnarray}
and the leptonic phase-space integral
\begin{eqnarray}
&&\int \frac{\rd^3 k_1}{(2\pi)^3 2E_{k_1}}\frac{\rd^3 k_2}{(2\pi)^3 2E_{k_2}}\, (2\pi)^4 \delta^{(4)}(q-k_1-k_2)\nn
&&\times
4\left(k_1^\mu k_2^\nu + k_2^\mu k_1^\nu - g^{\mu\nu}k_1 \sdt k_2 \right) \nn
&&\qquad = \frac{1}{6\pi} \left(q^\mu q^\nu - q^2 g^{\mu \nu}\right)
\,,\end{eqnarray}
leads to
\begin{eqnarray}
\rd \sigma &=& \frac{e^4Q_q^2}{12\pi s} \int \rd^4 q\ \delta^{(4)}(q+p_X-p_1-p_2)\\
&&\times \sum_X \braket{p p | J_\mu^\dagger | X} \braket{X | J_\nu | p p}
\frac{1}{q^2} \Big(\frac{q^\mu q^\nu}{q^2} - g^{\mu \nu}\Big)\,.\nonumber
\end{eqnarray}

\begin{figure}[b]
\includegraphics[width=3cm]{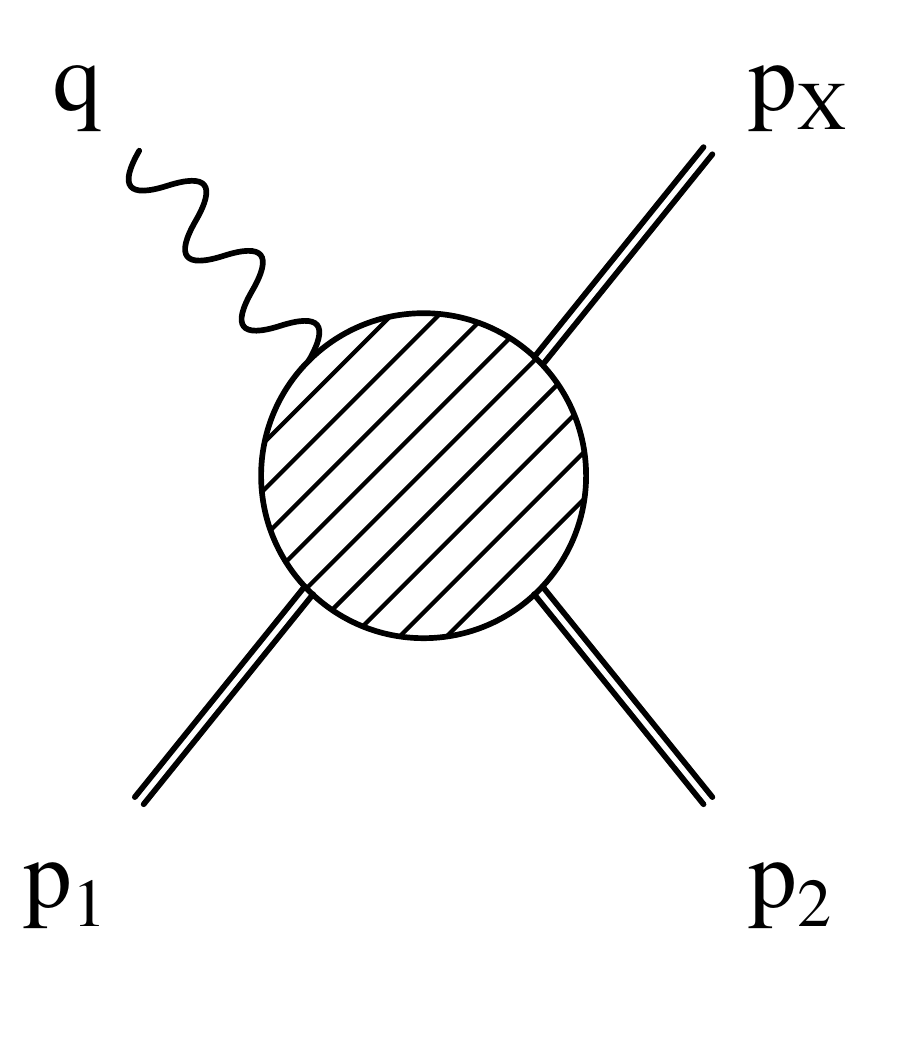}
\caption{Hadronic amplitude for single Drell-Yan.}
\label{fig:2}
\end{figure}

The hadronic amplitude is shown in \fig{2} and is
\begin{eqnarray}
&&(2\pi)^4 \delta^{(4)}(q+p_X-p_1-p_2)\,M^\mu_X \nn
&&= \int \rd^4 z\ e^{\img q \cdot z} \braket{X | J^\mu(z) | p_1 p_2} \nn
&&= \int \rd^4 z\ e^{\img q \cdot z} e^{-\img z \cdot (p_1+p_2-p_X)} \braket{X | J^\mu(0) | p_1 p_2} \nn
&&=(2\pi)^4 \delta^{(4)}(q+p_X-p_1-p_2) \braket{X | J^\mu(0) | p_1 p_2}
.\end{eqnarray}
The hadronic tensor can thus be rewritten as
\begin{eqnarray}
H^{\mu \nu} &=& \sum_X  (2\pi)^4\delta^{(4)}(q+p_X-p_1-p_2)M^{\mu*}_X M^\nu_X \nn
&=& \int\! \rd^4 z\, e^{-\img z \cdot q}\braket{p p | J^{\mu\dagger}(z)  J^\nu(0) | p p}
\,,\end{eqnarray}
such that
\begin{eqnarray} \label{eq:si_2}
\!\!\!\!\rd \sigma &=& \frac{4\pi \alpha^2 Q_q^2}{3 s} \!\!\int \! \frac{\rd^4 q}{(2\pi)^4}\, H_{\mu \nu}(p_1,p_2,q) \frac{1}{q^2} \Big(\frac{q^\mu q^\nu}{q^2} \!-\! g^{\mu \nu} \Big) 
\,.\nn\end{eqnarray}

The hadronic matrix element separates into two PDFs. It will be convenient to use light-cone coordinates where $p_1$ is in the $n$ direction and $p_2$ is in the $\bn$ direction, such that $p_1^- = \bn \cdot p_1$ and $p_2^+ = n \cdot p_2$ are large. Explicitly, $n^\mu = (1,0,0,1)$ and $\bn^\mu = (1,0,0,-1)$ for beams along the third spatial direction. We find it convenient to use the same indices for both spin and color, where $\ga_{ab}^\mu$ is $\de_{ab}$ for the color indices, to reduce the number of indices. Since we do not observe the transverse momentum of the lepton pair, we can integrate over $\mathbf{q}_\perp$,
\begin{widetext}
\begin{eqnarray} \label{eq:facthad}
\int \frac{\rd^2 \mathbf{q}_\perp}{(2\pi)^2} H^{\mu \nu} &=&\gamma^\mu_{ab} \gamma^\nu_{cd} \int \frac{\rd^2 \mathbf{q}_\perp}{(2\pi)^2}\int \rd^4 z\ e^{-\img z^+q^-/2-\img z^-q^+/2+\img \mathbf{q}_\perp \cdot \mathbf{z}_\perp}\braket{p_1 p_2 | \overline \psi_a(z) \psi_b(z) \overline \psi_c(0) \psi_d(0) | p_1 p_2} \nn
&=& \gamma^\mu_{ab} \gamma^\nu_{cd} \int \frac{\rd z^+ \rd z^- \rd^2 \mathbf{z}_\perp}{2}\,  e^{-\img z^+q^-/2-\img z^-q^+/2} \delta^{(2)}(\mathbf{z}_\perp) \braket{p_1 p_2 | \overline \psi_a(z) \psi_b(z) \overline \psi_c(0) \psi_d(0) | p_1 p_2} 
.\end{eqnarray}
At this point, one can contract the fields with the states in different ways. The momentum of the current at the point $z$, which by momentum conservation is equal in size to the momentum of the current at $0$, has large $-$ and $+$ components. Thus one of the fields at $z$ and $0$ must be contracted with $p_1$ and the other with $p_2$. Furthermore, quark number is conserved, so one cannot contract two $\psi$ fields with $p_1$, etc. There are two possible contractions which remain. The first one is
\begin{eqnarray} \label{eq:facthad2}
\int \frac{\rd^2 \mathbf{q}_\perp}{(2\pi)^2} H^{\mu \nu} 
&=& \gamma^\mu_{ab} \gamma^\nu_{cd} \int \frac{\rd z^+ \rd z^- \rd^2 \mathbf{z}_\perp}{2}\,  e^{-\img z^+q^-/2-\img z^-q^+/2} \delta^{(2)}(\mathbf{z}_\perp) \braket{p_1 | \overline \psi_a(z)  \psi_d(0) | p_1} \braket{p_2 |   \psi_b(z)  \overline \psi_c(0)|p_2}  \nn
&=& \gamma^\mu_{ab} \gamma^\nu_{cd} \int \frac{\rd z^+ \rd z^-}{2}\,  e^{-\img z^+q^-/2-\img z^-q^+/2} \braket{p_1 | \overline \psi_a(z^+,0,\mathbf{0}_\perp)  \psi_d(0) | p_1} \braket{p_2 |    \psi_b(0,z^-,\mathbf{0}_\perp) \overline \psi_c(0)|p_2}  \nn
&=&\frac{\pi^2}{8N_c^2} \gamma^\mu_{ab} \gamma^\nu_{cd}\, \slashed{n}_{da} \slashed{\bn}_{bc}\,
f_q\Big(\frac{q^-}{p_1^-}\Big) f_{\overline q}\Big(\frac{q^+}{p_2^+}\Big) 
= \frac{\pi^2}{2N_c}(n^\mu \bn^\nu+n^\nu \bn^\mu - 2 g^{\mu \nu})\,
f_q\Big(\frac{q^-}{p_1^-}\Big) f_{\overline q}\Big(\frac{q^+}{p_2^+}\Big)
\,.\end{eqnarray}
\end{widetext}
To derive the second line we note that $z^\pm \sim 1/q^{\mp} \sim 1/Q$, whereas the dependence on $z^-$ ($z^+$) of the matrix element of $p_1$ ($p_2$) is slowly varying and may be set to zero. In the third line we have used the definition of the PDFs,
\begin{eqnarray} \label{eq:PDFdef} 
&&\!\!\!\!\int\! \frac{\rd z^+}{4\pi}\, e^{- i z^+ q^-/2} \braket{p | [\overline\psi(z^+) W(z^+)]_a [W^\dagger(0)
\psi(0)]_b | p} \nn
&&\quad= \frac{\slashed{n}_{ba}}{4 N_c} \Big[ \theta\Big(\frac{q^-}{p^-}\Big) f_{q}\Big(\frac{q^-}{p^-}\Big) - \theta\Big(-\frac{q^-}{p^-}\Big) f_{\overline q}\Big(-\frac{q^-}{p^-}\Big) \Big]
\,.\nn
\end{eqnarray}
The Wilson line $W(z^+)$ goes to infinity and ensures gauge invariance, but was not written explicitly in \eq{facthad2} to avoid cumbersome notation. In momentum space the Wilson line is given by
\begin{equation} \label{eq:Wn}
  W_n = \sum_\mathrm{perms} \exp\Big[\frac{-g}{\img \partial^-} A_n^-(0)\Big] 
\,,\end{equation}
where $n$ denotes the direction of the energetic radiation.

Combining \eq{PDFdef} with \eq{si_2} yields the familiar result
\begin{eqnarray}
  \frac{\df \si}{\df q^2\, \df Y} &=& \frac{1}{2}\,\frac{\df \si}{\df q^0\, \df q^3} \\
  &=& \frac{4\pi\al^2 Q_q^2}{3N_c\, q^2 s}\, [f_q(x_1) f_{\overline q}(x_2) + f_{\overline q}(x_1) f_q(x_2)]
\,,\nonumber\end{eqnarray}
where we included the additional contribution to \eq{facthad} from the second contraction with $\psi \lra \overline \psi$, which exchanges the quark and antiquark PDFs. Here $x_i$ are the momentum fractions and $Y$ is the total rapidity of the lepton pair,
\begin{equation} \label{eq:Q2Y}
x_1 = \frac{q^-}{p_1^-} = \sqrt{\frac{q^2}{s}} e^Y
\,, \quad
x_2 = \frac{q^+}{p_2^+} = \sqrt{\frac{q^2}{s}} e^{-Y}
\,.\end{equation}

\subsection{Double Drell-Yan}
\label{subsec:DDY}

\begin{figure}[b]
\includegraphics[width=3cm]{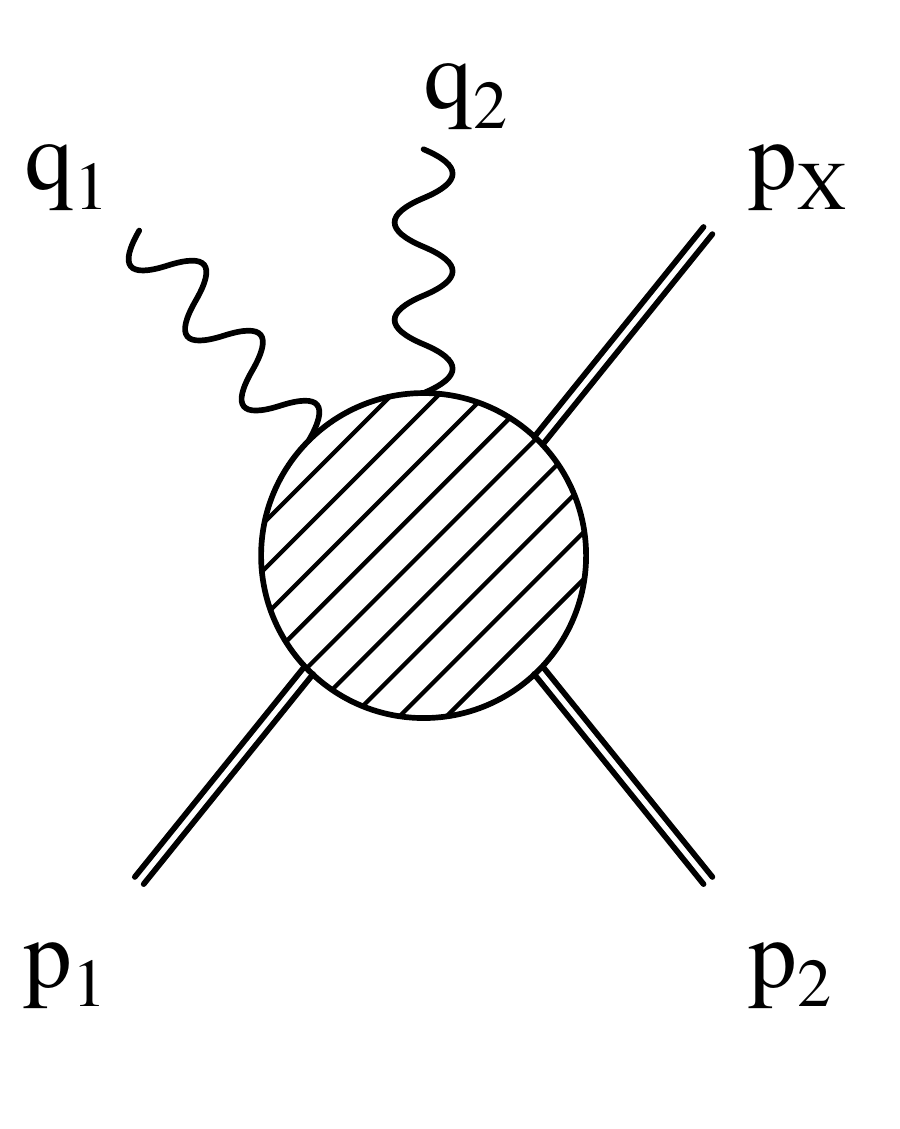}
\caption{Hadronic amplitude for double Drell-Yan.}
\label{fig:3}
\end{figure}
The leptonic amplitudes for double Drell-Yan are identical to that of single Drell-Yan, so the cross section is given by
\begin{eqnarray}
\rd \sigma &=& 2s \Big(\frac{4\pi \alpha^2 Q_q^2}{3 s}\Big)^2 \int \frac{\rd^4 q_1}{(2\pi)^4} \frac{\rd^4q_2}{(2\pi)^4}\, H_{\mu \nu \alpha \beta}\nn
&&  \times \frac{1}{q_1^2} \Big(\frac{q_1^\mu q_1^\nu}{q_1^2} - g^{\mu \nu}\Big)\, \frac{1}{q_2^2} \Big(\frac{q_2^\al q_2^\bt}{q_2^2} - g^{\al \bt}\Big)
\,,\end{eqnarray}
where the factor $2s$ ensures that we only count the flux factor once.
For simplicity we will for now assume identical quark flavors. 
The hadronic amplitude is shown in \fig{3} and equals
\begin{eqnarray}
&&(2\pi)^4 \delta^{(4)}(q_1+q_2+p_X-p_1-p_2)\,M^{\nu \bt}_X \\
&&\ = \!\int\! \rd^4 z_1\, \rd^4z_2\ e^{\img q_1 \cdot z_1+\img q_2 \cdot z_2} \braket{X | 
T\{ J_1^\nu(z_1) J_2^\bt(z_2) \} | p_1 p_2} 
,\nonumber \end{eqnarray}
and the corresponding hadronic tensor is
\begin{widetext}
\begin{eqnarray}
H^{\mu \nu \alpha \beta} &=& \sum_X  (2\pi)^4\delta^{(4)}(q_1+q_2+p_X-p_1-p_2)
M^{\mu \al*}_X M^{\nu \bt}_X \nn
&=&  \int\! \rd^4z_1\,  \rd^4 z_2\, \rd^4 z_3\, e^{-\img q_1 \cdot z_1-\img q_2 \cdot z_2+\img q_1 \cdot z_3}\, \braket{p_1p_2 | \overline T \{ J_1^{\mu\dagger}(z_1) J_2^{\alpha\dagger}(z_2)\} T \{ J_1^\nu(z_3) J_2^\bt(0) \}  | p_1 p_2}
\,.\end{eqnarray}

We now follow similar steps as for single Drell-Yan in order to factor the hadronic matrix element into dPDFs. The transverse momenta of the lepton pairs are not measured, so we can integrate over $\mathbf{q}_{1\perp}, \mathbf{q}_{2\perp}$,
\begin{eqnarray} \label{eq:HDDY}
\int\! \frac{\rd^2 \mathbf{q}_{1\perp}}{(2\pi)^2} \frac{\rd^2 \mathbf{q}_{2\perp}}{(2\pi)^2}  H^{\mu \nu \alpha \beta} 
&=& \frac18\int \rd z_1^+ \rd z_1^- \rd z_2^+ \rd z_2^- \rd z_3^+ \rd z_3^-  \rd^2 \mathbf{z}_\perp e^{-\img q_1^+ z_1^-/2-\img q_1^-  z_1^+/2}\,e^{-\img q_2^+ z_2^-/2-\img q_2^- z_2^+/2}\, e^{\img q_1^+ z_3^-/2+\img q_1^- z_3^+/2} \nn
&&\times \braket{p_1p_2 | \overline T \{ J^{\mu \dagger}_1(z_1,\mathbf{z}_\perp) J^{\al\dagger}_2(z_2,\mathbf{0}_\perp)\} T \{ J^\nu_1(z_3,\mathbf{z}_\perp) J^\bt_2(0)\}  | p_1 p_2} 
\,.\end{eqnarray}
Note that one transverse position integral remains, in contrast to single Drell-Yan. This integral would remain even if one measures the transverse momenta of the lepton pairs. This can be seen from Fig.~\ref{fig:dps_rperp}: the momentum $\mathbf{r}_\perp$ does not affect the transverse momentum of the final states, is arbitrary, and so is integrated over. The remaining $\rd^2 \mathbf{z}_\perp$ integral in Eq.~(\ref{eq:HDDY}) is the position space version of this $\mathbf{r}_\perp$ integral. It is important to note that $\mathbf{r}_\perp$ is not an observable quantity~\cite{Manohar:DPS2}.

The $\rd^2 \mathbf{z}_\perp$ integral means that the correlation function of four currents, Eq.~(\ref{eq:HDDY}), is not local in $\mathbf{z}_\perp$.
This allows soft radiation to contribute, since soft radiation can resolve the large distance $\mathbf{z}_\perp$. Soft radiation cannot resolve the  short-distance scales $z_i^\pm \sim 1/Q$. The soft radiation emitted by a quark field exponentiates into an eikonal (soft) Wilson line that can be factored out~\cite{Collins:2011zzd,Bauer:2001yt} 
\begin{equation}
  \psi_a'(x) = S_{n,a'a}(x) \psi_a(x) \,,
  \label{soft}
\end{equation}
where the soft Wilson line is in the direction $n$ of the quark and $a, a'$ are color indices. 

There are various ways of joining the fields to the incoming particles in the matrix elements, which allow for large $+$ and $-$ components of momentum
flowing into the four current vertices. For example,
\begin{eqnarray} \label{eq:dfact}
&&\int \frac{\rd^2 \mathbf{q}_{1\perp}}{(2\pi)^2}\int \frac{\rd^2 \mathbf{q}_{2\perp}}{(2\pi)^2}  H^{\mu \nu \alpha \beta} \\
&&= \frac18 \ga^\mu_{a'b'} \ga^\al_{c'd'} \ga^\nu_{e'f'} \ga^\bt_{g'h'} \int \rd z_1^+ \rd z_1^- \rd z_2^+ \rd z_2^- \rd z_3^+ \rd z_3^-  \rd^2 \mathbf{z}_\perp e^{-\img q_1^+ z_1^-/2-\img q_1^-  z_1^+/2}\,e^{-\img q_2^+ z_2^-/2-\img q_2^- z_2^+/2}\, e^{\img q_1^+ z_3^-/2+\img q_1^- z_3^+/2}  \nn
&&\quad \times \braket{p_1| \overline T \left\{ \overline \psi_a (z_1,\mathbf{z}_\perp)  \overline \psi_c (z_2,\mathbf{0}_\perp) \right\} T \left\{ \psi_f (z_3,\mathbf{z}_\perp) \psi_h (0)\right\}  | p_1} \nn
&&\quad \times \braket{p_2 | \overline T \left\{ \psi_b (z_1,\mathbf{z}_{\perp})  \psi_d (z_2,\mathbf{0}_\perp) \right\} T \left\{ \overline \psi_e (z_3,\mathbf{z}_{\perp}) \overline \psi_g (0)\right\}  |p_2} \nn
&&\quad \times \braket{0 | S_{n,aa'}^\dagger(z_1,\mathbf{z}_\perp) S_{\bn,b'b}(z_1,\mathbf{z}_\perp) S_{n,cc'}^\dagger(z_2,\mathbf{0}_\perp) S_{\bn,d'd}(z_2,\mathbf{0}_\perp) S_{\bn,ee'}^\dagger(z_3,\mathbf{z}_\perp) S_{n,f'f}(z_3,\mathbf{z}_\perp) S_{\bn,gg'}^\dagger(0) S_{n,h'h}(0) | 0} \nn
&&= \frac18 \ga^\mu_{a'b'} \ga^\al_{c'd'} \ga^\nu_{e'f'} \ga^\bt_{g'h'} \int  \rd^2 \mathbf{z}_\perp  \nn
&&\quad\times \int \rd z_1^+ \rd z_2^+ \rd z_3^+    e^{-\img q_1^-  z_1^+/2}\, e^{-\img q_2^- z_2^+/2}\, e^{\img q_1^- z_3^+/2} \braket{p_1| \overline T \left\{ \overline \psi_a (z_1^+,0,\mathbf{z}_\perp)  \overline \psi_c (z_2^+,0,\mathbf{0}_\perp) \right\} T \left\{ \psi_f (z_3^+,0,\mathbf{z}_\perp) \psi_h (0)\right\}  | p_1}\nn
&&\quad\times \int \rd z_1^- \rd z_2^- \rd z_3^- e^{-\img q_1^+  z_1^-/2}\, e^{-\img q_2^+ z_2^-/2}\, e^{\img q_1^+ z_3^-/2}  \braket{p_2 | \overline T \left\{ \psi_b (0,z_1^-,\mathbf{z}_\perp)  \psi_d (0,z_2^-,\mathbf{0}_\perp) \right\} T \left\{ \overline \psi_e (0,z_3^-,\mathbf{z}_\perp) \overline \psi_g (0)\right\}  |p_2} \nn
&&\quad \times \braket{0 | S_{n,aa'}^\dagger(0,0,\mathbf{z}_\perp) S_{\bn,b'b}(0,0,\mathbf{z}_\perp) S_{n,cc'}^\dagger(0) S_{\bn,d'd}(0) S_{\bn,ee'}^\dagger(0,0,\mathbf{z}_\perp) S_{n,f'f}(0,0,\mathbf{z}_\perp) S_{\bn,gg'}^\dagger(0) S_{n,h'h}(0) | 0}
\,.\nonumber\end{eqnarray}
Here we once again use that  $z_i^\pm \sim 1/Q$ varies rapidly, whereas in the matrix element of $p_1$ ($p_2$) the dependence on $z_i^-$ ($z_i^+$) is slow and can be set to zero. The dependence of the soft radiation is slow in both $z_i^+$ and $z_i^-$  and so only the dependence on $\mathbf{z}_\perp$ remains. The last line in \eq{dfact} is the soft function $S$ (we omit a subscript $qq\bq\bq$ since we do not consider soft functions with gluons in this paper). The soft function has a lot of indices but also has a lot of symmetry. When we contract indices below, only four independent soft functions will appear. We identify the dPDFs $F_{qq}$ and $F_{\bar q\bar q}$ in \eq{dfact} by
\begin{eqnarray} \label{eq:37}
F_{q q}(q_1^-,q_2^-,\mathbf{z}_\perp,p_1^-)_{acfh} &=& -4 \pi p_1^- \int \frac{\rd z_1^+}{4\pi} \frac{\rd z_2^+}{4\pi} \frac{\rd z_3^+}{4\pi}     e^{-\img q_1^-  z_1^+/2}\, e^{-\img q_2^- z_2^+/2}\, e^{\img q_1^- z_3^+/2}\nn
&& \times \braket{p_1| \overline T \left\{ \overline \psi_a (z_1^+,0,\mathbf{z}_\perp)  \overline \psi_c (z_2^+,0,\mathbf{0}_\perp) \right\} T \left\{ \psi_f (z_3^+,0,\mathbf{z}_\perp) \psi_h (0)\right\}  | p_1}
\,,\nn
F_{\bq \bq}(q_1^+,q_2^+,\mathbf{z}_\perp,p_2^+)_{bdeg} &=& -4 \pi p_2^+ \int \frac{\rd z_1^-}{4\pi} \frac{\rd z_2^-}{4\pi} \frac{\rd z_3^-}{4\pi}  e^{-\img q_1^+  z_1^-/2}\, e^{-\img q_2^+ z_2^-/2}\, e^{\img q_1^+ z_3^-/2} \nn
&& \times \braket{p_2 | \overline T \left\{ \psi_b (0,z_1^-,\mathbf{z}_\perp)  \psi_d (0,z_2^-,\mathbf{0}_\perp) \right\} T \left\{ \overline \psi_e (0,z_3^-,\mathbf{z}_\perp) \overline \psi_g (0)\right\}  |p_2} 
\,.\end{eqnarray}
The factors $p_1^-$ and $p_2^+$  are included to make the dPDFs boost invariant, indicating that each dPDF comes with a $\lqcd/Q$ factor in the cross-section. These dPDFs are dimension two objects and of order $\lqcd^2$ (including the $p_1^-$ factor). The overall minus signs in \eq{37} are due to the ordering of fermion fields. We should point out that we do not explicitly write out the collinear Wilson lines [as in \eqs{PDFdef}{Wn}] for the sake of brevity. Translation invariance gives
\begin{eqnarray}
F_{qq}(q_1^-,q_2^-,\mathbf{z}_\perp,p_1^-)_{acfh}
 &=&  F_{qq}( q_2^-,q_1^-,-\mathbf{z}_\perp,p_1^-)_{cahf}\,, \nn
F_{\bq\bq}(q_1^-,q_2^-,\mathbf{z}_\perp,p_1^-)_{bdeg} &=&  F_{\bq\bq}(q_2^-,q_1^-,-\mathbf{z}_\perp,p_1^-)_{dbge}
\,.\end{eqnarray}

The other possible ways of connecting the fields with the states in \eq{HDDY} that contribute are
\begin{eqnarray} \label{eq:dPDF_unc}
F_{q\bq}(q_1^-,q_2^-,\mathbf{z}_\perp,p_1^-)_{adfg} &=& -4\pi\, p_1^- \int \frac{\rd z_1^+}{4\pi} \frac{\rd z_2^+}{4\pi} \frac{\rd z_3^+}{4\pi}    e^{-i q_1^-  z_1^+/2}\ e^{-i q_2^- z_2^+/2}\ e^{i q_1^- z_3^+/2}  \nn
&&\times \braket{p_1| \overline T \left\{ \overline \psi_a (z_1^+,0,\mathbf{z}_\perp)   \psi_d (z_2^+,0,\mathbf{0}_\perp) \right\} T \left\{ \psi_f (z_3^+,0,\mathbf{z}_\perp) \overline \psi_g (0)\right\}  | p_1} 
\,,\nn
F_{\bq q}(q_1^-,q_2^-,\mathbf{z}_\perp,p_1^-)_{bceh} &=& -4\pi\, p_1^- \int \frac{\rd z_1^+}{4\pi} \frac{\rd z_2^+}{4\pi} \frac{\rd z_3^+}{4\pi}  e^{-i q_1^-  z_1^+/2}\ e^{-i q_2^- z_2^+/2}\ e^{i q_1^- z_3^+/2}  \nn
&&\times \braket{p_1| \overline T \left\{  \psi_b (z_1^+,0,\mathbf{z}_\perp)   \overline \psi_c (z_2^+,0,\mathbf{0}_\perp) \right\} T \left\{ \overline \psi_e (z_3^+,0,\mathbf{z}_\perp)  \psi_h (0)\right\}  | p_1} 
\,,\nn
I_{\bq q}(q_1^-,q_2^-,\mathbf{z}_\perp,p_1^-)_{adeh} &=& 4\pi\, p_1^- \int \frac{\rd z_1^+}{4\pi} \frac{\rd z_2^+}{4\pi} \frac{\rd z_3^+}{4\pi}    e^{-i q_1^-  z_1^+/2}\ e^{-i q_2^- z_2^+/2}\ e^{i q_1^- z_3^+/2}  \nn
&&\times \braket{p_1| \overline T \left\{ \overline \psi_a (z_1^+,0,\mathbf{z}_\perp)   \psi_d (z_2^+,0,\mathbf{0}_\perp) \right\} T \left\{ \overline \psi_e (z_3^+,0,\mathbf{z}_\perp)  \psi_h (0)\right\}  | p_1} 
\,,\nn
I_{q\bq} (q_1^-,q_2^-,\mathbf{z}_\perp,p_1^-)_{bcfg} &=& 4\pi\, p_1^- \int \frac{\rd z_1^+}{4\pi} \frac{\rd z_2^+}{4\pi} \frac{\rd z_3^+}{4\pi}  e^{-i q_1^-  z_1^+/2}\ e^{-i q_2^- z_2^+/2}\ e^{i q_1^- z_3^+/2}  \nn
&&\times \braket{p_1| \overline T \left\{  \psi_b (z_1^+,0,\mathbf{z}_\perp)  \overline \psi_c (z_2^+,0,\mathbf{0}_\perp) \right\} T \left\{  \psi_f (z_3^+,0,\mathbf{z}_\perp)  \overline \psi_g (0)\right\}  | p_1}
\,.\end{eqnarray}
$I_{q\bq}$ is the interference double PDF, corresponding to the contribution shown in \fig{int}. The origin of this name will become clearer in \subsec{int}.
Translation invariance leads to the relations
\begin{eqnarray}
F_{q\bq}(q_1^-,q_2^-,\mathbf{z}_\perp,p_1^-)_{adfg} &=& F_{\bq q}(q_2^-,q_1^-,-\mathbf{z}_\perp,p_1^-)_{dagf} 
\,,\nn
I_{\bq q} (q_1^-,q_2^-,\mathbf{z}_\perp,p_1^-)_{adeh} &=& I_{q \bq}(q_2^-,q_1^-, -\mathbf{z}_\perp,p_1^-)_{dahe}
\,.\end{eqnarray}
The soft functions that correspond to these other contractions can be expressed in terms of the soft function $S$. For $F_{q\bq} F_{\bq q}$ and $I_{\bq q} I_{q\bq}$ respectively, we have
\begin{eqnarray}
 \braket{0 | S_{n,aa'}^\dagger(\mathbf{z}_\perp) S_{\bn,b'b}(\mathbf{z}_\perp) S_{\bn,cc'}^\dagger(0) S_{n,d'd}(0) S_{\bn,ee'}^\dagger(\mathbf{z}_\perp) S_{n,f'f}(\mathbf{z}_\perp) S_{n,gg'}^\dagger(0) S_{\bn,h'h}(0) | 0} = S_{aa'bb'gg'hh'ee'ff'cc'dd'}
 , \nn
 \braket{0 | S_{n,aa'}^\dagger(\mathbf{z}_\perp) S_{\bn,b'b}(\mathbf{z}_\perp) S_{\bn,cc'}^\dagger(0) S_{n,d'd}(0) S_{n,ee'}^\dagger(\mathbf{z}_\perp) S_{\bn,f'f}(\mathbf{z}_\perp) S_{\bn,gg'}^\dagger(0) S_{n,h'h}(0) | 0} = S_{aa'bb'ee'ff'cc'dd'gg'hh'} 
.\end{eqnarray}

Combining all these ingredients, we find that the cross section is given by
\begin{eqnarray} 
\frac{\df \si^{\text{DPS}}}{\df q_1^2\, \df Y_1\, \df q_2^2\, \df Y_2}
&=&  \Big(\frac{4\pi \alpha^2 Q_q^2}{3 s}\Big)^2\, \frac{1}{q_1^2} \Big(\frac{q_1^\mu q_1^\nu}{q_1^2} - g^{\mu \nu}\Big)\, \frac{1}{q_2^2} \Big(\frac{q_2^\al q_2^\bt}{q_2^2} - g^{\al\bt}\Big)\, \gamma^\mu_{a'b'} \gamma^\al_{c'd'} \gamma^{\nu}_{e'f'} \gamma^\bt_{g'h'} \nn
&&\times \int\! \rd^2 \mathbf{z}_\perp\big[
F_{qq,acfh} F_{\bq\bq,bdeg} S_{aa'bb'cc'dd'ee'ff'gg'hh'} + 
F_{q\bq,adfg} F_{\bq q,bceh} S_{aa'bb'gg'hh'ee'ff'cc'dd'} \nn
&& + I_{\bq q,adeh}^{(1)} I_{q\bq,bcfg}^{(2)} S_{aa'bb'ee'ff'cc'dd'gg'hh'} + (q \lra \bq) \big]
\,,\end{eqnarray}
where the first dPDF is for the first hadron, and the second for the second hadron. This expression can be simplified using the color and spin decompositions of the dPDFs, discussed in \subsec{class}. (This is analogous to e.g.~$\braket{p_1 | \overline \psi_a \psi_b | p_1} \sim \delta_{ab}$ by color invariance, which we encountered in single parton scattering.) Using the decomposition in \subsec{class}, we find the result presented earlier,
\begin{eqnarray} \label{eq:si_ddy}
\frac{\df \si^{\text{DPS}}}{\df q_1^2\, \df Y_1\, \df q_2^2\, \df Y_2}
&=& \Big(\frac{4\pi \alpha^2 Q_q^2}{3N_c\, s}\Big)^2 \frac{1}{q_1^2 q_2^2} \int\! \rd^2 \mathbf{z}_\perp \bigg\{\big[(F^{1}_{qq}F^{1}_{\bq\bq} + F^{1}_{\Delta q \Delta q} F^{1}_{\Delta \bq \Delta \bq}) + (F^{1}_{q\bq}F^{1}_{\bq q} + F^{1}_{\Delta q \Delta \bq} F^{1}_{\Delta \bq \Delta q})\big] S^{11} \nn
&& + \frac{2N_c}{C_F} \big[(F^{T}_{qq}F^{T}_{\bq\bq} + F^{T}_{\Delta q \Delta q} F^{T}_{\Delta \bq \Delta \bq}) + (F^{T}_{q\bq}F^{T}_{\bq q} + F^{T}_{\Delta q \Delta \bq} F^{T}_{\Delta \bq \Delta q})\big] S^{TT} \nn
&& + \frac{1}{2} \Big[(I^{1}_{\bq q} + I^{1}_{\Delta \bq \Delta q})(I^{1}_{q \bq} + I^{1}_{\Delta q \Delta \bq}) + I^{1}_{\delta \bq \delta q} I^{1}_{\delta q \delta \bq}\Big] S_I^{11}
\nn
&& + \frac{N_c}{2} \Big[(I^{T}_{\bq q} + I^{T}_{\Delta \bq \Delta q})(I^{1}_{q \bq} + I^{1}_{\Delta q \Delta \bq}) + I^{T}_{\delta \bq \delta q} I^{1}_{\delta q \delta \bq} + (1 \lra T) \Big] S_I^{T1}
\nn
&& + \frac{N_c}{C_F} \Big[(I^{T}_{\bq q} + I^{T}_{\Delta \bq \Delta q})(I^{T}_{q \bq} + I^{T}_{\Delta q \Delta \bq}) + I^{T}_{\delta \bq \delta q} I^{T}_{\delta q \delta \bq}\Big] S_I^{TT} + (q \lra \bq) \bigg\}
\,, \end{eqnarray}
The soft functions are defined below in Eqs.~\eqref{eq:S1_def}-\eqref{eq:SI_def} and are normalized to be 1 at tree-level (except for $S_I^{T1}$ which vanishes at tree-level). We suppressed the arguments of the functions for brevity. The arguments of the dPDFs are the transverse separation $\mathbf{z}_\perp$ and the momentum fractions, given by equations analogous to \eq{Q2Y}. The soft functions only depend on $\mathbf{z}_\perp$. \eq{si_ddy} reduces to  the familiar result in the literature given in \eq{si_old}, when spin and color correlations as well as the interference dPDF are ignored. A rigorous proof of factorization requires one to show that Glauber gluons do not contribute, which we assume here.

The factorized form in \eq{si_ddy} contains soft functions from the soft Wilson lines in Eq.~(\ref{soft}). The soft function for the color-summed dPDF is
\begin{eqnarray} \label{eq:S1_def}
  S^{11} &=& \frac{1}{N_c^2}\, \de_{fa} \de_{hc}\, \de_{be} \de_{dg}\,
  \de_{a'b'} \de_{c'd'} \de_{e'f'} \de_{g'h'}\, S_{aa'bb'cc'dd'ee'ff'gg'hh'} \nn
  &=& \frac{1}{N_c^2}\,\braket{0 | \tr[S_n^\dagger(\mathbf{z}_\perp) S_{\bn}(\mathbf{z}_\perp) S_{\bn}^\dagger(\mathbf{z}_\perp) S_{n}(\mathbf{z}_\perp)]
  \tr[S_{n}^\dagger(0) S_{\bn}(0) S_{\bn}^\dagger(0) S_{n}(0)] | 0} = 1
  \,,
\end{eqnarray}
using unitarity of the soft Wilson line. The color-summed soft function $S^{11}$ is trivial and receives no QCD corrections.
The soft function for the color-correlated dPDF is
\begin{eqnarray} \label{eq:ST_def}
 S^{TT} &=& \frac{2}{C_F N_c}\, T^A_{fa} T^A_{hc}\, T^B_{be} T^B_{dg}\,
  \de_{a'b'} \de_{c'd'} \de_{e'f'} \de_{g'h'}\, S_{aa'bb'cc'dd'ee'ff'gg'hh'} \nn
  &=& \frac{2}{C_F N_c}\,\braket{0 | \tr[T^A S_n^\dagger(\mathbf{z}_\perp) S_{\bn}(\mathbf{z}_\perp) T^B S_{\bn}^\dagger(\mathbf{z}_\perp) S_{n}(\mathbf{z}_\perp)]
  \tr[T^A S_{n}^\dagger(0) S_{\bn}(0) T^B S_{\bn}^\dagger(0) S_{n}(0)] | 0} 
\,.\end{eqnarray}
The color-correlated soft function is nontrivial, and has been normalized to unity at tree-level,  $S^{TT} = 1 + \ord{\al_s}$. 
The soft functions for the interference terms are given by
\begin{eqnarray} \label{eq:SI_def}
  S_I^{11} &=& \frac{1}{N_c^2}\,\braket{0 | \tr[S_n^\dagger(\mathbf{z}_\perp) S_{\bn}(\mathbf{z}_\perp) S_{\bn}^\dagger(0) S_{n}(0)]
  \tr[S_{n}^\dagger(\mathbf{z}_\perp) S_{\bn}(\mathbf{z}_\perp) S_{\bn}^\dagger(0) S_{n}(0)] | 0}
  \,, \nn
  S_I^{T1} &=& \frac{2}{C_F N_c^2}\, \braket{0 | \tr[T^A S_n^\dagger(\mathbf{z}_\perp) S_{\bn}(\mathbf{z}_\perp) S_{\bn}^\dagger(0) S_{n}(0)]
  \tr[T^A S_{n}^\dagger(\mathbf{z}_\perp) S_{\bn}(\mathbf{z}_\perp) S_{\bn}^\dagger(0) S_{n}(0)] | 0} 
  \,, \nn
  S_I^{TT} &=& \frac{2}{C_F N_c}\,\braket{0 | \tr[T^A S_n^\dagger(\mathbf{z}_\perp) S_{\bn}(\mathbf{z}_\perp) T^B S_{\bn}^\dagger(0) S_{n}(0)]
  \tr[T^A S_{n}^\dagger(\mathbf{z}_\perp) S_{\bn}(\mathbf{z}_\perp) T^B S_{\bn}^\dagger(0) S_{n}(0)] | 0} 
\,.\end{eqnarray}
\end{widetext}
Defining
\begin{eqnarray}
\mathcal{W} &=& S_{\bn}^\dagger(\mathbf{z}_\perp) S_{n}(\mathbf{z}_\perp) S_{n}^\dagger(0) S_{\bn}(0)\nn
\mathcal{Y} &=& S_n^\dagger(\mathbf{z}_\perp) S_{\bn}(\mathbf{z}_\perp) S_{\bn}^\dagger(0) S_{n}(0)
\,,\end{eqnarray}
and using the identity
\begin{eqnarray}
T^A_{ab} T^A_{cd} &=& \frac 12 \delta_{ad} \delta_{cb} - \frac{1}{2N_c} \delta_{ab} \delta_{cd}
\,,\end{eqnarray}
they can be written as
\begin{eqnarray} 
  S^{TT} &=&  \frac{1}{C_F N_c}  \braket{0 | \tr[T^A \mathcal{W} \, T^A \mathcal{W}^\dagger] | 0}    \\
   &=& \frac{1}{2 C_F N_c}\, \braket{0 | \tr[\mathcal{W}]\tr[\mathcal{W}^\dagger] | 0}  - \frac{1}{2C_F N_c}\,, \nn
  S_I^{11} &=& \frac{1}{N_c^2}\,\braket{0 | \tr[\mathcal{Y}]^2 | 0}
  \,, \nn
  S_I^{T1} &=& \frac{2}{C_F N_c^2}\, \braket{0 | \tr[T^A \mathcal{Y}]\  \tr[T^A \mathcal{Y}] | 0}
  \,, \nn
  &=& - \frac{1}{C_F N_c^3}\,\braket{0 | \tr[\mathcal{Y}]^2 | 0} + \frac{1}{C_F N_c^2}\, \braket{0 | \tr [\mathcal{Y}^2]  | 0} 
  \,, \nn
  S_I^{TT} &=& \frac{1}{C_F N_c}  \braket{0 | \tr[T^A \mathcal{Y} \, T^A \mathcal{Y}] | 0}\nn
  && -
   \frac{1}{C_F N_c^2}  \braket{0 | \tr[T^A \mathcal{Y}]\ \tr[ T^A \mathcal{Y}] | 0} \nn
   &=& \frac{N_c^2+1}{2 C_F N_c^3}\, \braket{0 | \tr[\mathcal{Y}]^2 | 0}  - \frac{1}{C_F N_c^2}\,\braket{0 | \tr[\mathcal{Y}^2] | 0}\, .\nonumber
\end{eqnarray}

None of these soft functions are trivial. $S^{TT}$, $S_I^{11}$ and $S_I^{TT}$ are normalized to 1 at tree-level. The soft function $S_I^{T1}$ connects interference dPDFs with color structures $T^A \otimes T^A$ and $1 \otimes 1$, and only starts to contributes at order $\al_s$. The three interference soft functions are not independent, but satisfy
\begin{equation}  \label{eq:linear}
  S_I^{T1} = S_I^{11} - S_I^{TT}
\,,\end{equation}
so only two independent interference soft functions exist.

\subsection{WW}
\label{subsec:WW}

We now present analogous formulas for single $W$ production and $WW$ production through DPS. The cross section for $pp \to W$ (including its decay) is given by 
\begin{eqnarray} \label{eq:siW}
\!\!\!\!\frac{\rd \sigma^W}{\rd q^2\,\rd Y} &=& \frac{\rd \sigma_0^{W}}{\rd q^2\, \rd Y}\, [f_q(x_1) f_{\overline q}(x_2) + f_{\overline q}(x_1) f_q(x_2)]
\,,\nn
\!\!\!\!\frac{\rd \sigma_0^W}{\rd q^2\, \rd Y} &=& \frac{\pi \alpha^2 \abs{V_i V_f}^2 q^2 }{12 N_c \sin^4 \theta_W   [(q^2\!-\!m_W^2)^2 \!+\! m_W^2 \Ga_W^2] \, s}
\,,\end{eqnarray}
where $m_W$ and $\Ga_W$ are the mass and width of the $W$ boson. The quark flavors of the PDFs are suppressed in the above equation and are summed over.  The CKM matrix element $V_i$ depends on the flavor of the initial quarks. For example, the PDFs with CKM matrix element $V_i = V_{ud}$ and $V_{ud}^*$ are
\begin{eqnarray}
\frac{\rd \sigma^W}{\rd q^2 \rd Y} &=& \frac{\rd \sigma_0^{W}}{\rd q^2 \rd Y}\, [f_u f_{\overline{d}} + f_{\overline{d}} f_u + f_d f_{\overline{u}} + f_{\overline{u}} f_d]
\,.\end{eqnarray}
If the $W$ decays hadronically, their is also a CKM matrix element $V_f$ for the final state. [For a leptonically decaying $W$, one should set $V_f =1$ in \eq{siW}.]

The corresponding expression for $WW$ production (including decay of the $W$s) through double parton scattering is given by
\begin{widetext}
\begin{eqnarray}
\frac{\rd \sigma^{WW,\text{DPS}}}{\rd q_1^2\,\rd Y_1\, \rd q_2^2\, \rd Y_2}
&=& \frac{\rd \sigma_0^W}{\rd q_1^2\, \rd Y_1} \frac{\rd \sigma_0^W}{\rd q_2^2\, \rd Y_2}
\int\! \df^2 \mathbf{z}_\perp \,
\biggl\{ \Bigl[  \left(F_{qq}^1+F_{\Delta q \Delta q}^1\right)   \left(F_{\bar q\bar q}^1+F_{\Delta \bar q \Delta \bar q}^1\right) 
+ \left(F_{q \bar q}^1+F_{\Delta q \Delta \bar q}^1\right)   \left(F_{\bar q q}^1+F_{\Delta \bar q \Delta  q}^1\right) \Bigr]\nn
&&+\frac{2 N_c}{C_F} \Bigl[  \left(F_{qq}^T+F_{\Delta q \Delta q}^T\right)   \left(F_{\bar q\bar q}^T+F_{\Delta \bar q \Delta \bar q}^T\right)  
+ \left(F_{q \bar q}^T+F_{\Delta q \Delta \bar q}^T\right)   \left(F_{\bar q q}^T+F_{\Delta \bar q \Delta  q}^T\right)\Bigr] S^{TT} \nn
&&+ \Bigl[ \left(I_{q \bar q}^1+I_{\Delta q \Delta \bar q}^1\right)   \left(I_{\bar q q}^1+I_{\Delta \bar q \Delta  q}^1\right) \Bigr] S_I^{11} 
 + N_c \Bigl[ \left(I_{q \bar q}^T+I_{\Delta q \Delta \bar q}^T\right)   \left(I_{\bar q q}^1+I_{\Delta \bar q \Delta  q}^1\right) + (1 \leftrightarrow T) \Bigr] S_I^{T1} \nn
&&+\frac{2N_c}{C_F} \Bigl[ \left(I_{q \bar q}^T+I_{\Delta q \Delta \bar q}^T\right)   \left(I_{\bar q q}^T+I_{\Delta \bar q \Delta  q}^T\right)  \Bigr] S_I^{TT}
+ (q \leftrightarrow \bar q) \biggr\}
\,,\end{eqnarray}
\end{widetext}
where we suppress the quark flavors in the dPDFs.
Note that in contrast to \eq{si_ddy}, the spin structure $I_{\de q \de \bq}$ does not contribute.

\section{Double PDF}
\label{sec:dPDF}

In this section, we classify the allowed color and spin structures for the dPDF.  The $F_{qq}$ dPDF matrix element  has the schematic form
\begin{eqnarray}
\braket{p | \overline \psi_a \overline \psi_c  \psi_f \psi_h | p}\,,
\end{eqnarray}
where the subscripts represent color and spin indices, with similar expressions for  the other dPDFs given in \subsec{DDY}. The allowed color and Lorentz structures for these four-quark matrix elements are derived in this section. The possibility of nontrivial color and spin structures for unpolarized dPDFs was first discussed in Ref.~\cite{Mekhfi:1985dv}. Our decomposition in \subsec{class} is essentially the same as in Ref.~\cite{Diehl:2011yj}, but the normalizations are slightly different.

\subsection{Classification}
\label{subsec:class}

The $qq$ dPDF has  two possible color structures, since the proton is a color singlet state,
\begin{equation}
   \de_{af}\, \de_{ch}
   \,, \
   T^A_{af}\,T^A_{ch}
\,.\end{equation}
These are used to decompose the double PDF into the color-summed and color-correlated dPDFs as
\begin{eqnarray}
\left(F_{qq}\right)_{acfh} &=& \frac{1}{N_c^2} \delta_{fa} \delta_{hc}\, F_{qq}^{1} + \frac{2}{C_F N_c}T^A_{fa} T^A_{hc}\, F_{qq}^{T} \,,\nn
\left(F_{\bq\bq}\right)_{bdeg} &=& \frac{1}{N_c^2} \delta_{be} \delta_{dg}\, F_{\bq\bq}^{1} + \frac{2}{C_F N_c}T^A_{be} T^A_{dg}\,  F_{\bq\bq}^{T} \,,\nn
\left(F_{q\bq}\right)_{adfg} &=& \frac{1}{N_c^2} \delta_{fa} \delta_{dg}\, F_{q\bq}^{1} + \frac{2}{C_F N_c}T^A_{fa} T^A_{dg}\, F_{q\bq}^{T} \,,\nn
\left(F_{\bq q}\right)_{bceh} &=& \frac{1}{N_c^2} \delta_{be} \delta_{hc}\, F_{\bq q}^{1} + \frac{2}{C_F N_c}T^A_{be} T^A_{hc}\,  F_{\bq q}^{T}\,,
\end{eqnarray}
and similarly for $I_{q\bq}$ and $I_{\bq q}$. The double PDF with the $T^A \otimes T^A$ color structure can be interpreted as measuring diparton color correlations. This is clear from decomposing $F_{qq}$ into the $6$ (symmetric tensor) and $\overline{3}$ (antisymmetric tensor) and $F_{q\bq}$ into the 1 and 8 (adjoint) color structures:
\begin{eqnarray} \label{eq:col_int}
 F_{qq}^{(6)} &=& \frac{1}{N_c^2} F_{qq}^{1} + \frac{2}{N_c(N_c+1)} F_{qq}^{T} \,, \nn
 F_{qq}^{(\overline{3})} &=& \frac{1}{N_c^2} F_{qq}^{1} - \frac{2}{N_c(N_c-1)} F_{qq}^{T} \,, \nn
 F_{q\bq}^{(1)} &=& \frac{1}{N_c^2} F_{q\bar q}^{1} + \frac{2}{N_c} F_{q\bar q}^{T} \,, \nn
 F_{q\bq}^{(8)} &=& \frac{1}{N_c^2} F_{q\bar q}^{1} - \frac{2}{N_c(N_c^2-1)} F_{q\bar q}^{T} \,.
\end{eqnarray}

There are several spin structures that can appear in the double PDFs of unpolarized hadrons:
\begin{eqnarray} \label{eq:allspins}
 &&  \bnslash_{af} \bnslash_{ch}
   \,, \quad
 (\bnslash \ga_5)_{af} (\bnslash \ga_5)_{ch}
   \,, \quad
  \bnslash_{af} (\img \si_\perp^{\mu -} \ga_5)_{ch} 
   \,, \nn
 &&(\bnslash \ga_5)_{af} (\img \si_\perp^{\mu -} \ga_5)_{ch} 
   \,, \quad
  (\img \si_\perp^{\mu -} \ga_5)_{af} (\img \si_\perp^{\nu -} \ga_5)_{ch} 
   \,, 
\end{eqnarray}
where
\begin{equation}
  \img \si_\perp^{\mu -} \ga_5 = \bnslash \ga_\perp^\mu \ga_5
\,.\end{equation}
The free indices $\mu$ and $\nu$ in \eq{allspins} can be contracted with $z_\perp$ or with each other, as shown in \eq{spin_w_index}.
In SPS only the spin structure $\bnslash$ contributes for unpolarized hadrons. The structures $\bnslash \ga_5$ and $\img \si_\perp^{\mu -} \ga_5$ enter only for longitudinally and transversely polarized protons, respectively. They contribute to the $g_1(x)$ polarized structure function in deep-inelastic scattering, and the transversity distribution $h_1(x)$, respectively.
In the dPDF we have two partons, and the last two structures can appear for unpolarized protons due to diparton spin correlations. The decomposition of the dPDF into the various spin structures is given by (see also Ref.~\cite{Diehl:2011yj})
\begin{eqnarray}
&& \left(F_{qq}\right)_{acfh} =  \frac{1}{16} \big[ \nslash_{fa} \nslash_{hc} F_{qq} + (\nslash \ga_5)_{fa} (\nslash \ga_5)_{hc} F_{\De q \De q} 
\nn
&&- \nslash_{fa} (\img \si_{\perp \mu +}\gamma_5)_{hc}F^\mu_{q \delta q} - (\img \si_{\perp \mu +}\gamma_5)_{fa} \nslash_{hc}F^\mu_{ \delta q q} \nn
&&- (\nslash\gamma_5)_{fa} (\img \si_{\perp \mu +}\gamma_5)_{hc} F^\mu_{\Delta q \delta q} -  (\img \si_{\perp \mu +}\gamma_5)_{fa} (\nslash \gamma_5)_{hc}F^\mu_{ \delta q \Delta q} \nn
&&+ (\img \si_{\perp \mu +} \ga_5)_{fa} (\img \si_{\perp \nu +} \ga_5)_{hc} F_{\de q \de q}^{\mu\nu}\big]
\,,\end{eqnarray}
with~\footnote{
 Our sign convention for $\eps_\perp$ follows from $\epsilon_\perp^{\mu\nu} \equiv -\epsilon^{\mu\nu \alpha \beta}n_\alpha \bn_\beta /2$ and $\epsilon_{0123}=+1$. For $n^\mu = (1,0,0,1)$ and $\bn^\mu = (1,0,0,-1)$, we have $\epsilon_\perp^{12}=+1$. Note that interchanging $n$ and $\bn$ flips the sign of $\eps_\perp$. We also use $g_\perp^{11}=g_\perp^{22}=-1$, so raising or lowering $\perp$ indices gives a minus sign. }
\begin{eqnarray} \label{eq:spin_w_index}
F^\mu_{q \delta q} &=& M\epsilon^{\mu \nu}_\perp z_{\perp \nu} F_{q \delta q}\,,\quad
F^\mu_{\delta q q} = M\epsilon^{\mu \nu}_\perp z_{\perp \nu} F_{\delta q q}\,,\\
F^\mu_{\Delta q \delta q} &=& M z_{\perp}^\mu F_{\Delta q \delta q}\,,\quad \
F^\mu_{\delta q \Delta q} = M z_{\perp}^\mu F_{\delta q \Delta q}\,,\nn
F^{\mu\nu}_{\delta q \delta q} &=& \frac12 g_\perp^{\mu\nu} F_{\delta q\delta q} + M^2\Big(z_\perp^\mu z_\perp^\nu - \frac12 z_\perp \cdot z_\perp g_\perp^{\mu\nu}\Big)F^t_{\delta q \delta q}
\,,\nonumber
\end{eqnarray}
where $M$ is the proton mass. The resulting dPDFs $F_{q \delta q}$ etc.~depend on $z_\perp^2$ and the momentum fractions. The numerical factors have been chosen to match with the conventional normalization of the PDFs. Translating in the $\perp$ direction by $-z_\perp$ followed by $-z_\perp \to z_\perp$ gives the relations $F_{q \delta q}(x_1,x_2,\mathbf{z_\perp}) = - F_{\delta q q}(x_2,x_1,\mathbf{z_\perp})$, $F_{\Delta q \delta q}(x_1,x_2,\mathbf{z_\perp})  = - F_{\delta q \Delta q}(x_2,x_1,\mathbf{z_\perp})$. 

To clarify the spin structures, it is helpful to write out the double PDFs in terms of quark creation and annihilation operators. For example,
\begin{eqnarray} \label{eq:spin_int}
  F_{qq} &\sim& \braket{p | (a_{1R}^\dagger a_{1R} + a_{1L}^\dagger a_{1L}) (a_{2R}^\dagger a_{2R} + a_{2L}^\dagger a_{2L}) |p}
\,, \nn
  F_{\De q\De q} &\sim&\braket{p | (a_{1R}^\dagger a_{1R} - a_{1L}^\dagger a_{1L}) (a_{2R}^\dagger a_{2R} - a_{2L}^\dagger a_{2L}) |p}
\,, \nn
  F_{\de q \de q} &\sim& \braket{p | 2(a_{1R}^\dagger a_{1L} a_{2L}^\dagger a_{2R} + a_{1L}^\dagger a_{1R} a_{2R}^\dagger a_{2L}) |p}
\,, 
\end{eqnarray}
where we have assumed that the proton is in the $n^\mu = (1,0,0,1)$ direction. $F_{qq}$ measures the joint probability to find two quarks in the proton,
$F_{\Delta q\Delta q}$ measures the correlation between the longitudinal polarization of two quarks in the proton, and $F_{\delta q \delta q}$ measures the correlation between the transverse polarization of two quarks in the proton. $F_{\delta q\delta q}$ does not enter the leading-order factorization formula Eq.~(\ref{eq:si_new}), but can contribute at higher order in $\alpha_s$.

\subsection{Definitions}

Starting from the dPDFs with uncontracted indices in  \eqs{dfact}{dPDF_unc}, we now apply the spin and color decompositions of \subsec{class}. This leads to a large number of dPDFs, most of which contribute to the cross section in \eq{si_ddy}. We show a small sample of these for illustrative purposes below:
\begin{widetext}
\begin{eqnarray}
F_{qq}^1\Big(\frac{q_1^-}{p_1^-},\frac{q_2^-}{p_1^-},\mathbf{z}_\perp\Big) &=& -4\pi\, p_1^- \int \frac{\rd z_1^+}{4\pi} \frac{\rd z_2^+}{4\pi} \frac{\rd z_3^+}{4\pi}     e^{-\img q_1^-  z_1^+/2}\, e^{-\img q_2^- z_2^+/2}\, e^{\img q_1^- z_3^+/2} \nn
&&\times \braket{p_1| \overline T \left\{ \Big[\overline \psi (z_1^+,0,\mathbf{z}_\perp) \frac{\bnslash}{2} \Big]_a  \Big[\overline \psi (z_2^+,0,\mathbf{0}_\perp)\frac{\bnslash}{2} \Big]_b \right\} T \left\{ \psi_a (z_3^+,0,\mathbf{z}_\perp) \psi_b (0)\right\}  | p_1}
\,, \nn
F_{q\bq}^T\Big(\frac{q_1^-}{p_1^-},\frac{q_2^-}{p_1^-},\mathbf{z}_\perp\Big) &=& -4\pi\, p_1^- \int \frac{\rd z_1^+}{4\pi} \frac{\rd z_2^+}{4\pi} \frac{\rd z_3^+}{4\pi}    e^{-i q_1^-  z_1^+/2}\ e^{-i q_2^- z_2^+/2}\ e^{i q_1^- z_3^+/2}  \nn
&&\times \braket{p_1| \overline T \left\{ \Big[\overline \psi (z_1^+,0,\mathbf{z}_\perp)\frac{\bnslash}{2} T^A \Big]_a   \psi_b (z_2^+,0,\mathbf{0}_\perp) \right\} T \left\{ \psi_a (z_3^+,0,\mathbf{z}_\perp) \Big[\overline \psi (0)\frac{\bnslash}{2} T^A \Big]_b\right\}  | p_1} 
\,,\nn
F_{\De q\De \bq}^1\Big(\frac{q_1^-}{p_1^-},\frac{q_2^-}{p_1^-},\mathbf{z}_\perp\Big) &=& -4\pi\, p_1^- \int \frac{\rd z_1^+}{4\pi} \frac{\rd z_2^+}{4\pi} \frac{\rd z_3^+}{4\pi}    e^{-i q_1^-  z_1^+/2}\ e^{-i q_2^- z_2^+/2}\ e^{i q_1^- z_3^+/2}  \nn
&&\times \braket{p_1| \overline T \left\{ \Big[\overline \psi (z_1^+,0,\mathbf{z}_\perp)\frac{\bnslash}{2} \ga_5 \Big]_a   \psi_b (z_2^+,0,\mathbf{0}_\perp) \right\} T \left\{ \psi_a (z_3^+,0,\mathbf{z}_\perp) \Big[\overline \psi (0)\frac{\bnslash}{2} \ga_5 \Big]_b\right\}  | p_1} 
\,,\nn
I_{q \bq}^T\Big(\frac{q_1^-}{p_1^-},\frac{q_2^-}{p_1^-},\mathbf{z}_\perp\Big) &=& 4\pi\, p_1^- \int \frac{\rd z_1^+}{4\pi} \frac{\rd z_2^+}{4\pi} \frac{\rd z_3^+}{4\pi}  e^{-i q_1^-  z_1^+/2}\ e^{-i q_2^- z_2^+/2}\ e^{i q_1^- z_3^+/2}  \nn
&&\times \braket{p_1| \overline T \left\{ \psi_b (z_1^+,0,\mathbf{z}_\perp)  \Big[\overline \psi (z_2^+,0,\mathbf{0}_\perp)\frac{\bnslash}{2} T^A \Big]_a \right\} T \left\{  \psi_a (z_3^+,0,\mathbf{z}_\perp)  \Big[\overline \psi (0)\frac{\bnslash}{2} T^A \Big]_b \right\}  | p_1} 
\,,\nn
I_{\de q \de \bq}\Big(\frac{q_1^-}{p_1^-},\frac{q_2^-}{p_1^-},\mathbf{z}_\perp\Big)&=& 4\pi\, p_1^- \int \frac{\rd z_1^+}{4\pi} \frac{\rd z_2^+}{4\pi} \frac{\rd z_3^+}{4\pi}  e^{-i q_1^-  z_1^+/2}\ e^{-i q_2^- z_2^+/2}\ e^{i q_1^- z_3^+/2}  \nn
&&\times \braket{p_1| \overline T \bigg\{ \psi_b (z_1^+,0,\mathbf{z}_\perp)  \Big[\overline \psi (z_2^+,0,\mathbf{0}_\perp) \frac{\img \si_\perp^{\mu-}}{2} \ga_5 \Big]_a \bigg\} T \bigg\{  \psi_a (z_3^+,0,\mathbf{z}_\perp)  \Big[\overline \psi (0) \frac{\img \si^\perp_{\mu-}}{2} \ga_5 \Big]_b \bigg\}  | p_1} 
.\end{eqnarray}
\end{widetext}
We have written the arguments as $q_1^-/p_1^-$ and $q_2^-/p_2^-$, since by boost invariance the dPDFs can only depend on these combinations. In addition to the color and spin structures, we also have different quark flavors, which have been suppressed. For example, one can have dPDFs $F_{uu}$, $F_{ud}$, $F_{\Delta u \Delta d}$, etc. Taking moments of the dPDF with respect to the momentum fractions $x_i = q_i^-/p_1^-$ turns the dPDF into the matrix element of a bilocal operator.

\subsection{Discrete symmetries}
\label{subsec:sym}

In this section we study the properties of the double PDFs under discrete symmetries, for which it will be convenient to use the dPDFs with uncontracted indices. We start by considering the complex conjugate of the double PDF:
\begin{widetext}
\begin{eqnarray}
F_{qq}(q_1^-,q_2^-,\mathbf{z}_\perp,p_1^-)_{acfh}^* &=& -4\pi\, p_1^- \int \frac{\rd z_1^+}{4\pi} \frac{\rd z_2^+}{4\pi} \frac{\rd z_3^+}{4\pi}\,  e^{\img q_1^-  z_1^+/2}\, e^{\img q_2^- z_2^+/2}\, e^{-\img q_1^- z_3^+/2} \nn
&& \times \braket{p_1| \overline T \left\{ [\overline \psi (0) \ga^0]_h [\overline \psi (z_3^+,0,\mathbf{z}_\perp) \ga^0]_f \right\} T \left\{ [\ga^0 \psi(z_2^+,0,\mathbf{0}_\perp)]_c [\ga^0\psi (z_1^+,0,\mathbf{z}_\perp)]_a \right\}  | p_1} \nn
&=& -4\pi\, p_1^- \int \frac{\rd z_1^+}{4\pi} \frac{\rd z_2^+}{4\pi} \frac{\rd z_3^+}{4\pi}\,    e^{-\img q_1^-  z_1^+/2}\, e^{-\img q_2^- z_2^+/2}\, e^{\img q_1^- z_3^+/2} \nn
&& \times \braket{p_1| \overline T \left\{ [\overline \psi (z_1^+,0,\mathbf{z}_\perp) \ga^0]_f  [\overline \psi (z_2^+,0,\mathbf{0}_\perp) \ga^0]_h \right\} T \left\{ [\ga^0\psi (z_3^+,0,\mathbf{z}_\perp)]_a [\ga^0 \psi(0)]_c \right\}  | p_1} 
\,.\end{eqnarray}
The last line was obtained by interchanging $z_1 \lra z_3$ and using momentum conservation to change the field in which the $z_2^+$ coordinate appears. From
\begin{equation} \label{eq:conjspin}
\ga^0 (\bnslash)^\dagger \ga^0 = \bnslash\,, \quad
\ga^0 (\bnslash \ga_5)^\dagger \ga^0 = \bnslash \ga_5\,, \quad
\ga^0 (\bnslash \ga_\perp^\mu \ga_5)^\dagger \ga^0 = \bnslash \ga_\perp^\mu \ga_5\,, \quad
(T^A)^\dagger = T^A
\,,\end{equation}
it follows that spin and color structures are unaffected. We thus conclude that $F_{qq}$ is real. This is true for the other double PDFs, except for $I_{\bq q}$ and $I_{q\bq}$ which satisfy
\begin{equation}
 I_{\bq q}(q_1^-,q_2^-,\mathbf{z}_\perp,p_1^-)^* =  I_{q\bq}(q_1^-,q_2^-,\mathbf{z}_\perp,p_1^-)
\,.\end{equation}
Thus the individual interference double PDFs are not necessarily real, but their contribution to the cross section is. This is not surprising, given their interpretation in the context of a quark model in \subsec{int}.

Under parity
\begin{eqnarray}
F_{qq}(q_1^-,q_2^-,\mathbf{z}_\perp,p_1^-)_{acfh}^P &=& -4\pi\, p_1^- \int \frac{\rd z_1^+}{4\pi} \frac{\rd z_2^+}{4\pi} \frac{\rd z_3^+}{4\pi}\,  e^{-\img q_1^-  z_1^+/2}\, e^{-\img q_2^- z_2^+/2}\, e^{\img q_1^- z_3^+/2} \\
&& \times \braket{p_1^P| \overline T \left\{ [\overline \psi (0,z_1^+,-\mathbf{z}_\perp) \ga^0]_a  [\overline \psi (0,z_2^+,\mathbf{0}_\perp) \ga^0]_c \right\} T \left\{ [\ga^0\psi (0,z_3^+,-\mathbf{z}_\perp)]_f [\ga^0 \psi(0)]_h \right\}  | p_1^P} \nonumber
\,.\end{eqnarray}
Using equations analogous to \eq{conjspin}, we find that the spin and color structures are unchanged (with $\bn \lra n$), yielding $F_{qq}$ for a proton in the $\bn$ direction with $\mathbf{z}_\perp \to -\mathbf{z}_\perp$. This is not particularly useful, since the double PDF for a proton in the $n$ and $\bn$ direction are already related by a rotation.

Under charge conjugation
\begin{eqnarray}
F_{qq/P}(q_1^-,q_2^-,\mathbf{z}_\perp,p_1^-)_{acfh}^C &=& -4\pi\, p_1^- \int \frac{\rd z_1^+}{4\pi} \frac{\rd z_2^+}{4\pi} \frac{\rd z_3^+}{4\pi}\, e^{-\img q_1^-  z_1^+/2}\, e^{-\img q_2^- z_2^+/2}\, e^{\img q_1^- z_3^+/2} \nn
&& \times \braket{\overline p_1| \overline T \left\{ [\ga^0 \ga^2 \psi (z_1^+,0,\mathbf{z}_\perp)]_a^T  [\ga^0 \ga^2 \psi (z_2^+,0,\mathbf{0}_\perp) \ga^0]^T_c \right\} T \left\{ [\overline \psi (z_3^+,0,\mathbf{z}_\perp) \ga^0 \ga^2]^T_f [\overline \psi(0) \ga^0 \ga^2]^T_h \right\}  | \overline p_1} \nn
&=& F_{\bq \bq/\overline{P}}(q_1^-,q_2^-,\mathbf{z}_\perp,p_1^-)_{fhac}
\,.\end{eqnarray}
relating the double PDF for the proton and the antiproton. Using equations analogous to \eq{conjspin}, we find that the spin and color structures are almost unchanged: For each $\De q \lra \De \bq$ the overall sign flips. For example, $F_{\De q \De q/P}^{T} = F_{\De \bq \De \bq/\overline{P}}^{T}$ and $F_{\De q \de q/P}^{T} = - F_{\De \bq \de \bq/\overline{P}}^{T}$.  This is because $\Delta q$ and $\Delta \bq$ are defined in terms of chirality rather than helicity. For massless particles, chirality is the same as helicity, whereas for anti-particles they are opposite.
\end{widetext}

\subsection{Interpretation}
\label{subsec:int}

We will now provide an interpretation of the double PDF using a nonrelativistic quark model. Alternatively, the interpretation of dPDFs becomes clearer when they are written in terms of light-cone wave functions of the colliding hadrons~\cite{Blok:2010ge}. As a warm up we start with the single PDF in \eq{PDFdef}. Since it is boost invariant, we can work in a frame where $p$ is at rest,
\begin{eqnarray}
\!\!\!\!\ket{p} &\to& \sqrt{2E_p} \sum_s \int\! \frac{\rd^3 \mathbf{k}}{(2\pi)^3} \phi_q(\mathbf{k},s) \ket{\mathbf{k},s} \otimes \ket{ \mathbf{p-k}}
.\end{eqnarray}
Here $\ket{\mathbf{k},s}$ is a quark with momentum $\mathbf{k}$ and spin/color $s$, and $\ket{ \mathbf{p-k}}$ represents everything else that makes up the total momentum of the proton. We have switched to a nonrelativistic normalization for the states. The normalization of the wave function $\phi_q$ is given by
\begin{equation}
  \sum_s \int\! \frac{\df^3 \mathbf{k}}{(2\pi)^3} |\phi_q(\mathbf{k},s)|^2 = 1
\,.\end{equation}
Writing the fields in the PDF in terms of creation and annihilation operators,
\begin{eqnarray}
f_q\Big(\frac{q^-}{p^-}\Big) &=& 2E_p \int\! \frac{\rd z^+}{4\pi}\, e^{-\img z^+ q^-/2} \nn
&& \times \sum_s \int\! \frac{\rd^3 \mathbf{k}}{(2\pi)^3}\, \phi_q(\mathbf{k},s) \sum_{s'} \int\! \frac{\rd^3 \mathbf{k}^\prime}{(2\pi)^3}\, \phi_q(\mathbf{k}^\prime,s^\prime)^* \nn
&&
\times \sum_r \int\! \frac{\rd^3 \mathbf{l}}{(2\pi)^3}\, \sum_{r'} \int\! \frac{\rd^3 \mathbf{l}'}{(2\pi)^3} e^{\img z^+ l^{\prime -}/2}  
 \nn
&& \times\braket{\mathbf{k}^\prime,s^\prime \otimes  \mathbf{p-k}^\prime | \frac{a^\dagger_{l',r'}}{\sqrt{2 E_{l'}}} \frac{a_{l,r}}{\sqrt{2 E_l}} | \mathbf{k},s  \otimes  \mathbf{p-k}}
\nn
&& \times\overline u(\mathbf{l}^\prime,r') \frac{\slashed{\bn}}{2} u(\mathbf{l},r) 
\,.\end{eqnarray}
The matrix element is
\begin{eqnarray}
\frac{\de_{r,s} \de_{r',s'} (2\pi)^3 \delta^{(3)}(\mathbf{k}^\prime-\mathbf{k})(2\pi)^3 \delta^{(3)}(\mathbf{k}^\prime-\mathbf{l}^\prime)
(2\pi)^3 \delta^{(3)}(\mathbf{k}-\mathbf{l})}{2 E_k} 
\,,\nn \end{eqnarray}
leading to
\begin{eqnarray}
f\Big(\frac{q^-}{p^-}\Big) &=& 2E_{p} \!\int\! \frac{\rd z^+}{4\pi} e^{-\img z^+ q^-/2} \!
\int\!\! \frac{\rd^3 \mathbf{k}}{(2\pi)^3} \abs{\phi_q(\mathbf{k})}^2  e^{i z^+ k^-/2}\nn
&&\times \frac{1}{2E_k} \sum_{s,s'} \overline u(\mathbf{k},s') \frac{\slashed{\bn}}{2} u(\mathbf{k},s) \nn
&=& \sum_s \int\! \frac{\rd^3 \mathbf{k}}{(2\pi)^3} \abs{\phi_q(\mathbf{k},s)}^2   \delta\Big(\frac{q^-}{p^-} - \frac{k^-}{p^-}\Big)
\,.\end{eqnarray}
This gives the single PDF its probabilistic interpretation. It is the probability to find a parton in the proton for a given value of $x=q^-/p^-$.

We will now show that a similar probabilistic interpretation is possible for the double PDF. Boost invariance allows us to go to the rest frame again, where we write the wave function for two quarks in a proton as 
\begin{eqnarray}
\ket{p} &\to& \frac{\sqrt{2E_p}}{2} \sum_{s_1,s_2} \int\! \frac{\rd^3 \mathbf{k}_1}{(2\pi)^3} \int\! \frac{\rd^3 \mathbf{k}_2}{(2\pi)^3} \phi_{qq}(\mathbf{k}_1,s_1,\mathbf{k}_2,s_2) 
\nn
&&\times \ket{\mathbf{k}_1, s_1, \mathbf{k}_2, s_2, \mathbf{p-k_1-k_2}}
\,.\end{eqnarray}
We are assuming a single quark flavor for simplicity, and accordingly included an overall factor of $1/2$ in the above equation. The normalization of the wave function $\phi_{qq}$ is
\begin{equation}
\frac{1}{2} \sum_{s_1,s_2} \int\! \frac{\rd^3 \mathbf{k}_1}{(2\pi)^3} \int\! \frac{\rd^3 \mathbf{k}_2}{(2\pi)^3} |\phi_{qq}(\mathbf{k}_1,s_1,\mathbf{k}_2,s_2)|^2 = 1 
\,.\end{equation}
Following the same steps as for the single PDF, we find
\begin{widetext}
\begin{eqnarray}
  F_{qq}^{1}(x_1,x_2,\mathbf{r}_\perp)
  &=& \sum_{s_1,s_2} \int\! \frac{\rd^3 \mathbf{k}_1}{(2\pi)^3}\,\frac{\rd^3 \mathbf{k}_2}{(2\pi)^3}\,
 \phi_{qq}\Big(\mathbf{k}_1+\frac{1}{2}\mathbf{r}_\perp,s_1;\mathbf{k}_2-\frac{1}{2}\mathbf{r}_\perp,s_2\Big)^* 
 \phi_{qq}\Big(\mathbf{k}_1-\frac{1}{2}\mathbf{r}_\perp,s_1;\mathbf{k}_2+\frac{1}{2}\mathbf{r}_\perp,s_2\Big) \nn
 &&\times \delta(x_1 - k_1^-/p^-) \delta(x_2 - k_2^-/p^-) 
\,.\end{eqnarray}
We see that $\mathbf{r}_\perp$ corresponds to a transfer of transverse momentum between the two double PDFs (see \fig{dps_rperp}). In this form the double PDF does not have a probabilistic interpretation. However, if we first switch to a wave function that depends on $k^-$ and $\mathbf{k}_\perp$ (in which we absorb the Jacobian) and then Fourier transform to position space for all transverse variables, the dPDF does have a probabilistic interpretation,
\begin{eqnarray}
  F_{qq}^{1}(x_1,x_2,\mathbf{z}_\perp)
  &=& \sum_{s_1,s_2} \int\! \frac{\df k_1^- \rd^2 \mathbf{k}_{1\perp}}{(2\pi)^3}\,\frac{\df k_2^- \rd^2 \mathbf{k}_{2\perp}}{(2\pi)^3}\,
 \phi_{qq}\Big(k_1^-,\mathbf{k}_{1\perp}+\frac{1}{2}\mathbf{r}_\perp,s_1;k_2^-,\mathbf{k}_{2\perp}-\frac{1}{2}\mathbf{r}_\perp,s_2\Big)^* \nn
 &&\times \phi_{qq}\Big(k_1^-,\mathbf{k}_{1\perp}-\frac{1}{2}\mathbf{r}_\perp,s_1;k_2^-,\mathbf{k}_{2\perp}+\frac{1}{2}\mathbf{r}_\perp,s_2\Big) 
 \delta(x_1 - k_1^-/p^-) \delta(x_2 - k_2^-/p^-) 
 \nn  
  &=& \frac{(p^-)^2}{4\pi^2} \sum_{s_1,s_2} \int\! \df^2 \mathbf{y}_\perp\,
 \Big|\phi_{qq}\Big(x_1 p^-,\mathbf{y}_\perp+\mathbf{z}_\perp,s_1;x_2 p^-,\mathbf{y}_\perp,s_2\Big)\Big|^2
\,.\end{eqnarray}

Likewise, for the other double PDFs we find
\begin{eqnarray}
  F_{q\bq}^{1}(x_1,x_2,\mathbf{r}_\perp)
  &=& \sum_{s_1,s_2} \int\! \frac{\rd^3 \mathbf{k}_1}{(2\pi)^3}\,\frac{\rd^3 \mathbf{k}_2}{(2\pi)^3}\,
 \phi_{q\bq}\Big(\mathbf{k}_1+\frac{1}{2}\mathbf{r}_\perp,s_1;\mathbf{k}_2-\frac{1}{2}\mathbf{r}_\perp,s_2\Big)^* 
 \phi_{q\bq}\Big(\mathbf{k}_1-\frac{1}{2}\mathbf{r}_\perp,s_1;\mathbf{k}_2+\frac{1}{2}\mathbf{r}_\perp,s_2\Big) \nn
 &&\times \delta(x_1 - k_1^-/p^-) \delta(x_2 - k_2^-/p^-) 
 \,, \nn
  I_{q\bq}^{1}(x_1,x_2,\mathbf{r}_\perp)
  &=& \sum_{s_1,s_2} \int\! \frac{\rd^3 \mathbf{k}_1}{(2\pi)^3}\,\frac{\rd^3 \mathbf{k}_2}{(2\pi)^3}\,
 \phi_{q\bq}\Big(\mathbf{k}_2-\frac{1}{2}\mathbf{r}_\perp,s_2;\mathbf{k}_1+\frac{1}{2}\mathbf{r}_\perp,s_1\Big)^* 
 \phi_{q\bq}\Big(\mathbf{k}_1-\frac{1}{2}\mathbf{r}_\perp,s_1;\mathbf{k}_2+\frac{1}{2}\mathbf{r}_\perp,s_2\Big) \nn
 &&\times \delta(x_1 - k_1^-/p^-) \delta(x_2 - k_2^-/p^-) 
\,.\end{eqnarray}
\end{widetext}
By swapping the momenta, we can write  $I_{q\bq}$ as $\phi_{\bq q}^* \phi_{q \bq}$, which makes it clear why it is called the interference double PDF. The interpretation of the various spin and color correlations was given in \eqs{col_int}{spin_int}.

\section{Renormalization Group Evolution}
\label{sec:RGE}

In this section, we compute the renormalization group evolution of the dPDFs and soft functions using a recently introduced rapidity regulator~\cite{Chiu:2011qc,Chiu:2012ir}.
 
\subsection{Rapidity RGE}
\label{subsec:RRGE}

In the calculation of the RG evolution of the dPDFs and soft functions, we will encounter so-called rapidity divergences. These arise because the collinear degrees of freedom (the dPDFs) and soft degrees of freedom (soft functions) are only separated in rapidity and not in invariant mass. Typically, rapidity divergences arise in integrals such as
\begin{eqnarray} \label{eq:rap_eg}
  \int_{k_\perp}^Q \frac{\df k^-}{k^-} &=& \int_{k_\perp}^\rho \frac{\df k^-}{k^-}
  + \int_\rho^Q \frac{\df k^-}{k^-}  \nn
  &\to& \int_{k_\perp}^\infty \frac{\df k^-}{k^-}
  + \int_0^Q \frac{\df k^-}{k^-}
\,.\end{eqnarray}
In the factorized cross section this integral gets split into a contribution from the soft and from the collinear region, as shown on the first line. On the second line we systematically expand in the power counting, where in the soft region $\rho \to \infty$ and in the collinear region $\rho \to 0$. Both the soft and collinear contributions have a rapidity divergence. 
Rapidity divergences only appear when one factorizes the cross section, and must cancel between the collinear and soft contributions. These divergences need to be regulated and the corresponding series of large (single) logarithms of $k_\perp/Q$ in the cross section need to be resummed for a reliable prediction. We achieve this using the recently developed rapidity renormalization group~\cite{Chiu:2011qc,Chiu:2012ir}, which was introduced in the framework of soft-collinear effective theory (SCET)~\cite{Bauer:2000ew,Bauer:2000yr,Bauer:2001ct,Bauer:2001yt}. 

We emphasize that QCD and SCET are equivalent ways of describing double parton scattering and we have purposefully kept our discussion as general as possible. There are a lot of similarities between either approach: For example, the graphs in this section may be calculated using QCD or SCET Feynman rules, since each collinear sector in SCET is essentially a boosted copy of QCD. The two approaches can differ in that terms are moved between the collinear, soft, and hard contributions, even though the total is the same, which is essentially a scheme dependence. One difference, compared to the rapidity regulator of Ref.~\cite{Chiu:2011qc,Chiu:2012ir}, is that in the usual Collins-Soper-Sterman formalism~\cite{Collins:1984kg} the rapidity divergences do not cancel between the soft and collinear contributions and thus appear in the hard subprocess as well (see the discussion in Section 5.8 of Ref.~\cite{Chiu:2012ir}).

Note that factorization already splits momentum integrals by invariant mass, 
\begin{eqnarray} 
  \int_{\Lambda}^Q \frac{\df^4 k}{k^4} &=&   \int_{\Lambda}^\rho \frac{\df^4 k}{k^4} +    \int_{\rho}^Q \frac{\df^4 k}{k^4}\nn
  &\to& \int_{\Lambda}^\infty \frac{\df^4 k}{k^4} +    \int_{0}^Q \frac{\df^4 k}{k^4}
\,,\end{eqnarray}
where $\Lambda$ is an infrared scale of order $\lqcd$. One may think of the first and second term as (roughly) corresponding to the PDF and the partonic cross section. In dimensional regularization the ultraviolet divergence of the first term cancels the IR divergence of the second term. The ultraviolet divergence leads to an anomalous dimension (for the PDF this would yield the usual splitting functions), which can then be used to sum the large logarithms of $Q/\Lambda$. The problem with rapidity divergences is that they are not regulated by dimensional regularization.
\begin{figure}[b]
\includegraphics[width=2.25cm]{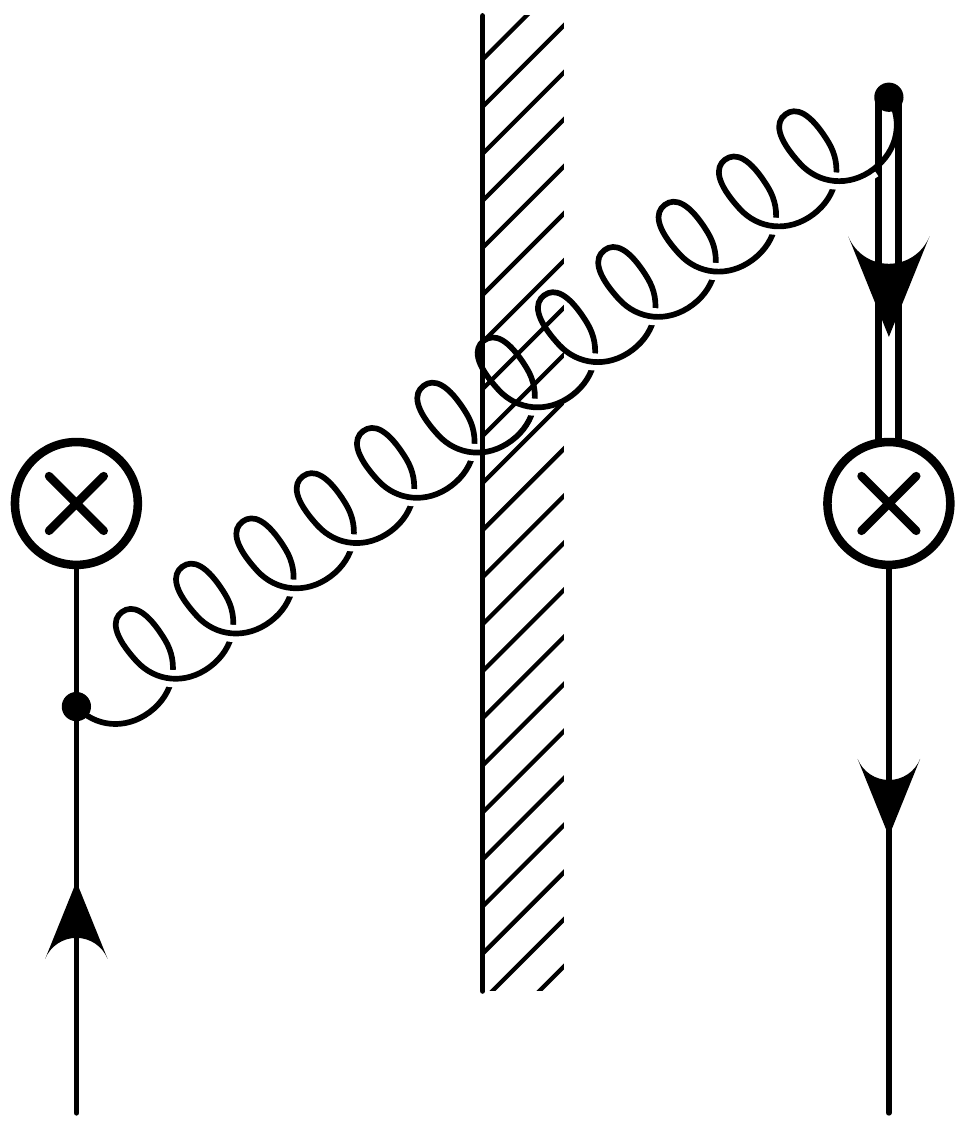} \hspace{2cm}
\includegraphics[width=2.25cm]{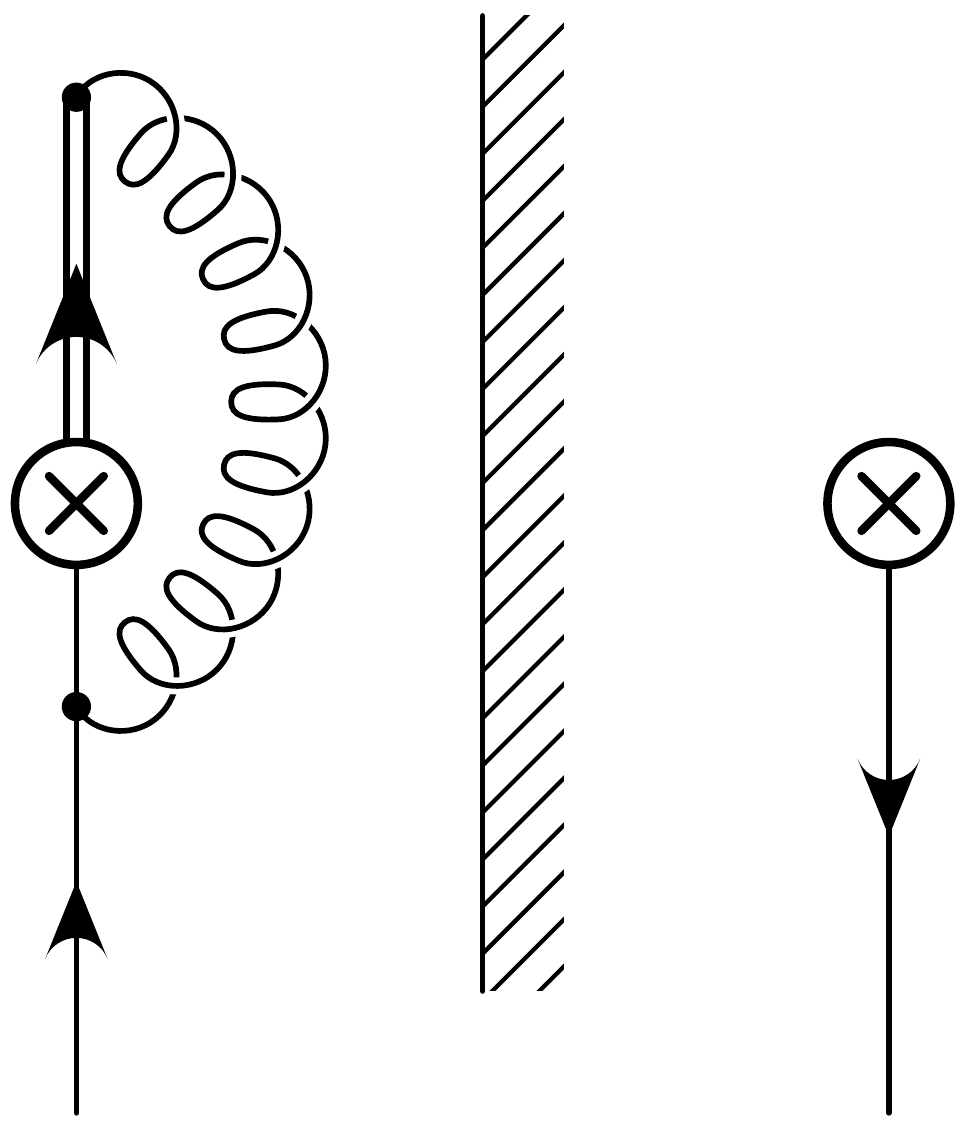}
\caption{Real and virtual contributions to the PDF evolution. The Wilson line is denoted by a double line.}
\label{fig:PDFreal}
\end{figure}

The rapidity renormalization group modifies the collinear Wilson lines [see \eq{Wn}] and the soft Wilson lines,
\begin{eqnarray} \label{eq:WSeta}
  W_n &=& 1 - \frac{g w^2\, \nu^\eta}{(\img \partial^-)^{1+\eta}} A_n^-(0) + \dots \,,\nn
  S_n &=& 1-  \frac{g w\, \nu^{\eta/2}}{\img \partial^+ (2\img\partial^3)^{\eta/2}} A_s^+(0) + \dots
\,.\end{eqnarray}
Here $\eta$ is the rapidity regulator, $\nu$ a new renormalization scale and $w$ a bookkeeping parameter. The proper extension of \eq{WSeta} beyond one-loop, where you can have multiple emissions, is discussed in Refs.~\cite{Chiu:2011qc,Chiu:2012ir}.
Rapidity divergences arise as $1/\eta$ poles in calculations and will lead to a $\nu$-anomalous dimension, in complete analogy to $1/\eps$ poles and the $\mu$ anomalous dimension. The parameter $w$ plays a role in deriving the $\nu$-anomalous dimensions analogous to the running coupling $\al_s$ for the $\mu$-anomalous dimensions,
\begin{eqnarray} \label{eq:run_coupl}
\!\!\!\!\!\! \mu \frac{\df \al_s}{\df \mu} = -2 \eps\, \al_s + \ord{\eps^0}
 \,,\ \
 \nu \frac{ \df w}{\df \nu} = - \frac{\eta}{2}\, w + \ord{\eta^0}
\,.\end{eqnarray}
In contrast to $\al_s$, the running of $w$ does not have a finite term and at the end of the calculation one takes $w=1$. The rapidity logarithms will be summed using the $\nu$-RGE~\cite{Chiu:2011qc,Chiu:2012ir}. One advantage of the rapidity regulator is that the zero-bin subtraction~\cite{Manohar:2006nz} vanishes for our calculation.

We will illustrate the use of the rapidity regulator by calculating two diagrams that arise for the single PDF. The PDF has no rapidity divergences, so the rapidity divergences cancel when you add these diagrams. However, these same diagrams appear with different color factors for the color-correlated dPDF. In that case the rapidity divergences will no longer cancel.
We start by considering the left graph in \fig{PDFreal}. Summing over the external polarization and introducing a gluon mass $M$ to regulate the IR divergences, 
\begin{widetext}
\begin{eqnarray} \label{eq:1A}
I_{A} &=& -\Big(\frac{\mu^2 e^{\ga_E}}{4\pi} \Big)^\eps \frac{1}{2p^-} \int\! \frac{\df^d k}{(2\pi)^d}\, 2\pi \de(k^2-M^2) \de\Big(x - \frac{p^- - k^-}{p^-} \Big) 
 \tr\Big[\frac{-gw^2 \bn_\rho \nu^\eta T^A}{(k^-)^{1+\eta}} \,\frac{\bnslash}{2} \,\frac{\img(\slashed{p}-\slashed{k})}{(p-k)^2}\, \img g \ga^\rho T^A \slashed{p}\Big]\nn
 &=& \frac{g^2 w^2C_F}{2\pi} \Big(\frac{\mu^2 e^{\ga_E}}{4\pi} \Big)^\eps \Big(\frac{\nu}{p^-}\Big)^\eta \frac{x}{(1-x)^{1+\eta}}\, \int\! \frac{\df^{d-2} \mathbf{k}_\perp}{(2\pi)^{d-2}}\, \frac{1}{\mathbf{k}_\perp^2 + x\,m^2} \nn
 &=& \frac{\al_s w^2 C_F}{2\pi}\, \Gamma(\eps) e^{\eps \ga_E} \Big(\frac{\mu^2}{m^2}\Big)^{\eps}  \Big(\frac{\nu}{p^-}\Big)^\eta \frac{x^{1-\eps}}{(1-x)^{1+\eta}} \nn
 &=& \frac{\al_s w^2C_F}{2\pi}\, \bigg\{\Big[-\frac{1}{\eta} \Gamma(\eps) e^{\eps \ga_E} \Big(\frac{\mu^2}{m^2}\Big)^{\eps} - \frac{1}{\eps} \ln \frac{\nu}{p^-}\Big] \de(1-x) + \frac{1}{\eps} \frac{x}{(1-x)_+} + \ord{\eps^0 \eta^0} \bigg\}
\,.\end{eqnarray}
\end{widetext}
To obtain the expanded expression on the last line, we used the distribution identity
\begin{equation}
  \frac{1}{(1-x)^{1+\eta}} = -\frac{1}{\eta} \de(1-x) + \frac{1}{(1-x)_+} + \ord{\eta}
\,.\end{equation}
[The definition of the (standard) plus distribution was given in \eq{plusdef}.]

Absorbing the divergences on the last line of \eq{1A} into the renormalization factor $Z_{A}$, the contribution from this diagram to the anomalous dimension is
\begin{eqnarray} \label{eq:ga1A}
\gamma_{A,\mu} &=& - \mu \frac{\df Z_{A}}{\df \mu} 
= \frac{\al_s C_F}{\pi} \Big[\frac{x}{(1-x)_+} - \delta(1-x) \ln \frac{\nu}{p^-} \Big] \,,\nn
\gamma_{A,\nu} &=& - \nu \frac{\df Z_{A}}{\df \nu} 
= -\frac{\al_s C_F}{2\pi} \ln \frac{\mu^2}{M^2}\, \delta(1-x)
\,.\end{eqnarray}
Here we used \eq{run_coupl}, illustrating the role of $w$.
The left graph in \fig{PDFreal} also has a mirror image and thus contributes with a combinatorial weight of 2.

Note that the anomalous dimension contains an explicit dependence on $p^-$, indicating the breaking of boost invariance. This is required, since any rapidity regulator must break boost invariance [see e.g.~\eq{rap_eg}]. Of course the cross section is still boost invariance.

The second graph in \fig{PDFreal} is
\begin{widetext}
\begin{eqnarray}
I_{B} &=& \Big(\frac{\mu^2 e^{\ga_E}}{4\pi} \Big)^\eps \frac{1}{2p^-}\, \de(1-x) \int\! \frac{\df^d k}{(2\pi)^d}\, 
 \tr\Big[\frac{\bnslash}{2}\, \frac{gw^2 \bn_\rho \nu^\eta T^A}{(k^-)^{1+\eta}}\, \frac{\img(\slashed{p}-\slashed{k})}{(p-k)^2}\, \img g \ga^\rho T^A\, \slashed{p}\Big] \frac{-\img}{k^2-M^2}
 \nn
 &=& 2\img g^2 w^2 C_F\, \de(1-x)  \Big(\frac{\mu^2 e^{\ga_E}}{4\pi} \Big)^\eps \nu^\eta
 \int\! \frac{\df^d k}{(2\pi)^d}\, \frac{p^- - k^-}{(k^-)^{1+\eta}[(k^- - p^-)k^+ - \mathbf{k}_\perp^2 +\img 0] [k^- k^+ - \mathbf{k}_\perp^2 - M^2 + \img 0]} \nn 
 &=& -\frac{g^2 w^2 C_F}{2\pi}\, \de(1-x)  \Big(\frac{\mu^2 e^{\ga_E}}{4\pi} \Big)^\eps \nu^\eta \int_0^{p^-}\!\! \frac{\df k^-}{p^-}\, \frac{p^- - k^-}{(k^-)^{1+\eta}}  \int \frac{\df^{d-2} \mathbf{k}_\perp}{(2\pi)^{d-2}}  \frac{1}{\mathbf{k}_\perp^2 + x\,m^2} \nn
 &=& -\frac{\al_s w^2 C_F}{2\pi}\, \de(1-x)\, \Ga(\eps) e^{\eps \ga_E} \Big(\frac{\mu^2}{M^2} \Big)^\eps \Big(\frac{\nu}{p^-}\Big)^\eta \int_0^{1}\! \df u\, \frac{1-u}{u^{1+\eta}} \nn
 &=& \frac{\al_s w^2 C_F}{2\pi}\, \de(1-x)\, \Ga(\eps) e^{\eps \ga_E} \Big(\frac{\mu^2}{M^2} \Big)^\eps  \frac{1}{\eta(1-\eta)} \Big(\frac{\nu}{p^-}\Big)^\eta \nn
 &=& \frac{\al_s w^2 C_F}{2\pi}\, \de(1-x)\, \Big[\frac{1}{\eta} \Gamma(\eps) e^{\eps \ga_E} \Big(\frac{\mu^2}{m^2}\Big)^{\eps} + \frac{1}{\eps} \ln \frac{\nu}{p^-} + \frac{1}{\eps} + \ord{\eps^0 \eta^0} \Big] 
\,.\end{eqnarray}
\end{widetext}
In the second step we performed the $k^+$ integral by contours. This leads to $0 \leq k^- \leq p^-$, since the poles are otherwise on the same side of the real axis. We then subsequently perform the $\mathbf{k}_\perp$ and $k^-$ integrals, and expand in $\eta$ and $\eps$.
The contribution of this diagram to the anomalous dimension is given by
\begin{eqnarray}
\gamma_{B,\mu} &=& \frac{\al_s C_F}{\pi} \Big( \ln \frac{\nu}{p^-} +1\Big)\delta(1-x) \nn
\gamma_{B,\nu} &=& \frac{\al_s C_F}{2\pi} \ln \frac{\mu^2}{M^2}\delta(1-x)
\end{eqnarray}
This graph also has a mirror image and so has a combinatorial weight of 2. As we anticipated, $\ga_{A,\nu}+\ga_{B,\nu} = 0$. However, this will no longer be true for color-correlated and interference double PDFs. In those cases, we  get essentially the same graphs as discussed here. The first graph still has color factor $C_A$, but the second graph has a different color factor, e.g.~$C_F-C_A/2$ for the color-correlated dPDF. The $\nu$ anomalous dimension no longer vanishes for the dPDF, but cancels against the $\nu$ anomalous dimension of the soft function.

\subsection{Double PDF}
\label{subsec:dPDF_RGE}

In the previous section we gave a brief introduction to rapidity regulator and illustrated its use for two explicit examples. Here we simply tabulate our results for the various diagrams. The diagrams involving a single quark line are shown in table~\ref{tab:sPDF}, with separate columns listing the color factors for the $1\otimes 1$ and $T^A \otimes T^A$ color structures. Since no transverse momentum can be transferred these graphs are all proportional to $\de^{(2)}(\mathbf{r}_\perp)$. The diagrams connecting both quark lines are shown in table~\ref{tab:dPDF}. In principle these diagrams allow for mixing between color structures, but this cancels in the sum over diagrams. Here we used the color factors $C_d$ and $C_1$,
\begin{equation}
  C_d = \frac{N^2-4}{N}
  \,, \quad
  C_1 = \frac{N^2-1}{4N^2}
\,,\end{equation}
and the standard plus distribution defined in \eq{plusdef}.

\renewcommand{\arraystretch}{1.9}
\setlength{\arraycolsep}{3pt}
\begin{table*}
\centering
\begin{eqnarray*}
\begin{array}{cc|c|c|c|c|c}
& \text{Graph} & 1\otimes 1 & T^A\otimes T^A &\hat \gamma_\mu &\hat \gamma_\nu
&\text{Wt.} \\
 \hline
IA & \begin{minipage}{3cm} \includegraphics[height=3cm]{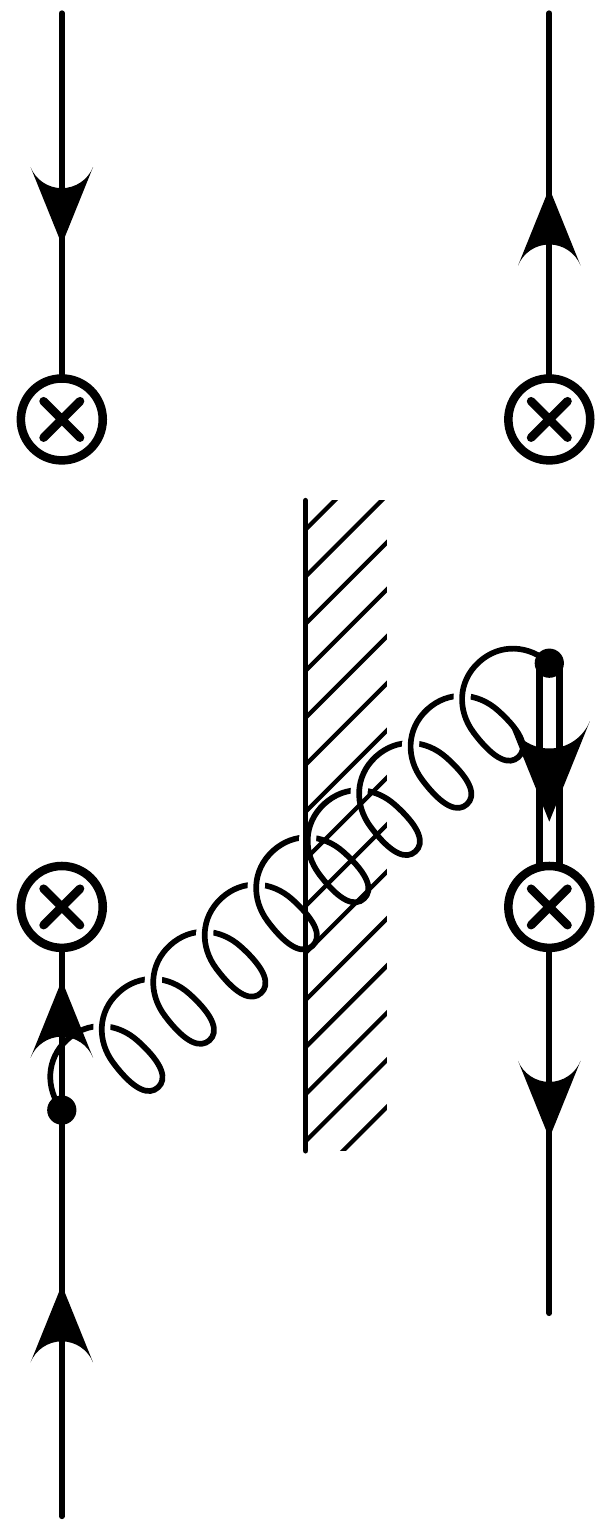} \end{minipage}
& C_F & C_F - \frac{1}{2} C_A &   \frac{x}{(1-x)_+} - \delta(1-x) \ln \frac{\nu}{p^-} & - \de^{(2)}(\mathbf{r}_\perp) \ln \frac{\mu^2}{M^2}& 2  \\ \hline
IB & \begin{minipage}{3cm} \includegraphics[height=3cm]{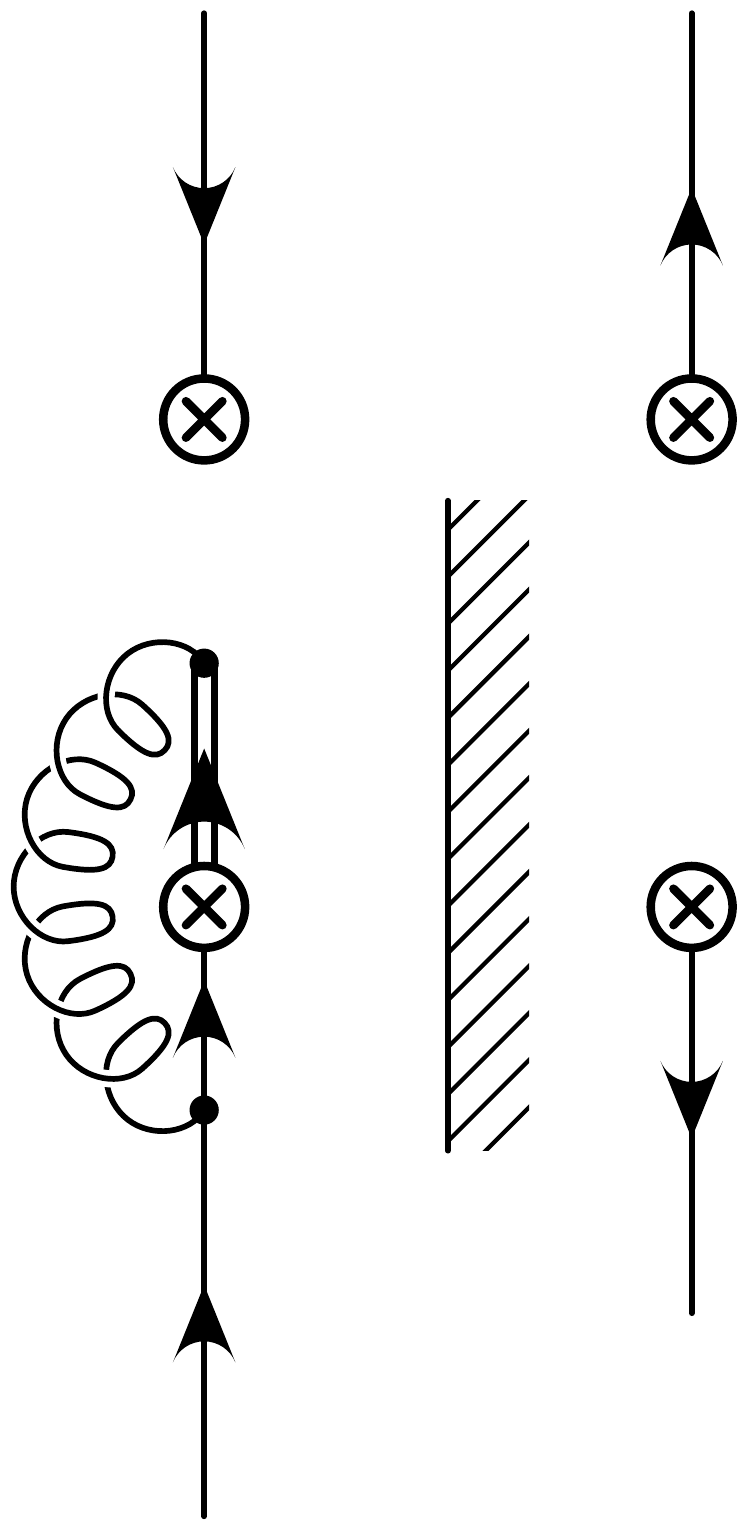} \end{minipage}
 & C_F & C_F & \big( \ln \frac{\nu}{p^-} +1\big)\delta(1-x)  & \de^{(2)}(\mathbf{r}_\perp) \ln \frac{\mu^2}{M^2} & 2 \\ \hline
  II & \begin{minipage}{3cm} \includegraphics[height=3cm]{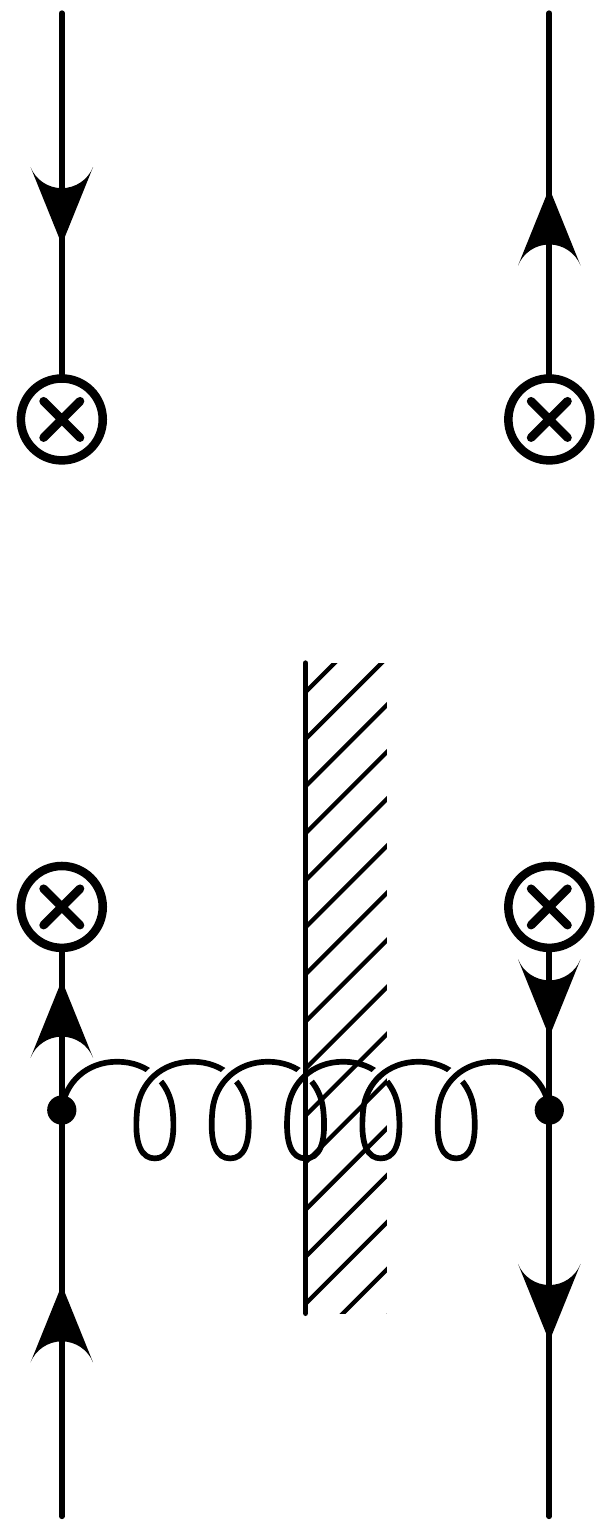} \end{minipage}
 & C_F  & C_F - \frac{1}{2} C_A &  (1-x) \zeta  & 0 & 1 \\ \hline
W & \begin{minipage}{2cm} \includegraphics[width=2cm]{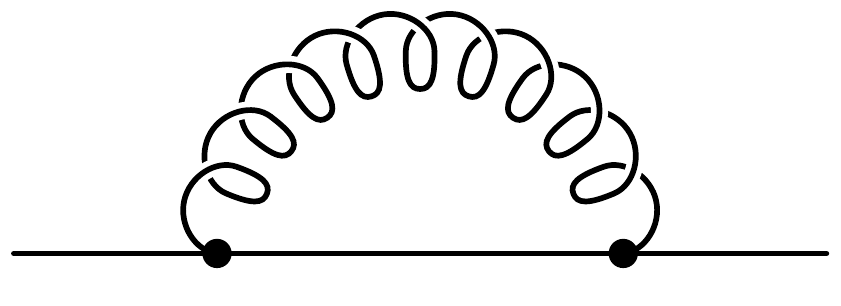} \end{minipage}
& C_F  & C_F &  \frac12\delta(1-x) & 0 & -1 \\
\end{array}
\end{eqnarray*}
\caption{Double PDF renormalization from diagrams involving a single quark line. The columns show the graph, the color factors for the $1\otimes 1$ and $T^A \otimes T^A$ color structures, the $\mu$ and $\nu$ anomalous dimension and the weight factor for each contribution. Only Graph $II$ depends on the spin structure through $\zeta$, where $\zeta = 1$ for $q,\bq, \Delta q, \Delta \bq$ and $\zeta = 0$ for $\de q, \de \bq$.}
\label{tab:sPDF}
\end{table*}

\renewcommand{\arraystretch}{1.9}
\setlength{\arraycolsep}{3pt}
\begin{table*}
\centering
\begin{eqnarray*}
\begin{array}{cc|c|c|c|c|cc}
& \text{Graph} & 1\otimes 1 & T^A \otimes T^A & \hat\gamma_\mu & \hat \gamma_\nu
&\text{Wt.} \\
\hline
 & \multirow{2}{*}{\begin{minipage}{3cm} \includegraphics[height=3cm]{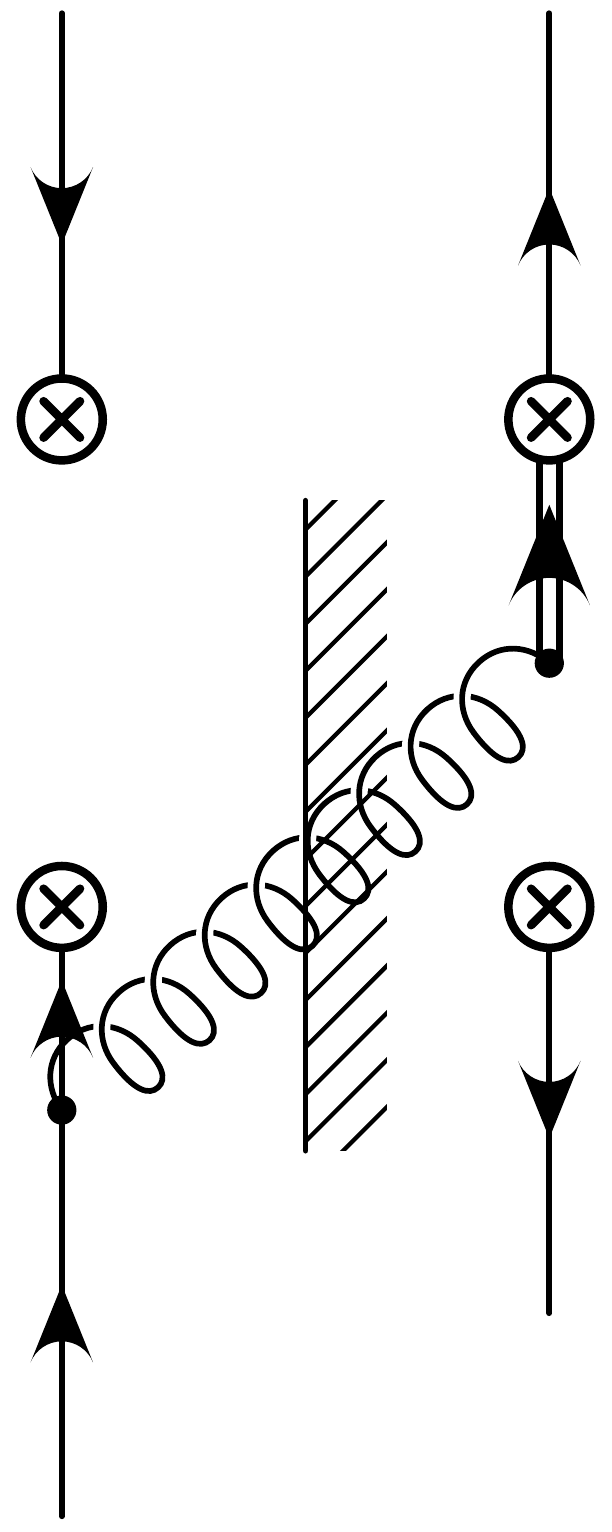}  \end{minipage}} & 0 \left( 1 \otimes 1 \right) &  \frac14(C_d+C_A) T^A \otimes T^A & 0  &
-\frac{1}{\pi} \frac{1}{\mu^2} \frac{1}{(\mathbf{r}_\perp^2/\mu^2)_+} - \de^{(2)}(\mathbf{r}_\perp) \ln \frac{\mu^2}{M^2} & 2 \\[0.2cm]
 IA^\prime &  & T^A \otimes T^A &  C_1\, 1\otimes 1 &  & \\[0.5cm]
 & & & & & & & \\
 \hline
 & \multirow{2}{*}{\begin{minipage}{3cm} \includegraphics[height=3cm]{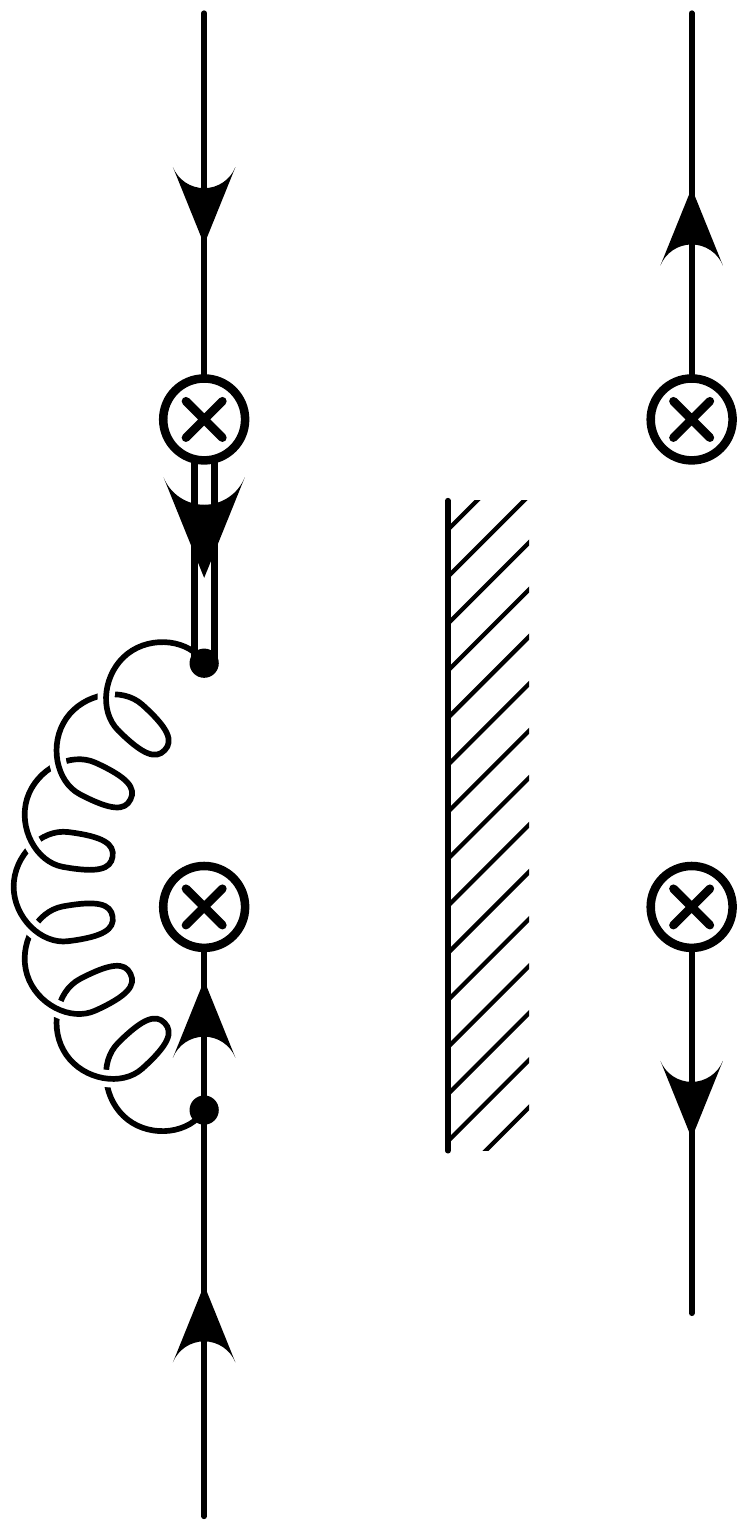} \end{minipage}} & 0\left( 1\otimes 1 \right) &  \frac14(C_d-C_A) T^A \otimes T^A & 0 &
\frac{1}{\pi} \frac{1}{\mu^2} \frac{1}{(\mathbf{r}_\perp^2/\mu^2)_+} + \de^{(2)}(\mathbf{r}_\perp) \ln \frac{\mu^2}{M^2} & 2\\[0.2cm]
  IB^\prime & & T^A \otimes T^A  &  C_1\, 1 \otimes 1  &  &
&   \\[0.5cm]
 & & & & & & & \\
\end{array}
\end{eqnarray*}
\caption{Double PDF renormalization from diagrams involving both quark lines. The columns show the graph, the color factors for the $1\otimes 1$ and $T^A \otimes T^A$ color structures, the $\mu$ and $\nu$ anomalous dimension and the weight factor. The graphs do not depend on spin. The diagrams are not diagonal in color, and the anomalous dimensions are weighted by the sum of both  color factors for a given graph.}
\label{tab:dPDF}
\end{table*}

Adding up these diagrams, we obtain the anomalous dimensions of the dPDFs. The corresponding RGE is 
\begin{eqnarray}
\mu \frac{\rd}{\rd \mu} F(x_1,x_2,\mathbf{r}_\perp) &=& \int\! \frac{\rd y_1}{y_1} \frac{\rd y_2}{y_2}\, \rd^2 \mathbf{k}_\perp \gamma_\mu\Big(\frac{x_1}{y_1},\frac{x_2}{y_2},\mathbf{r}_\perp\!-\!\mathbf{k}_\perp\Big) \nn
&& \times F(y_1,y_2,\mathbf{k}_\perp)
\,,\nn
\nu \frac{\rd}{\rd \nu} F(x_1,x_2,\mathbf{r}_\perp) &=& \int\! \frac{\rd y_1}{y_1} \frac{\rd y_2}{y_2}\, \rd^2 \mathbf{k}_\perp \gamma_\nu\Big(\frac{x_1}{y_1},\frac{x_2}{y_2},\mathbf{r}_\perp\!-\!\mathbf{k}_\perp\Big) \nn
&& \times F(y_1,y_2,\mathbf{k}_\perp)
\,,\end{eqnarray}
where we have chosen not to put any factors of $2\pi$ in the convolution integral. It is convenient to write our one-loop results  for the dPDF anomalous dimensions using
$\hat \gamma_{\mu,\nu}$ defined by
\begin{eqnarray}
\gamma_\mu(x_1,x_2,\mathbf{r}_\perp) 
&=&\frac{\alpha_s(\mu)}{\pi}\, \delta^{(2)}(\mathbf{r}_\perp) \big[\hat \gamma_\mu(x_1) \delta(1-x_2) \nn
&&+\delta(1-x_1) \hat \gamma_\mu(x_2) \big] \,,\nn
\gamma_\nu(x_1,x_2,\mathbf{r}_\perp) 
&=&\frac{\alpha_s(\mu)}{\pi}\, \hat \gamma_\nu(\mathbf{r}_\perp) \delta(1\!-\!x_1) \delta(1\!-\!x_2)\,.
\end{eqnarray}

For the color-summed dPDF $F^{1}$, the total anomalous dimensions are
\begin{eqnarray}
\hat \gamma_\mu^{F^{1}}(x) &=&C_F\Big[\frac{2x}{(1-x)_+} +  (1-x)\zeta + \frac32 \delta(1-x) \Big]  \nn
&=& C_F P_{qq}(x)\,, \nn
\hat \gamma_\nu^{F^{1}}(\mathbf{r}_\perp) &=& 0
\,.\end{eqnarray}
which is the usual evolution for each of the individual quarks. It is understood that the splitting function is modified, depending on the spin structure. The splitting function for the longitudinal polarization is the same as for the unpolarized case, $P_{\De q \De q}(x) = P_{qq}(x)$. For the transverse polarized case, $P_{\de q \de q}(x) = P_{qq}(x) - (1-x)$. For mixed spin structures such as $F_{q \delta q}(x_1,x_2,\mathbf{r}_\perp)$, the $P_{qq}$ splitting function is used for $x_1$ and the $P_{\delta q \delta q}$ spitting function is used for $x_2$.

For the color-correlated dPDF we then find
\begin{eqnarray} \label{eq:ga_F_T}
\hat \gamma_\mu^{F^{T}} (x)
&=&  \left(C_F-\frac12C_A\right) P_{qq}(x) \nn
&&+  C_A \Big( \ln \frac{\nu}{p^-} + \frac34\Big) \delta(1-x) \,,\nn
\hat \gamma_\nu^{F^{T}}(\mathbf{r}_\perp) &=& -\frac{C_A}{\pi} \frac{1}{\mu^2} \frac{1}{(\mathbf{r}_\perp^2/\mu^2)}_+
\,.\end{eqnarray}
Note that the gluon mass $M$ drops out in the sum over diagrams and does not appear in the anomalous dimensions, as it should. A simple cross-check on these results is provided by
\begin{equation} \label{eq:gamu_ganu}
 \nu \frac{\df \ga_\mu}{\df \nu} = \mu \frac{\df \ga_\nu}{\df \mu}
\,,\end{equation}
which can be verified using
\begin{equation}
 \mu \frac{\df}{\df \mu} \frac{1}{\mu^2} \frac{1}{(\mathbf{r}_\perp^2/\mu^2)}_+ = - 2 \delta(\mathbf{r}_\perp^2)=-2 \pi \delta^{(2)}(\mathbf{r}_\perp)\,.
\end{equation}

From the anomalous dimension in \eq{ga_F_T} we read off that the natural scales to evaluate $F^{T}$ are $\mu \sim |\mathbf{r}_\perp| \sim \lqcd$ and $\nu \sim p^- \sim Q$. By evaluating $F^{T}$ at this scale and running it to some common scale $(\mu,\nu)$ for all the functions in the factorization theorem, the large logarithms are summed. 

\subsection{Soft Function}
\label{subsec:soft_RGE}

We now calculate the anomalous dimension of the soft function $S^{TT}$ in momentum space. We remind the reader that the soft function $S^{11} = 1$ and so does not receive any QCD corrections.

\begin{figure}
 \includegraphics[width=1.5cm]{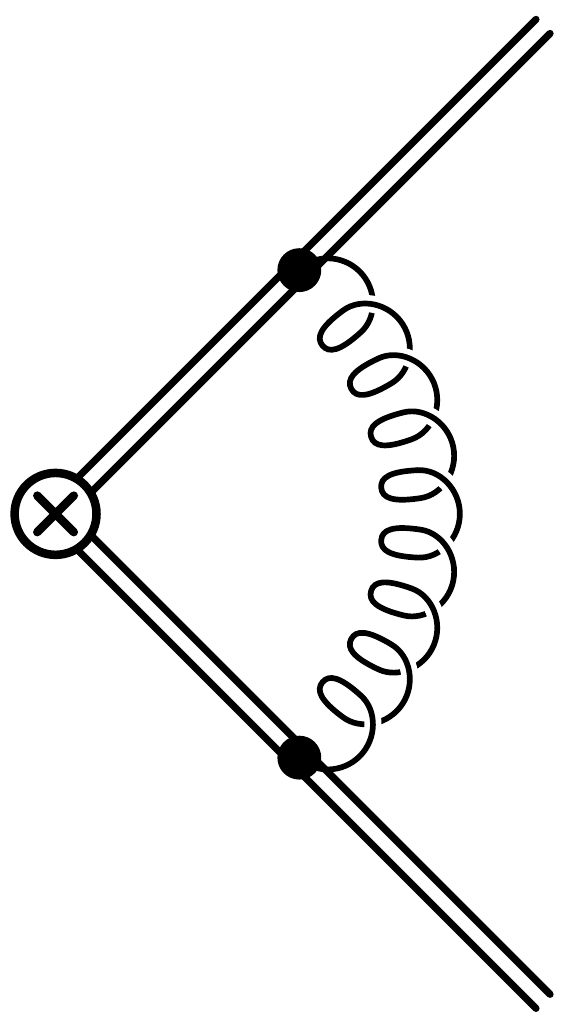}\hspace{2cm}
 \includegraphics[width=1.75cm]{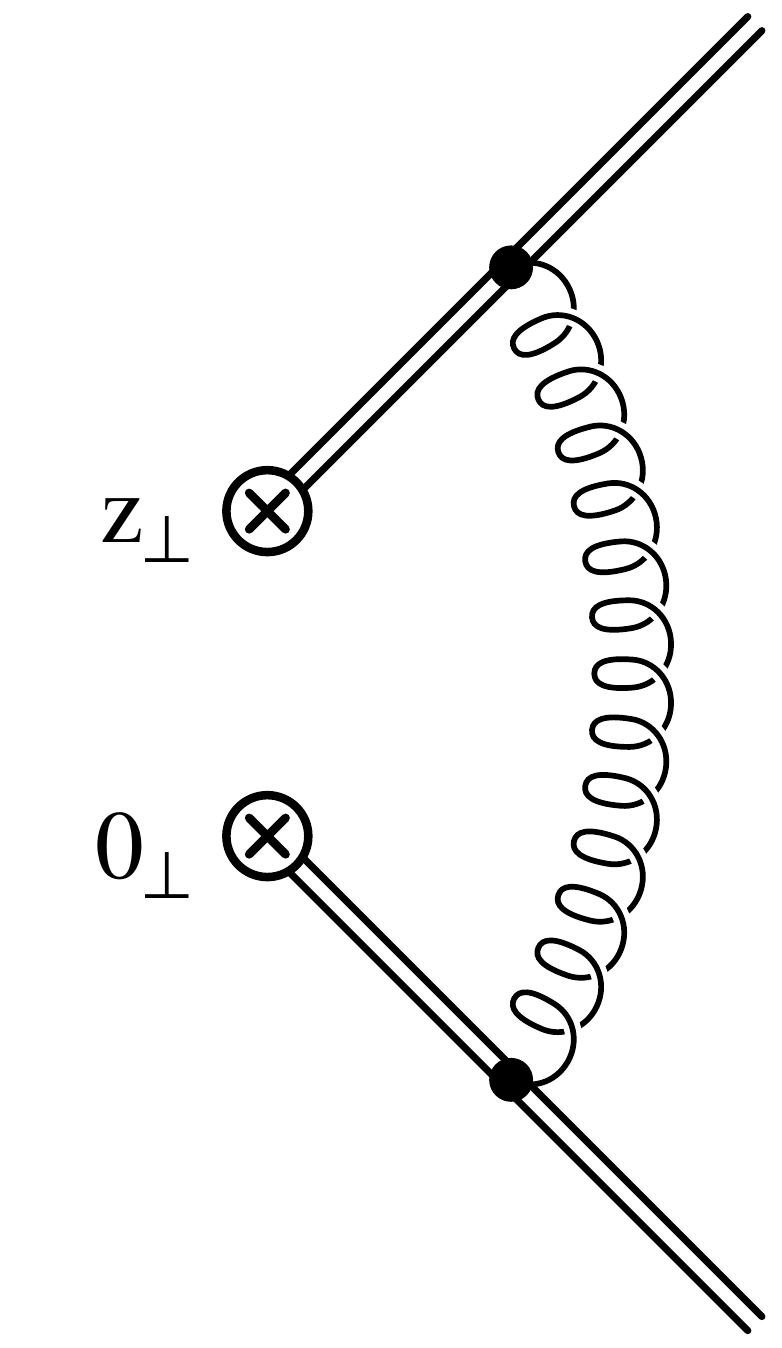}
 \caption{One-loop soft function diagrams. Only the Wilson lines with gluons attached are shown.}
 \label{fig:soft}
\end{figure}

The one-loop contribution from gluon exchange between Wilson lines at the same position, e.g. $S_{\bn}^\dagger(0)$ and $S_n(0)$, is the left graph shown in \fig{soft} and is given by
\begin{eqnarray}\label{eq:Is1}
 I_{S1} &=& -2(2\pi)^2 \img g^2 w^2 \de^{(2)}(\mathbf{r}_\perp)\, \Big(\frac{e^{\ga_E} \mu^2}{4\pi}\Big)^\eps \nu^\eta \int\! \frac{\df^d k}{(2\pi)^d} \nn
 &&\times  \frac{|2k^3|^{-\eta}}{(-k^-+\img 0)(k^2-M^2+\img 0)(-k^+ +\img 0)} \nn
  &=& \frac{\al_s w^2}{\pi}\, (2\pi)^2 \de^{(2)}(\mathbf{r}_\perp) \bigg[ -\frac{1}{\eta} e^{\eps \ga_E} \Ga(\eps) \Big(\frac{\mu^2}{M^2}\Big)^\eps
  + \frac{1}{2\eps^2}  \nn
  && + \frac{1}{2\eps} \ln \frac{\mu^2}{\nu^2} + \ord{\eta^0 \eps^0}\bigg]
  \,,
\end{eqnarray}
where we left out the color factor. We calculated this loop integral by first performing the $k^0$ integral by contours, followed by the $k^3$ integral and the standard $\mathbf{k}_\perp$ integral. We have checked that the result is independent of the $\img 0$-prescription of the soft Wilson lines (the difference is a scaleless integral), implying that the result for this graph is the same for incoming quarks or outgoing anti-quarks, and vice versa.

The exchange between Wilson lines at different positions, e.g. $S_{\bn}^\dagger(\mathbf{z}_\perp)$ and $S_n(0)$, is the right graph shown in \fig{soft} and is given by
\begin{eqnarray} \label{eq:Is2}
 I_{S2} &=& -2(2\pi)^2 \img g^2 w^2 \Big(\frac{e^{\ga_E} \mu^2}{4\pi}\Big)^\eps \nu^\eta \int\! \frac{\df^d k}{(2\pi)^d}\, \de^{(2)}(\mathbf{k}_\perp - \mathbf{r}_\perp) \nn
 &&\times \frac{|2k^3|^{-\eta}}{(-k^-+\img 0)(k^2-M^2+\img 0)(-k^+ +\img 0)}\,  \nn
  &=& \frac{\al_s w^2}{\pi} \bigg[- 4\pi \frac{1}{\eta} \frac{1}{\mathbf{r}_\perp^2 + M^2} + \ord{\eta^0,\eps^0} \bigg] \nn
  &=& \frac{\al_s w^2}{\pi} \bigg\{ \!-\!\frac{1}{\eta} \bigg[4\pi \frac{1}{\mu^2} \frac{1}{(\mathbf{r}_\perp^2/\mu^2)}_+ \!\!\!+\! \ln \frac{\mu^2}{M^2}\, (2\pi)^2 \de^{(2)}(\mathbf{r}_\perp) \bigg] \nn
  &&+ \ord{\eta^0,\eps^0} \bigg\}\,.
\end{eqnarray}
The calculation is very similar, except that this time the $\mathbf{k}_\perp$ integral is performed using the delta function. In the last step we take the limit $M^2 \to 0$ to isolate the IR divergences.

The RG equations for the soft function are given by
\begin{eqnarray} \label{eq:S_RGE}
  \mu \frac{\df S(\mathbf{r}_\perp)}{\df \mu} &=& \int\!\df^2 \mathbf{k}_\perp\, \ga_\mu^{S}(\mathbf{k}_\perp) S(\mathbf{r}_\perp-\mathbf{k}_\perp)
\,,\nn 
  \nu \frac{\df S(\mathbf{r}_\perp)}{\df \nu} &=& \int\!\df^2 \mathbf{k}_\perp\, \ga_\nu^{S}(\mathbf{k}_\perp) S(\mathbf{r}_\perp-\mathbf{k}_\perp)
\,.\end{eqnarray}
For $S^{TT}$ the color factors for graphs $S1$ and $S2$ are
\begin{eqnarray}
 \frac{2}{C_F N_c} \times 4\, \tr[T^A T^C [T^C,T^B]]\, \tr[T^A T^B] &=& 2C_A
 \,, \\
 \frac{2}{C_F N_c} \times 4\, \tr[T^A T^C T^B]\, \tr[T^A [T^C,T^B]] &=& -2C_A
\,,\nonumber \end{eqnarray} 
leading to the anomalous dimensions
\begin{eqnarray} \label{eq:ga_S}
  \ga_\mu^{S^{TT}}(\mathbf{r}_\perp) &=& \frac{2\al_s(\mu)C_A}{\pi} \ln \frac{\mu^2}{\nu^2}\, \de^{(2)}(\mathbf{r}_\perp)\,,\nn
  \ga_\nu^{S^{TT}}(\mathbf{r}_\perp) &=& \frac{2\al_s(\mu) C_A}{\pi^2}\, \frac{1}{\mu^2} \frac{1}{(\mathbf{r}_\perp^2/\mu^2)}_+ 
.\end{eqnarray}
Note that the IR divergences again cancel in the sum of the diagrams. From \eq{ga_S} we conclude that the natural scales for evaluating the soft function are $\nu \sim \mu \sim |\mathbf{r}_\perp| \sim \lqcd$. An important cross check is provided by
\begin{equation}
  2\ga_\nu^{F^{T}} + \ga_\nu^{S^{TT}} \de(1-x_1) \de(1-x_2) = 0
\,,\end{equation}
since the rapidity divergences should cancel between the collinear and soft sectors.

\subsection{Rapidity Resummation}
\label{subsec:nu_evo}

We now turn to solving the $\nu$-RGE. For definiteness we discuss this for the color-correlated soft function, whose RGE and anomalous dimension are given in \eqs{S_RGE}{ga_S}.
This is most easily solved in Fourier space, where the convolution becomes a product. The anomalous dimension in position space is given by
\begin{eqnarray}
\ga_\nu^{S^{TT}}(\mathbf{z}_\perp)  &\equiv&\int\! \df^2 \mathbf{r}_\perp \, e^{\img \mathbf{r}_\perp \cdot\, \mathbf{z}_\perp}\, \ga_\nu^{S^{TT}}(\mathbf{r}_\perp) 
  \nn
  &=&
  \frac{2\al_s C_A}{\pi^2} \int\! \df^2 \mathbf{r}_\perp\,e^{\img \mathbf{r}_\perp \cdot\, \mathbf{z}_\perp}\, \frac{1}{\mu^2} \frac{1}{(\mathbf{r}_\perp^2/\mu^2)}_+ 
  \nn
 &=&  - \frac{2\al_s C_A}{\pi} \ln \frac{\mathbf{z}_\perp^2 \mu^2 e^{2\ga_E}}{4} 
\,,\end{eqnarray}
where there is no $1/(2\pi)^2$ in the Fourier transform since there is none in \eq{S_RGE}.
It is now straightforward to solve the RGE, since it is local in $\mathbf{z}_\perp$,
\begin{equation}
  \nu \frac{\df S^{TT}(\mathbf{z}_\perp)}{\df \nu} = \ga_\nu^{S^{TT}}(\mathbf{z}_\perp) S^{TT}(\mathbf{z}_\perp)
\,,\end{equation}
yielding the evolution kernel
\begin{eqnarray} \label{eq:bsol}
  U_\nu^{S^{TT}}(\mathbf{z}_\perp,\nu,\nu_0) &=& \Big(\frac{\mathbf{z}_\perp^2 \mu^2 e^{2\ga_E}}{4}\Big)^{-\om}
  \,,\nn
  \om &=& \frac{2 \al_s(\mu) C_A}{\pi} \ln \frac{\nu}{\nu_0}
\,.\end{eqnarray}
Transforming back to momentum space, we find
\begin{eqnarray} \label{eq:ptold}
 &&S^{TT}(\mathbf{r}_\perp,\nu) = \int\!\df^2 \mathbf{k}_\perp\, U_\nu^{S^{TT}}(\mathbf{k}_\perp,\nu,\nu_0) S^{TT}(\mathbf{r}_\perp-\mathbf{k}_\perp,\nu_0) 
 \,,\nn
 &&U_\nu^{S^{TT}}(\mathbf{k}_\perp,\nu,\nu_0) = \frac{\om\, e^{-2\ga_E \om}}{\pi}\, \frac{\Ga(1-\om)}{\Ga(1+\om)} \frac{1}{\mu^{2\om}\, (\mathbf{k}_\perp^2)^{1-\om}}
.\end{eqnarray}
This solution is only valid for $\om < \frac{3}{4}$, though it can be analytically continued for $\om < 1$. The singularity for $\om \to 1$ has been observed before in transverse momentum resummation for Drell-Yan like processes~\cite{Frixione:1998dw}. The problem is that \eq{bsol} develops a singularity at $\mathbf{z}_\perp = 0$. This UV region should not contribute and in Ref.~\cite{Bozzi:2003jy} it was essentially cut off. We resolve this issue through our choice of renormalization scale: Rather than $\nu_0 \sim |\mathbf{r}_\perp|$, we choose $\nu_0 = 2e^{-\ga_E}/|\mathbf{z}_\perp|$. This scale choice leads to
\begin{eqnarray}
  U_\nu^{S^{TT}}\Big(\mathbf{z}_\perp,\nu,\frac{2e^{-\ga_E}}{|\mathbf{z}_\perp|}\Big) &=& \exp\!\Big[\!-\!\frac{\al_s(\mu) C_A}{\pi} \Big(L^2 \!+\! 2L \ln \frac{\nu}{\mu}\Big)\Big] 
  \,,\nn
  L &=& \ln \frac{\mathbf{z}_\perp^2 \mu^2 e^{2\ga_E}}{4}
\,.\end{eqnarray}
Its Fourier transform is given by
\begin{eqnarray} \label{eq:ptnew}
S^{TT}(\mathbf{r}_\perp,\nu) &=& \int\! \df \mathbf{k}_\perp\, U_\nu^{S^{TT}}\Big(\mathbf{k}_\perp,\nu,\frac{2e^{-\ga_E}}{|\mathbf{z}_\perp|}\Big) \\
&& \times S^{TT}\Big(\mathbf{r}_\perp-\mathbf{k}_\perp,\frac{2e^{-\ga_E}}{|\mathbf{z}_\perp|}\Big) 
\,,\nn
  U_\nu^{S^{TT}}\Big(\mathbf{k}_\perp,\nu,\frac{2e^{-\ga_E}}{|\mathbf{z}_\perp|}\Big) &=& \frac{1}{4\pi} \!\int\!\! \df \mathbf{z}_\perp^2 J_0(|\mathbf{z_\perp}||\mathbf{k}_\perp|) \nn
  && \times \exp\!\Big[\!-\!\frac{\al_s(\mu) C_A}{\pi} \Big(L^2 \!+\! 2L \ln \frac{\nu}{\mu}\Big)\Big] 
,\nn
S^{TT}\Big(\mathbf{r}_\perp,\frac{2e^{-\ga_E}}{|\mathbf{z}_\perp|}\Big) &=& \int\! \df^2 \mathbf{z}_\perp\,e^{-\img \mathbf{r}_\perp \cdot\, \mathbf{z}_\perp}\, S^{TT}\Big(\mathbf{z}_\perp,\frac{2e^{-\ga_E}}{|\mathbf{z}_\perp|}\Big) 
.\nonumber\end{eqnarray}
The argument $2e^{-\ga_E}/|\mathbf{z}_\perp|$ is kept in momentum space evolution kernel and soft function as a reminder of our original scale choice for $\nu_0$, though $\mathbf{z}_\perp$ is of course no longer a variable, since it has been integrated over. In the integral for $U_\nu^{S^{TT}}$ in \eq{ptnew}, only the region where $|\mathbf{z}_\perp| \sim 1/\mu \sim 1/|\mathbf{r}_\perp|$ contributes. We emphasize that the scale choice for $\nu_0$ cancels between the evolution and the soft function up to the order in $\al_s$ that one is working at. Our choice for $\nu_0$ rearranges the resummed perturbation theory, which is required to make the evolution in \eq{ptold} well behaved.

Our solution is similar to the one proposed in Ref.~\cite{Becher:2010tm}. There an analytic expression for the leading $(1-\om)$ piece of $U_\nu^{S^{TT}}$ in \eq{ptnew} was also obtained, using Borel resummation.

\subsection{Interference Double PDF}

In \subsec{dPDF_RGE} we calculated the anomalous dimension for the regular dPDF, which we repeat here for the interference dPDF. The results are shown in table \ref{tab:iPDF}.
\begin{table*}
\centering
\renewcommand{\arraystretch}{2}
\setlength{\arraycolsep}{3pt}
\begin{eqnarray*}
\begin{array}{cc|c|c|c|c|cc}
& \text{Graph} & \text{Singlet} & \text{Octet} &\hat \gamma_\mu &\hat \gamma_\nu
&\text{Wt.} \\
 \hline
 & \multirow{2}{*}{\begin{minipage}{3cm} \includegraphics[height=3cm]{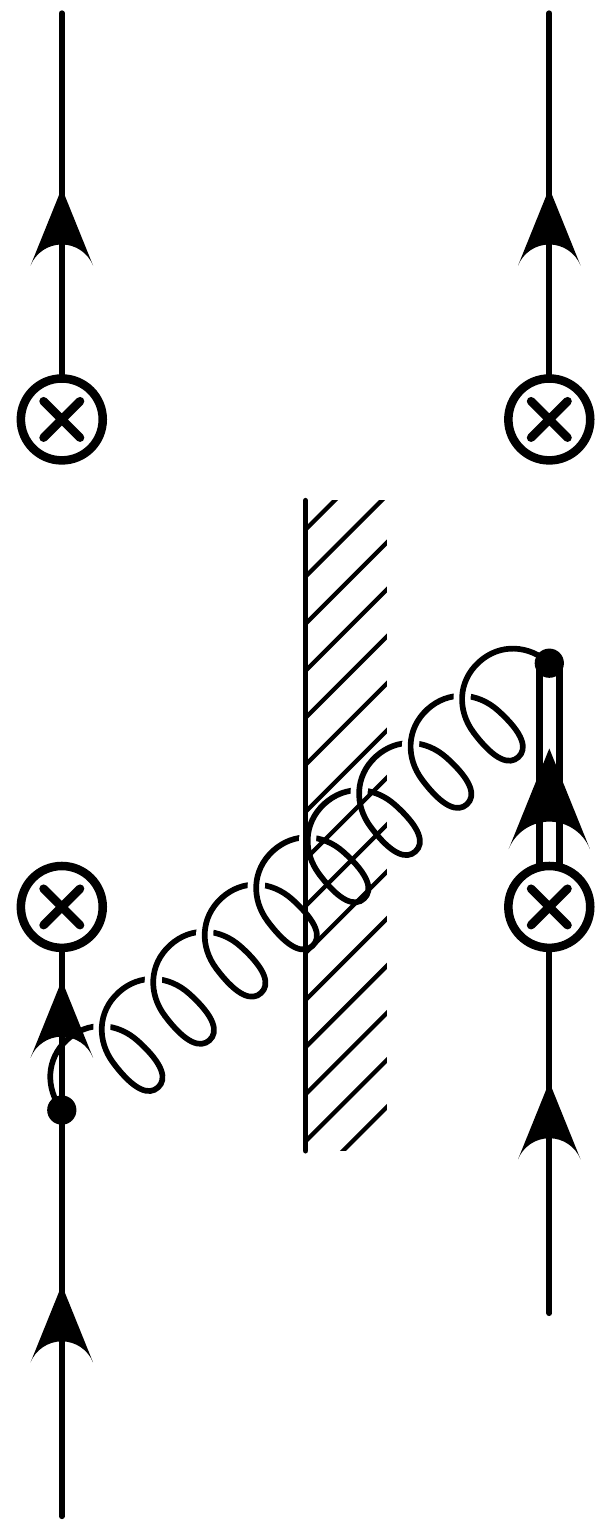} \end{minipage}}
& 0\, ( 1 \otimes 1 )& \frac14(C_A-C_d) T^A \otimes T^A &    \frac{x}{(1-x)_+} - \delta(1-x) \ln \frac{\nu}{p^-} & -\delta^2(\mathbf{r}_\perp)  \ln \frac{\mu^2}{M^2}& 2  \\[0.2cm]
IA & & -T^A \otimes T^A & - C_1 1 \otimes 1 & & & \\[0.5cm]
 & & & & & & & \\  
\hline
 IA^\prime & \begin{minipage}{3cm} \includegraphics[height=3cm]{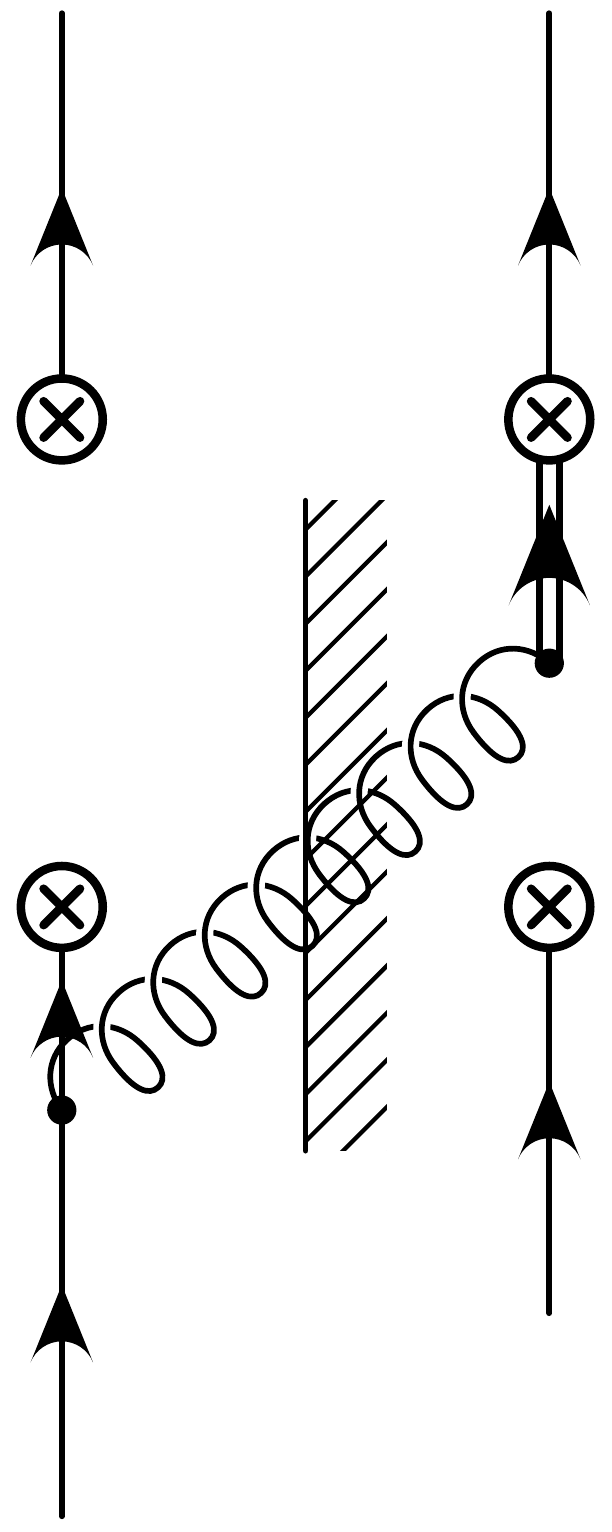} \end{minipage} & C_F  & C_F - \frac12 C_A& 0 &
\parbox{20ex}{$-\frac{1}{\pi} \frac{1}{\mu^2} \frac{1}{(\mathbf{r}_\perp^2/\mu^2)_+}$ \\[5pt]
$-  \de^{(2)}(\mathbf{r}_\perp)\ln \frac{\mu^2}{M^2} $}  & 2   \\
\hline
IB & \begin{minipage}{3cm} \includegraphics[height=3cm]{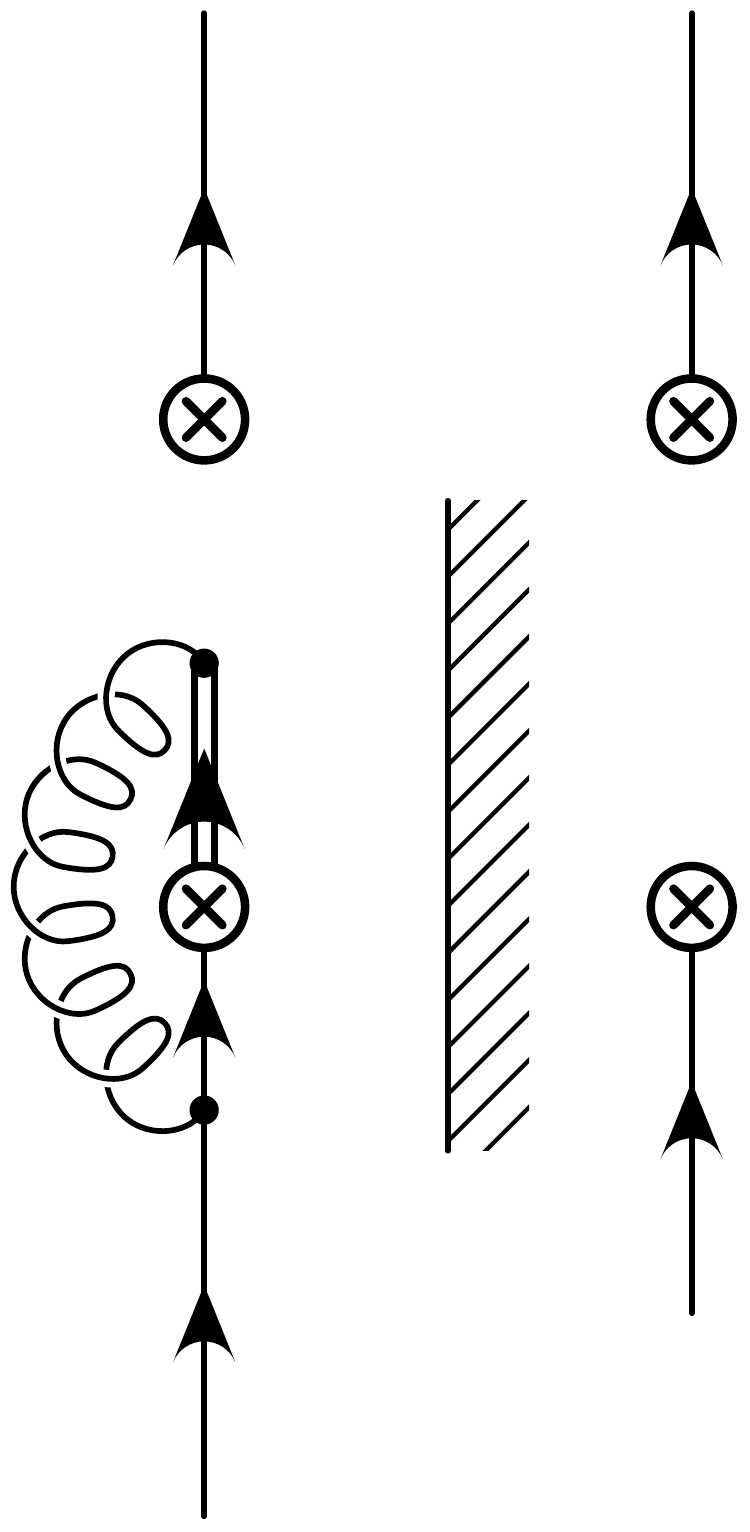} \end{minipage}
 & C_F & C_F &  \big( \ln\frac{\nu}{p^-}+1\big) \delta(1-x)  & \de^2(\mathbf{r}_\perp) \ln \frac{\mu^2}{M^2} & 2 \\
 \hline
 & \multirow{2}{*}{\begin{minipage}{3cm} \includegraphics[height=3cm]{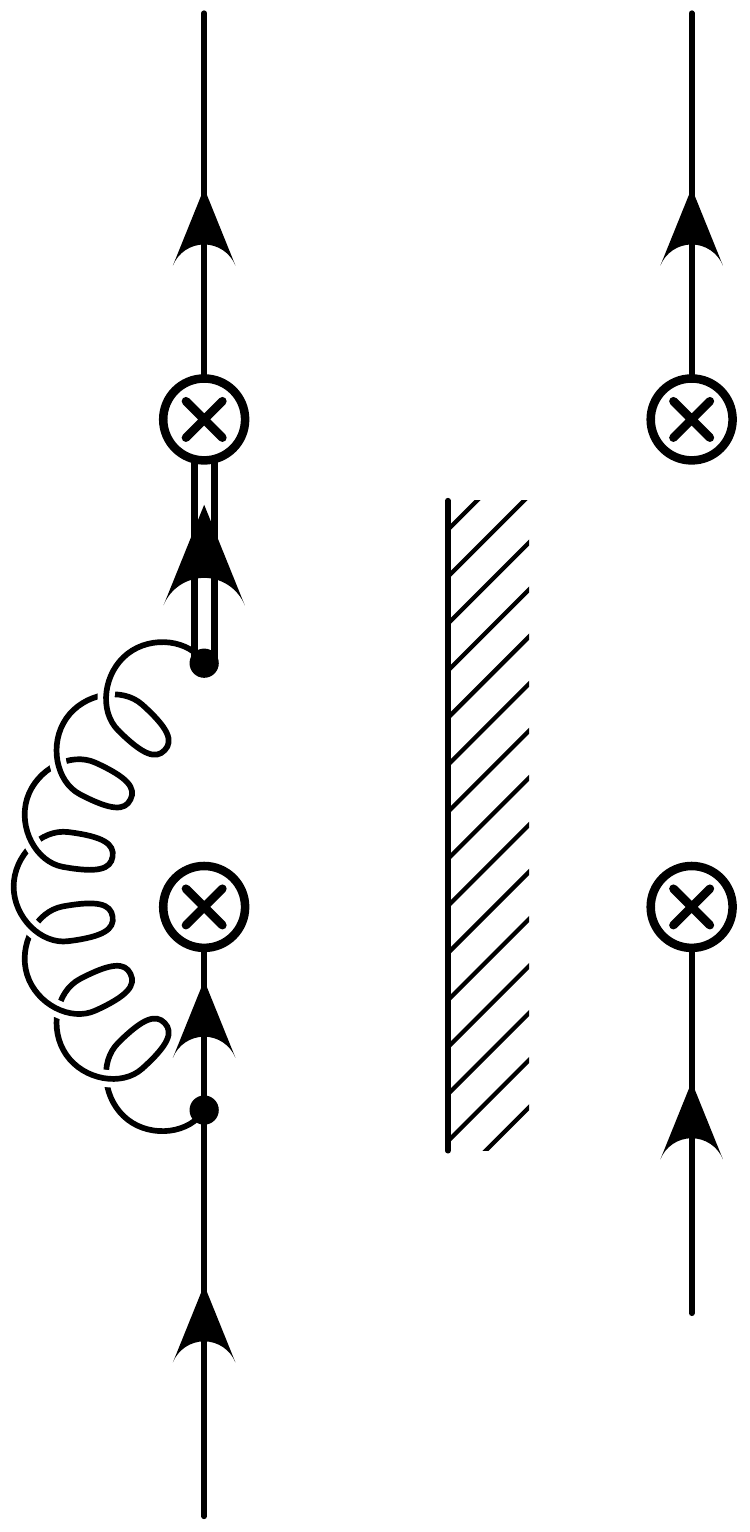}  \end{minipage}} & 0\, ( 1\otimes 1 )&  \frac14(C_A+C_d) T^A \otimes T^A & 0  &  \parbox{20ex}{\ \\$- \frac{1}{\pi} \frac{1}{\mu^2} \frac{1}{(\mathbf{r}_\perp^2/\mu^2)_+}$ \\[5pt]
$- \de^{(2)}(\mathbf{r}_\perp)\ln \frac{\mu^2}{M^2}$}  & 2 \\[0.2cm]
 IB^\prime &  & T^A \otimes T^A &  C_1 1 \otimes 1 &  &
 \\[0.5cm]
 & & & & & & & \\  
\hline 
  II & \multirow{2}{*}{\begin{minipage}{3cm} \includegraphics[height=3cm]{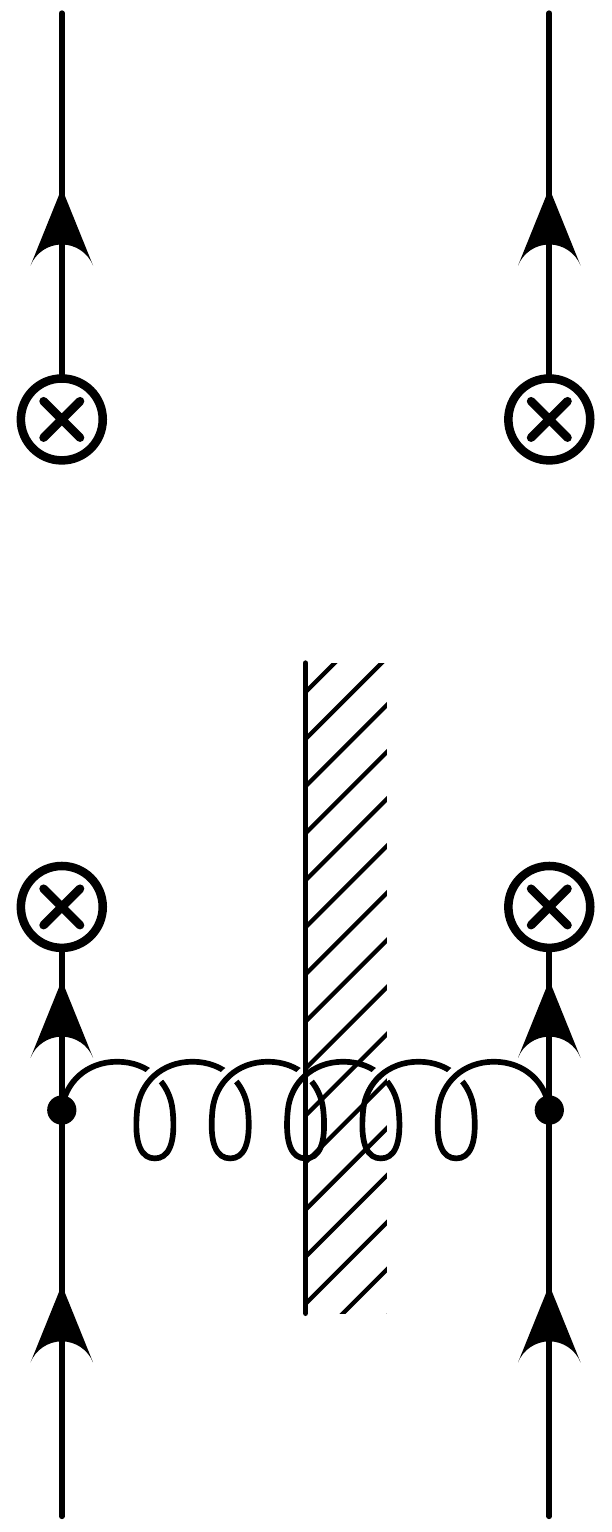} \end{minipage}}
 & 0\, ( 1\otimes 1 )& \frac14(C_A-C_d) T^A \otimes T^A &  \frac12 (1-x)  X & 0 & 1 \\[0.2cm]
 & & -T^A \otimes T^A & -C_1 1 \otimes 1  & & & \\[0.5cm]
  & & & & & & & \\  
\hline
W & \begin{minipage}{2cm} \includegraphics[width=2cm]{fd2} \end{minipage}
& C_F & C_F & \frac12\delta(1-x) & 0 & -1
\end{array}
\end{eqnarray*}
\caption{Diagrams for the interference dPDF. The columns show the graph, the color factors for the $1\otimes 1$ and $T^A \otimes T^A$ color structures (which mix), the $\mu$ and $\nu$ anomalous dimension and the weight factor. The only spin dependence is in the matrix $X$, given in \eq{ispin}.}
\label{tab:iPDF}
\end{table*}
Graph $II$ contains a matrix in spin space, 
\renewcommand{\arraystretch}{1}
\setlength{\arraycolsep}{2pt}
\begin{eqnarray} \label{eq:ispin}
X &=& \left(\begin{array}{ccc|cccc|c}
1 & -1 & 0 & 0 & 0 & 0 & 0 & 0 \\ 
-1 & 1 & 0 & 0 & 0 & 0 & 0 & 0 \\
0 & 0 & 2 & 0 & 0 & 0 & 0 & 0 \\
\hline
0 & 0 & 0 & 1 & 0 & i  & 0 & 0 \\
0 & 0 & 0 & 0 & 1 & 0 & i & 0 \\
0 & 0 & 0 & -i  & 0 & 1 & 0 & 0   \\
0 & 0 & 0 & 0 &  -i  & 0 & 1  & 0   \\
\hline
0 & 0 & 0 & 0 & 0 & 0 & 0 & 0   \\
\end{array}\right)
\end{eqnarray}
where the rows and columns correspond to the spin structures 
$I_{q\bar q}$, $I_{\Delta q \Delta \bar q}$, $I_{\delta q \delta \bar q}$, $I_{q \delta \bar q}$,
$I_{\delta q \bar q}$, $I_{ \Delta q \delta \bar q}$, $I_{ \delta q \Delta \bar q}$, and $I_{\de q \de \bar q}^t$.

Summing up these diagrams, we find that the anomalous dimension is given by
\begin{widetext}
\begin{eqnarray} \label{eq:gaiPDF}
\hat \gamma_\mu^I(x) &=& \Big[2 \ln \frac{\nu}{p^-}  + \frac{3}{2}\Big]\, \de(1-x)
\begin{pmatrix} C_F & 0 \\ 0 & C_F \end{pmatrix}
+\Big[2 \ln \frac{\nu}{p^-} \de(1-x) - \frac{2x}{(1-x)_+} - \frac12 (1-x) X \Big]
\begin{pmatrix} 0 & 1 \\ C_1 & \tfrac{1}{4}(C_d - C_A) \end{pmatrix}\,,  \nn
\hat \gamma_\nu^I(\mathbf{r}_\perp) &=& -\frac{2}{\pi} \frac{1}{\mu^2} \frac{1}{(\mathbf{r}_\perp^2/\mu^2)_+}  
\begin{pmatrix} C_F & 1 \\ C_1 & C_F + \tfrac{1}{4}(C_d - C_A) \end{pmatrix}
\,, 
\end{eqnarray}
\end{widetext}
as a matrix in  color space $\{1\otimes1,T^A \otimes T^A\}$. Repeating the analysis around \eq{gamu} and noting that for $N_c>2$ the matrix multiplying $\ln (\nu/p_1^-)$ has positive eigenvalues, we infer that all the interference contributions are Sudakov suppressed.

\subsection{Interference Soft Function}

In this section we calculate the renormalization of the interference soft functions. The graphs are the same as in \subsec{soft_RGE}, but the color structures are different. We find that the the anomalous dimensions of the interference soft function are
\begin{eqnarray} 
  \ga_\mu^{S_I}(\mathbf{r}_\perp) &=& \frac{\al_s(\mu)}{\pi} \ln \frac{\mu^2}{\nu^2}\, \de^{(2)}(\mathbf{r}_\perp) \begin{pmatrix} 6C_F & -2C_F \\ 2C_F & 10C_F -4C_A \end{pmatrix} 
  \,,\nn 
  \ga_\nu^{S_I}(\mathbf{r}_\perp) &=& \frac{\al_s(\mu)}{\pi^2}\, \frac{1}{\mu^2} \frac{1}{(\mathbf{r}_\perp^2/\mu^2)}_+ \begin{pmatrix} 6C_F & -2C_F \\ 2C_F & 10C_F -4C_A\end{pmatrix} 
.\nn\end{eqnarray}

We will now verify the correctness of this result by checking that the interference contribution to the cross section is $\nu$ independent. Ignoring irrelevant factors,
\begin{widetext}
\begin{eqnarray} \label{eq:gaiS}
  \frac{\df \si}{\df \nu} &\propto& \frac{\df}{\df \nu} 
  \begin{pmatrix} I^{1} & I^{T} \end{pmatrix}
  \begin{pmatrix} S^{11} & C_A(S^{11} - S^{TT}) \\ C_A(S^{11} - S^{TT}) & 2\frac{C_A}{C_F}\, S^{TT} \end{pmatrix}  
  \begin{pmatrix} I^{1} \\ I^{T} \end{pmatrix}  
  \nn
  &\propto& 2\times (-2) \begin{pmatrix} I^{1} & I^{T} \end{pmatrix}
  \begin{pmatrix} S^{11} & C_A(S^{11} - S^{TT}) \\ C_A(S^{11} - S^{TT}) & 2\frac{C_A}{C_F}\, S^{TT} \end{pmatrix}  
  \begin{pmatrix} C_F & 1 \\ C_1 & C_F + \tfrac{1}{4}(C_d - C_A) \end{pmatrix}
  \begin{pmatrix} I^{1} \\ I^{T} \end{pmatrix}  \nn
  && +
 \begin{pmatrix} I^{1} & I^{T} \end{pmatrix}
  \begin{pmatrix} 6C_F S^{11} -2C_F S^{TT}  & C_A[4 C_F S^{11} + (4C_A - 12C_F) S^{TT}]  \\ C_A[4 C_F S^{11} + (4C_A - 12C_F)S^{TT}] & 2\frac{C_A}{C_F}[2 C_F S^{11} + (10C_F -4C_A)S^{TT}] \end{pmatrix}  
  \begin{pmatrix} I^{1} \\ I^{T} \end{pmatrix}  
  = 0
\,,\end{eqnarray}
\end{widetext}
using the anomalous dimensions for the interference dPDFs and soft functions in \eqs{gaiPDF}{gaiS}. This provides a cross-check of our results.

\subsection{One-Loop Soft Function}

We now give expressions for the perturbative one-loop soft functions, which are valid for $\mathbf{r}_\perp^2 \gg \lqcd^2$. From the finite terms in \eqs{Is1}{Is2}, we find
\begin{eqnarray}
S(\mathbf{r}_\perp) &=& (2\pi)^2 \delta^{(2)}(\mathbf{r}_\perp)- \frac{\alpha_s C}{\pi } \bigg\{ 2\pi \Big[ \frac{\ln \mathbf{r}_\perp^2/\nu^2} { \mathbf{r}_\perp^2} \Big]_{\mu^2}
\nn
&&+\frac{\pi^2}{24}\,(2\pi)^2 \delta^{(2)}(\mathbf{r})\bigg\}
\,. \end{eqnarray}
The $[\ ]_{\mu^2}$ distribution is defined so that the integral from $0$ to $\mu^2$ vanishes.
In terms of standard plus distributions,
\begin{eqnarray}
\!\!\!\!\!\!\!\!\Big[ \frac{\ln \mathbf{r}_\perp^2/\nu^2}{ \mathbf{r}_\perp^2} \Big]_{\mu^2} = 
\frac{1}{\mu^2} \Big[ \frac{\ln \mathbf{r}_\perp^2/\mu^2}{ \mathbf{r}_\perp^2/\mu^2} \Big]_+ \!\!\!+\!
\frac{1}{\mu^2} \frac{\ln \mu^2/\nu^2}{(\mathbf{r}_\perp^2/\mu^2)}_+
\!\!\!.\end{eqnarray}
The color factor $C$ for the soft functions in this paper is
\begin{eqnarray}
S^{TT} &:& 2 C_A\,, \nn
S_I^{11} &:& 4 C_F\,, \nn
S_I^{TT} &:& 12 C_F-4 C_A
\,.\end{eqnarray}
The soft function $S_I^{T1} = S_I^{11} - S_I^{TT}$ and only starts at one loop order.

\section{Conclusions}
\label{sec:conc}

In this paper we studied double parton scattering from the point of view of QCD factorization and renormalization. We presented a detailed derivation of the cross section for double Drell-Yan production in terms of double PDFs. Flavor, spin and color correlations, as well as interference effects, lead to a large number of different contributions. In the color-correlated and interference terms, the effects of soft radiation are nontrivial, and are given by  soft functions in the factorization formula. We also derived the QCD evolution of the dPDFs and soft functions, treating both the ultraviolet and rapidity divergences. The solution of these equations shows that color-correlated and interference contributions are Sudakov suppressed, and thus small for double parton scattering at high energies.  We also discussed several basic properties of double PDFs, such as their classification, properties under discrete symmetries and interpretation. In a forthcoming publication~\cite{Manohar:DPS2}, we will discuss the mixing of single PDFs into double PDFs and the related issue of double counting between single and double parton scattering.

\begin{acknowledgments}
We would like to thank C.~Campagnari, F.~Golf, and A.~Yagil for introducing us to double parton scattering, and for helpful discussions, and  A.~Jain and D.~Neill for discussions on the rapidity renormalization group.  We also thank A.~Jain and F.~Tackmann for comments on the manuscript.
This work is supported by DOE grant DE-FG02-90ER40546. 
\end{acknowledgments}

\bibliography{dps}

\end{document}